\documentclass[12pt]{article}
\usepackage{graphics,amssymb,float}
\usepackage{graphicx} 
\usepackage{rotating} 
\usepackage{placeins}
\usepackage{dcolumn}% Align table columns on decimal point
\usepackage{bm}% bold math
\usepackage{cite}
\usepackage{amsmath}
\usepackage{indentfirst} % activate indent in the first paragraph
\usepackage{epstopdf}
\usepackage{color}

\usepackage{graphicx,amssymb,mathrsfs,epstopdf}
\def\met{\mbox{${\hbox{$E$\kern-0.6em\lower-.1ex\hbox{/}}}_T$}}

\def\fbi{~{\rm fb}^{-1}}
\def\f\pbi{~{\rm pb}^{-1}}
\def\eg{{\it e.g.}}
\def\ie{{\it i.e.}}
\def\etc{{\it etc.}}
\def\anti{\overline}

\def\chitz{\widetilde{\chi}^0_2}
\def\chiz{\widetilde{\chi}^0_1}
\def\chipm{\widetilde{\chi}^\pm_1}

\def\st{\tilde{t}}

\def\gev{~{\rm GeV}}

\def\beq{\begin{equation}}
\def\eeq{\end{equation}}
\def\bea{\begin{eqnarray}}
\def\eea{\end{eqnarray}}
\def\gam{\gamma}

\def\omhsq{\Omega_{\tilde \chi_1^0}h^2}
\def\sigsi{\sigma^{\rm SI}(\chiz p)}

\def\br{{\rm BR}}

\def\tanb{\tan\beta}

\def\cnone{\tilde \chi_1^0}

\def\lsim{\mathrel{\raise.3ex\hbox{$<$\kern-.75em\lower1ex\hbox{$\sim$}}}}
\def\gsim{\mathrel{\raise.3ex\hbox{$>$\kern-.75em\lower1ex\hbox{$\sim$}}}}

\begin{document}

\thispagestyle{empty}
\renewcommand{\thefootnote}{\fnsymbol{footnote}}

\begin{center}

\vspace*{-1cm}
\begin{flushright}
LPSC14011\\
UCD-2014-001\\
\end{flushright}

\vspace*{1.4cm}
{\LARGE\bf The phenomenological MSSM\\[2mm] in view of the 125~GeV Higgs data}\\[12mm]

{\large 
B.~Dumont$^{\,a}$\footnote[1]{Email: dumont@lpsc.in2p3.fr}, 
J.\,F.~Gunion$^{\,b}$\footnote[2]{Email: jfgunion@ucdavis.edu},  
S.~Kraml$^{\,\,a}$\footnote[3]{Email: sabine.kraml@lpsc.in2p3.fr}, 
%S.~Sekmen$^{\,c}$\footnote[4]{Email: ssekmen@mail.cern.ch}
}\\[6mm]

{\it
$^{a}$\,Laboratoire de Physique Subatomique et de Cosmologie, UJF Grenoble 1,\\ 
CNRS/IN2P3, INPG, 53 Av.\ des Martyrs, F-38026 Grenoble, France\\[2mm]
$^{b}$\,Department of Physics, University of California, Davis, CA 95616, USA\\[2mm]
%$^{c}$\,CERN, CH-1211 Geneva 23, Switzerland\\[2mm]
}

%\vspace*{1.4cm}
%{\sc Draft}\\
%\today

\vspace*{1.4cm}

\abstract{
The parameter space of the phenomenological MSSM (pMSSM) is explored by means of Markov Chain Monte Carlo (MCMC) methods, taking into account the latest LHC results on the Higgs signal at 125 GeV in addition to relevant low-energy observables and LEP constraints. We use a Bayesian approach to derive posterior densities for the parameters and observables of interests. We find in particular that the Higgs measurements have a significant impact on the parameters $\mu$ and $\tan \beta$ due to radiative corrections to the bottom Yukawa coupling. We show moreover the impact of the most recent dark matter measurements on the probability distributions, and we discuss prospects for the next run of the LHC at 13--14~TeV.
}

\end{center}

\renewcommand{\thefootnote}{\arabic{footnote}}

\newpage

%==========================================================================
\section{Introduction}
%==========================================================================

The recent discovery~\cite{Aad:2012tfa,Chatrchyan:2012ufa} of a new particle with mass of 125~GeV 
and properties consistent with the Standard Model (SM) Higgs boson is clearly the most significant news 
from the Large Hadron Collider (LHC).
Thanks to the various production and decay modes that are accessible for such a light Higgs, 
and thanks to the excellent LHC operation, 
many complementary measurements of signal strengths, 
defined as $\mu\equiv (\sigma\times {\rm BR})/(\sigma\times {\rm BR})_{\rm SM}$  
for each production$\times$decay mode, 
are available from the 7--8~TeV LHC run~\cite{Aad:2013wqa,ATLAS-CONF-2013-014,CMS-PAS-HIG-13-005}. 
These make it possible to determine the properties of the observed new state  
with good precision (see \cite{Belanger:2013xza} for a global fit to the latest data).

So the Higgs has been found---but where is new physics? 
Indeed, the other significant news from the LHC is, unfortunately, the absence of any compelling sign of new physics. 
In particular, there is no hint of supersymmetry (SUSY), one of the most thoroughly studied ideas for physics beyond the SM (see, for example,  Refs.~\cite{Martin:1997ns,Chung:2003fi} for recent reviews). 

As is well known, in the Minimal Supersymmetric Standard Model (MSSM) a light Higgs mass of the order of 125~GeV requires that stops be either very heavy or near-maximally mixed.
In addition to a modification of the electroweak symmetry breaking (EWSB) sector by the presence of a second Higgs doublet, the MSSM predicts a wealth of new particles that couple to the light Higgs boson. These can, depending on their masses and mixings, modify the Higgs couplings and consequently the production and decay rates of the Higgs boson in various channels. It is thus interesting to ask whether, besides the measured Higgs mass, the Higgs {\em signal strengths} provide constraints on the MSSM and may thus be used as a guide for where to look for SUSY. 

Indeed, the apparent excess in the diphoton channel reported by both ATLAS and CMS 
in 2012~\cite{Aad:2012tfa,Chatrchyan:2012ufa} 
motivated scenarios with light staus in the MSSM \cite{Carena:2012gp} or 
small $\tan\beta$/large $\lambda$ in the next-to-MSSM \cite{Ellwanger:2011aa,Gunion:2012zd} 
(see also \cite{Kraml:2013wna,Arbey:2012bp}). 
This drastically changed with the updated results presented at the Moriond 2013 conference and thereafter, 
which point towards a very SM-like Higgs boson, without the need of any modifications of the couplings due 
to new, beyond-the-SM particles.  

The implications of the latest Higgs data for the MSSM were discussed recently in \cite{Djouadi:2013uqa,Cahill-Rowley:2013vfa}. Ref.~\cite{Djouadi:2013uqa} concentrated on describing (the consequences for) the heavy Higgs 
states in the limit of heavy SUSY particles; the best coupling fit was found at low $\tan\beta$, $\tan\beta\approx 1$,  
with a not too high CP-odd Higgs mass of $m_A\approx 560$~GeV. 
Ref.~\cite{Cahill-Rowley:2013vfa} analyzed the consequences of the SUSY null-searches on the one hand 
and of the measurements of the Higgs properties on the other hand based on flat random scans of the so-called phenomenological MSSM (pMSSM) with the conclusion that SUSY searches and Higgs boson properties are to a very good approximation orthogonal.  More concretely, Ref.~\cite{Cahill-Rowley:2013vfa}
concluded that Higgs coupling measurements at the 14~TeV LHC, and particularly at a 500~GeV ILC, will be sensitive to regions of the pMSSM space that are not accessible to direct SUSY searches.  

In this paper, we follow a different approach. Performing a Bayesian analysis of  the pMSSM parameter space by means of a Markov Chain Monte Carlo analysis, we investigate how the latest LHC results on the properties of the 125 GeV Higgs state impact the probability distributions of the pMSSM parameters, masses
and other observables.
In doing so, we carefully take into account all available information on the 
production$\times$decay processes\footnote{The importance of considering the distinct production$\times$decay processes---instead of just the decay modes---was recently emphasized in \cite{Heinemeyer:2013tqa,Boudjema:2013qla}.} on top of constraints from LEP searches and low-energy observables.
In addition, we explore consequences for our probability distributions from the latest dark matter constraints 
and discuss prospects for measurements of the Higgs signal at the next run of the LHC at 13--14~TeV.
Our results are orthogonal and directly comparable to the pMSSM interpretation of the CMS SUSY searches~\cite{SUS-12-030,SUS-13-020}.

%==========================================================================
\section{Analysis}
%==========================================================================

%-------------------------------------------------------------------------------------------
\subsection{Definition of the phenomenological MSSM (pMSSM)}
\label{sec:pmssm}
%-------------------------------------------------------------------------------------------

The purpose of this study is to assess what current Higgs data tell us, and do not tell us, about the 
MSSM at the weak scale, without any assumption as to the SUSY-breaking scheme.  
A priori, the weak-scale MSSM has 120 free parameters, assuming that $R$-parity is conserved (to avoid proton decay and to ensure that the lightest SUSY particle, the lightest supersymmetric particle (LSP), is stable) and assuming that the gravitino is heavy. 
This is clearly too much for any phenomenological study. 
However, most of these parameters are associated with CP-violating phases and/or flavor changing neutral currents (FCNC), which are severely constrained by experiment.  A few reasonable assumptions about the flavor and CP structure therefore allow us to reduce the number of free parameters by a factor 6, without imposing any SUSY-breaking scheme. Working with parameters defined at the weak scale is indeed of great advantage for our purpose, because models of SUSY breaking always introduce relations between the soft terms that need not  hold in general. 

Concretely, the only generic way to satisfy very strong constraints on CP violation is to take all parameters to be real. 
FCNC constraints are satisfied in a generic way by taking all sfermion mass matrices and trilinear couplings to be 
flavor-diagonal. As a further simplification, the various independent  sfermion masses for the 2nd generation  are taken to be equal to their counterparts for the 1st generation. 
Regarding the trilinear $A$-terms of the first two generations, these only enter phenomenology multiplied by the associated very small Yukawa couplings and are thus not experimentally relevant unless unreasonably large. Only the 3rd generation
parameters $A_t$, $A_b$ and $A_\tau$ have observational impact.

This leaves us with 19 real, weak-scale SUSY Lagrangian parameters---the
so-called p(henomenological) MSSM~\cite{Djouadi:1998di}. 
As mentioned, the pMSSM captures most of the phenomenological 
features of the R-parity conserving MSSM and, most importantly, encompasses and goes beyond a broad 
range of more constrained SUSY models.  
The free parameters of the pMSSM are the following: 
\begin{itemize}
   \item the gaugino mass parameters $M_1$, $M_2$, and $M_3$; 
   \item the ratio of the Higgs vacuum expectation values (VEVs), $\tan\beta=v_2/v_1$;
   \item the higgsino mass parameter $\mu$ and 
            the pseudo-scalar Higgs mass $m_A$;
    \item 10 sfermion mass parameters $m_{\tilde{F}}$, where 
         $\tilde{F} = \tilde{Q}_1, \tilde{U}_1, \tilde{D}_1, 
                      \tilde{L}_1, \tilde{E}_1, 
                      \tilde{Q}_3, \tilde{U}_3, \tilde{D}_3, 
                      \tilde{L}_3, \tilde{E}_3$\\ 
(with 2nd generation sfermion masses equal to their 1st generation counterparts, {\it i.e.} $m_{\tilde{Q}_1}\equiv m_{\tilde{Q}_2}$, 
           $m_{\tilde{L}_1}\equiv m_{\tilde{L}_2}$, {\it etc.}), and          
   \item the trilinear couplings $A_t$, $A_b$ and $A_\tau$\,,               
\end{itemize}
in addition to the SM parameters.  
To minimize theoretical uncertainties in the Higgs sector, these parameters are conveniently defined 
at the scale $M_{\rm SUSY} \equiv \sqrt{m_{\tilde t_1}m_{\tilde t_2}}$, often also referred to as the 
EWSB scale.

The pMSSM parameter space is constrained by a number of theoretical requirements.  
In particular, the Higgs potential must be bounded from below and lead to consistent EWSB, and 
the sparticle spectrum must be free of tachyons.
Moreover, in this study, we require that the LSP is the lightest neutralino, $\tilde\chi^0_1$. 
These requirements we refer to as theoretical constraints. 
Note that we do not check for charge and/or color breaking minima beyond warnings from the spectrum generator; this could be done, \eg, using {\tt Vevacious}~\cite{Camargo-Molina:2013qva}, but would require too much CPU time for this  study.

%--------------------------------------------------------------------------------------
\subsection{Construction of the pMSSM prior} 
\label{sec:prior}
%--------------------------------------------------------------------------------------

We perform a global Bayesian analysis that yields posterior probability densities of model parameters, masses and observables.  
We allow the pMSSM parameters to vary within the following ranges:
\begin{eqnarray}
\nonumber &-3\,{\rm TeV} \le M_1,\, M_2,\, \mu \le 3\,{\rm TeV}\,;& \\
\nonumber &0 \le M_3,  m_{\tilde{F}}, m_A \le 3\,{\rm TeV}\,;& \\
\nonumber &-7\,{\rm TeV} \le A_t, A_b, A_\tau \le 7\,{\rm TeV}\,;& \\
 &2 \le \tan\beta \le 60\,.& 
\label{eq:subspace}
\end{eqnarray}
A point in this space will be denoted by $\theta$. In addition, we treat the 
SM parameters $m_t$, $m_b(m_b)$ and $\alpha_s(M_Z)$ as nuisance parameters, 
constrained with a likelihood.
For each pMSSM point, we use  
{\tt SoftSUSY\_3.3.1}~\cite{Allanach:2001kg} to compute the SUSY spectrum,
{\tt SuperIso\_v3.3}~\cite{Mahmoudi:2008tp} to compute the low-energy constraints, and 
{\tt micrOMEGAs\_2.4.5}~\cite{Belanger:2001fz} to compute the neutralino relic density $\omhsq$, direct detection cross sections and to check compatibility with various pre-LHC sparticle mass limits.
Moreover, we use 
{\tt SDECAY\_1.3b}~\cite{Muhlleitner:2003vg} and {\tt HDECAY\_5.11}~\cite{Djouadi:1997yw} to produce SUSY and Higgs decay tables.
The various codes are interfaced using the SUSY Les Houches Accord~\cite{Skands:2003cj}. 

The posterior density of $\theta$ given data $D$ is given by
\begin{equation}
p(\theta | D) \sim L(D | \theta)\, p_0(\theta)\,,
\label{pchain}
\end{equation}
where $L(D | \theta)$ is the likelihood and $p_0(\theta)$ is the prior probability density, or
prior for short. Beginning with a flat distribution in the parameters within the ranges defined by Eq.~(\ref{eq:subspace}),  $p_0(\theta)$ is obtained by incorporating  the theoretical constraints noted above.  In other words, $p_0(\theta)$ is the result of sculpting the flat parameter distributions by the requirements related to theoretical consistency and $\tilde{\chi}^0_1$ being the LSP. 
This $p_0(\theta)$ defines the starting prior, which will be modified by actual data using  Eq.~(\ref{pchain}).    
Since we consider multiple independent 
measurements $D_i$, the combined likelihood is given by $L(D | \theta) = \prod_{i}L(D_i | \theta)$.  

We partition the data into two parts: 
\begin{enumerate}
\item a set of constraints, listed in Table~\ref{tab:preHiggs}, which are independent of the Higgs measurements; these constraints are used for the MCMC sampling and are collectively referred to by the label ``preHiggs'', and
\item the Higgs  measurements, which include the Higgs mass 
window, $m_h=123-128$~GeV, and the signal strength likelihood as derived in \cite{Belanger:2013xza}.
\end{enumerate}
With this partitioning, the posterior density becomes
\begin{equation}
p(\theta | D) \sim L(D^{\rm Higgs} | \theta) \, L(D^{\rm preHiggs} | \theta) \, p_0(\theta) = L(D^{\rm Higgs} | \theta) \, p^{\rm preHiggs}(\theta)\,,
\end{equation}
where $p_0(\theta)$ is the prior (as defined earlier) at the start of the inference chain and $p^{\rm preHiggs}(\theta) \sim L(D^{\rm preHiggs} | \theta) \, p_0(\theta)$ can be viewed as a prior that encodes the information from the preHiggs-measurements as well as the theoretical consistency requirements.  This partitioning allows us to assess the impact of the Higgs  results on the pMSSM parameter space while being consistent with constraints from the previous measurements. Note that at this stage we do not consider the direct limits from SUSY 
searches from ATLAS or CMS.

\begin{sidewaystable}[tbp]
\caption{The measurements that are the basis of our pMSSM prior $p^{\rm preHiggs}(\theta)$.  
All measurements were used to sample points from the pMSSM parameter space via MCMC methods. 
The likelihood for each point was reweighted post-MCMC based on better determinations of 
$\br(b \rightarrow s\gamma)$, $\br(B_s \rightarrow \mu \mu)$, $R(B_u \rightarrow \tau \nu)$, and $m_t$. }
\begin{center}
\begin{tabular}{|c|c|c|c|c|}
\hline
$i$     & Observable    & Constraint   & Likelihood function & MCMC / \\
        & $\mu_j(\theta)$               & $D^{\rm preHiggs}_j$             &  $L(D^{\rm preHiggs}_j|\mu_j(\theta))$  & post-MCMC \\
\hline\hline
1a & $\br(b \rightarrow s\gamma)$~\cite{Amhis:2012bh,Misiak:2006zs} & $(3.55 \pm 0.24^{\rm stat} \pm 0.23^{\rm th} \pm 0.09^{\rm sys})\times 10^{-4}$ & Gaussian & MCMC \\
1b & $\br(b \rightarrow s\gamma)$~\cite{HFAG2013} & $(3.43 \pm 0.21^{\rm stat} \pm 0.23^{\rm th} \pm 0.07^{\rm sys})\times 10^{-4}$ & Gaussian & reweight \\
\hline
2a & $\br(B_s \rightarrow \mu \mu)$~\cite{oldbsmm} & observed ${\rm CL}_s$ curve from \cite{oldbsmm} & $d(1 - {\rm CL}_s)/d(BR(B_s \rightarrow \mu\mu))$ & MCMC \\
2b & $\br(B_s \rightarrow \mu \mu)$~\cite{CMSandLHCbCollaborations:2013pla} & $(2.9 \pm 0.7 \pm 0.29^{\rm th})\times 10^{-9}$ & Gaussian & reweight \\
\hline
3a & $R(B_u \rightarrow \tau \nu)$\cite{Nakamura:2010zzi} & $1.63\pm 0.54$ & Gaussian & MCMC \\
3b & $R(B_u \rightarrow \tau \nu)$\cite{HFAG2013} & $1.04\pm 0.34$ & Gaussian & reweight \\
\hline
4 & $\Delta a_\mu$~\cite{Hagiwara:2011af} & $(26.1 \pm 8.0^{\rm exp}\pm 10.0^{\rm th})\times 10^{-10}$ & Gaussian & MCMC \\
\hline
5a & $m_t$~\cite{topmasslhc} & $173.3\pm0.5^{\rm stat}\pm1.3^{\rm sys}$~GeV & Gaussian & MCMC \\
5b & $m_t$~\cite{CDF:2013jga} & $173.20\pm0.87$~GeV & Gaussian & reweight \\
\hline
6 & $m_b(m_b)$~\cite{Nakamura:2010zzi} & $4.19^{+0.18}_{-0.06}$~GeV & Two-sided Gaussian & MCMC \\
\hline
7 & $\alpha_s(M_Z)$~\cite{Nakamura:2010zzi} & $0.1184 \pm 0.0007$ & Gaussian & MCMC \\
\hline
8 & sparticle & LEP \cite{lepsusy} & 1 if allowed & MCMC \\
   & masses & (via micrOMEGAs~\cite{Belanger:2001fz}) & 0 if excluded & \\
\hline
%9 & prompt $\tilde{\chi}_1^\pm$ & $c\tau(\tilde{\chi}_1^\pm) < 10$~mm & 1 if allowed & post-MCMC \\
%     &  &  & 0 if excluded & \\
%\hline
\end{tabular}
\label{tab:preHiggs}
\end{center}
\end{sidewaystable}

In addition to the experimental results included in our calculation of the prior $p^{\rm preHiggs}
(\theta)$, Table~\ref{tab:preHiggs}  lists the corresponding likelihood $L(D^{\rm preHiggs}_{j} | \mu_j(\theta))$ for each observable $j$, where $\mu_j(\theta)$ denotes the model prediction for the observable $j$,  such as $\br(b \rightarrow s\gamma)$ for a given $\theta$.    
We obtained a discrete representation of the prior $p^{\rm preHiggs}(\theta)$ within the
sub-space defined in Eq.~(\ref{eq:subspace}) by sampling points from $p^{\rm preHiggs}(\theta)$ using a MCMC method (for an introduction see, \eg, \cite{Trotta:2008qt}).  
By construction, this method produces a sample of points whose density in the neighborhood  of 
$\theta$ is $\propto p^{\rm preHiggs}(\theta)$, \ie\ the sampled points will constitute a
discrete representation of the preHiggs likelihood as a function of the pMSSM parameters
$\theta$. 

Our study is based on approximately $2\times 10^6$ MCMC points, which were originally sampled  
for the CMS study \cite{SUS-12-030} in which some of us participated. 
(The CMS study then used a random sub-sample of 7205 points from this data.) 
In the meanwhile, several experimental constraints that enter the preHiggs likelihood function have been updated. 
For example, first evidence for the decay $B_s \rightarrow \mu \mu$ was reported by the LHCb collaboration in~\cite{Aaij:2012nna} and recently new improved measurements have become available by CMS and LHCb~\cite{CMSandLHCbCollaborations:2013pla}.  We have taken the up-to-date value into account by reweighting each sampled point by the ratio of 
the new $\br(B_s \rightarrow \mu \mu)$ likelihood, 2b,  to the old likelihood, 2a, in Table~\ref{tab:preHiggs}. 
Analogous reweighting was performed to take into account the updated values of $\br(b \rightarrow s\gamma)$,
$R(B_u \rightarrow \tau \nu)$, and $m_t$.

%--------------------------------------------------------------------------------------
\subsection{Higgs likelihood} 
\label{sec:higgslikeli}
%--------------------------------------------------------------------------------------

For fitting the properties of the observed Higgs boson, 
we use all the publicly available results on the signal strengths $ \mu(X,Y)$ relative to SM expectations,\footnote{In the following, signals strengths are always denoted as $\mu({\rm process})$ in order to avoid confusion with the supersymmetric Higgs mass parameter, denoted as $\mu$ alone.} 
\begin{equation}
   \mu(X,Y)\equiv {\sigma(X)\br(H\to Y)\over \sigma(X_{\rm SM} )\br(H_{\rm SM}\to Y)}\,,
   \label{eq:mu}
\end{equation}
published by the ATLAS and CMS collaborations.\footnote{Later the generic $H$ of Eq.~(\ref{eq:mu}) will of course be the light MSSM Higgs, $h$.}  
Here, $X$ denotes the fundamental production mechanisms: gluon fusion (ggF, $gg\to H$),  
vector-boson fusion (VBF, $WW/ZZ \to H$), and production in association with a $Z$ or $W$ boson (VH) or with a pair of top quarks (ttH);
and $Y$ denotes the Higgs decay final states ($Y=\gam\gam$, $ZZ^*$, $WW^*$, $b\bar b$ and $\tau\tau$ are currently accessible). 
Concretely, we consider the results in the (ggF+ttH) versus  (VBF+VH) production plane for the $H\to \gamma\gamma$, $VV^* \equiv (ZZ^*, WW^*)$, and $\tau\tau$ decay modes.
Moreover, we consider the results in the $b\bar b$ final state from vector boson associated production, as well as the ATLAS limits on invisible decays from $ZH$ associated production with $Z\to \ell^+\ell^-$ and $H \to {\rm invisible}$. 
All these results are combined into the ``Higgs signal likelihood'' $L(D^{\rm Higgs} | \theta)$ in the form of 
$e^{-\chi^2_h/2}$, with the total $\chi^2$ from the Higgs signal, $\chi^2_h$, computed using the global fitting program developed in \cite{Belanger:2013xza}.  (For details on the computation, we refer the interested reader to \cite{Belanger:2013xza}.)

For the concrete calculation, we use {\tt HDECAY\_5.11} and approximate $\sigma(gg \to h)/\sigma(gg \to H_{\rm SM}) \simeq \Gamma(h \to gg)/\Gamma(H_{\rm SM} \to gg)$. Moreover,  for computing the SM results entering  Eq.~(\ref{eq:mu}), we use the MSSM decoupling limit with $m_A$ and the relevant SUSY masses set to 4~TeV. This ensures completely SM-like Higgs boson couplings at tree-level, as well as vanishing  
radiative contributions from the SUSY particles (including non-decoupling effects). We choose this procedure in order to guarantee that the radiative corrections being included are precisely the same for the numerator and denominator in Eq.~(\ref{eq:mu}).

For completeness, we also take into account the limits  
from the $H,A\to\tau\tau$ searches in the MSSM~\cite{CMS-PAS-HIG-13-021}. These limits are implemented in a binary fashion: we set the likelihood from each of these constraints to 1 when the 95\% CL limit is obeyed and to 0 when it is violated. (Including or not including this limit however has hardly any visible effect on the posterior distributions.)

%--------------------------------------------------------------------------------------
\subsection{Dark matter constraints} 
\label{sec:dmconst}
%--------------------------------------------------------------------------------------

The calculation of the properties of the neutralino LSP as a thermal cold dark matter (DM) candidate 
(or one of the cold DM components) depends on a number of cosmological assumptions, like 
complete thermalization, no non-thermal production, no late entropy production, \etc\ 
In order to be independent of these assumptions, 
we will show results with and without requiring consistence with DM constraints. 
When we do apply DM constraints, we adopt the following procedure.
For the relic density, we apply an upper bound as a smoothed step function at the Planck 
value of $\Omega h^2=0.1189$~\cite{Ade:2013zuv}, accounting for a 10\% theory-dominated uncertainty. 
Concretely, we take
\begin{equation}
L=
\left\lbrace
\begin{array}{ccc}
1 & \mbox{if} & \Omega h^2 < 0.119\,, \\
\exp[(0.119-\Omega h^2)/0.012)^2 / 2] & \mbox{if} & \Omega h^2 > 0.119\,.\end{array}\right.
\end{equation}
For  the spin-independent scattering cross section off protons, we use the 90\%~CL limit from LUX~\cite{Akerib:2013tjd}, 
rescaling the computed $\sigsi$ by a factor $\xi=\omhsq/0.119$ to account for the lower local density when the neutralino is only part of the DM. (The alternative would be to assume that the missing amount of $\omhsq$ is substituted by non-thermal production, which would make the direct detection constraints more severe. 
Our approach is more conservative in the sense of not being overly restrictive.)

%--------------------------------------------------------------------------------------
\subsection{Prompt chargino requirement} 
\label{sec:promptcharg}
%--------------------------------------------------------------------------------------

Before presenting the sampled distributions, another comment is in order. 
Letting $M_1$, $M_2$ and $\mu$, vary freely over the same range implies that about $2/3$ of the time 
$M_2$ or $\mu$ will be the smallest mass parameter in the neutralino mass matrix. 
This implies that  in a considerable portion of the pMSSM parameter space the 
$\tilde{\chi}_1^{\pm}$ and $\tilde{\chi}_2^0$ are close in mass or almost degenerate 
with the LSP, $\tilde{\chi}_1^0$~\cite{Gunion:1987yh}.
When the $\tilde{\chi}_1^{\pm}$--$\tilde{\chi}_1^0$ mass difference becomes very small, 
below about 300~MeV,  
the charginos are long-lived and can traverse the detector before they decay.  
This typically occurs for wino-LSP scenarios with $|M_2|\ll |M_1|,\, |\mu|$. 
Since long-lived heavy charged particles were not considered in the SUSY searches used in \cite{SUS-12-030}, 
charginos were required to decay promptly; in practice this means a cut on the average proper lifetime of $c\tau < 10$~mm. 
In order to be able to directly compare our results (based on the Higgs measurements) 
with the CMS study (based on SUSY search results) \cite{SUS-12-030} and its up-coming update \cite{SUS-13-020}, 
we also require ``prompt'' chargino decays, \ie\  $c\tau < 10$~mm. 
Most of our conclusions are insensitive to this requirement.
Wherever it matters, we will however also show the results obtained without imposing the 
%the prompt chargino requirement.
$c\tau$ cut.

%==========================================================================
\section{Results}
%==========================================================================

%----------------------------------------------------------------------------------------------------------------------------------
\subsection{Pre-Higgs distributions and impact of the Higgs mass}\label{sec:YellowPlots}
%----------------------------------------------------------------------------------------------------------------------------------

\begin{figure}[htbp]
\begin{center}
\includegraphics[width=0.24\linewidth]{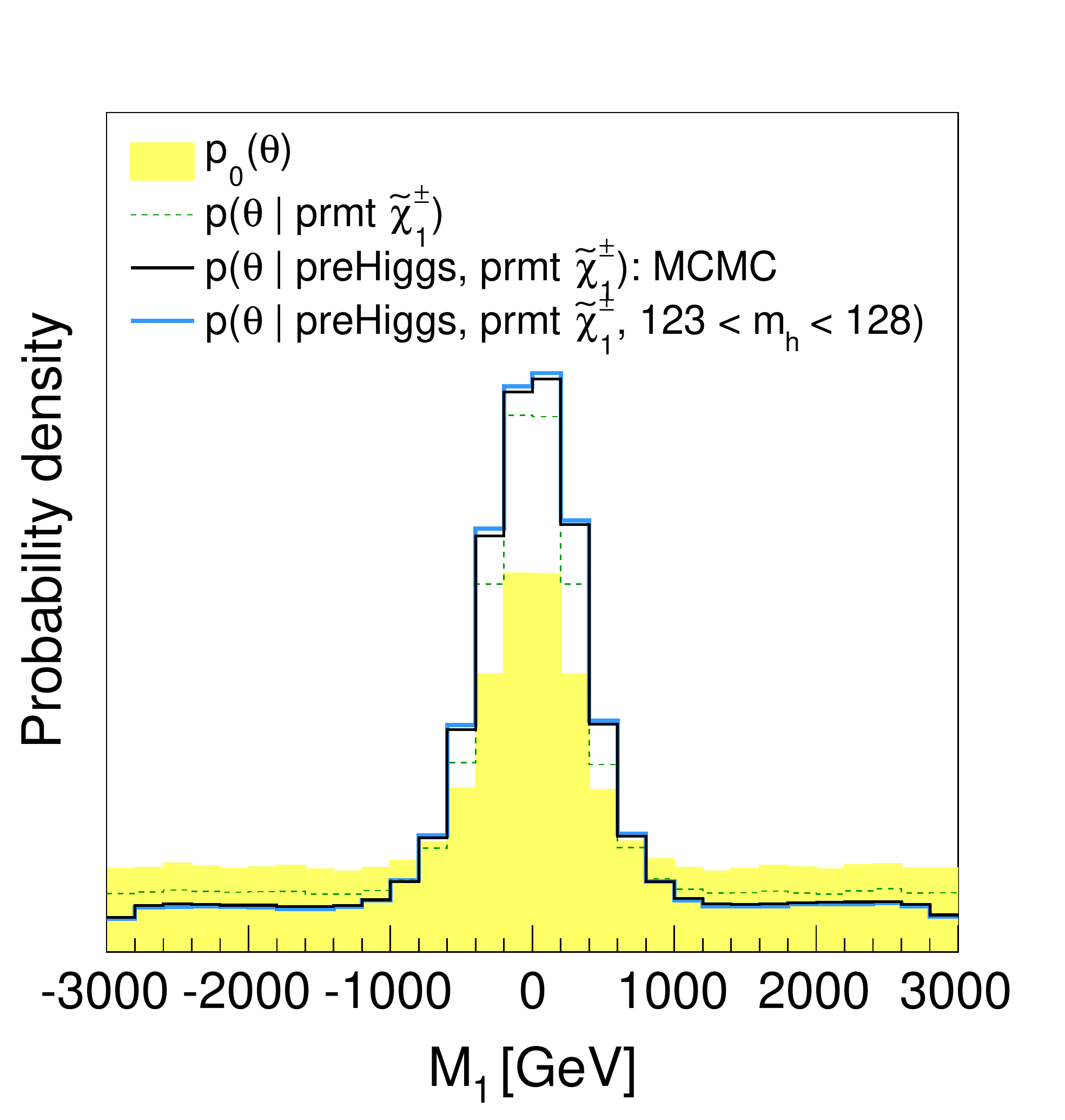}
\includegraphics[width=0.24\linewidth]{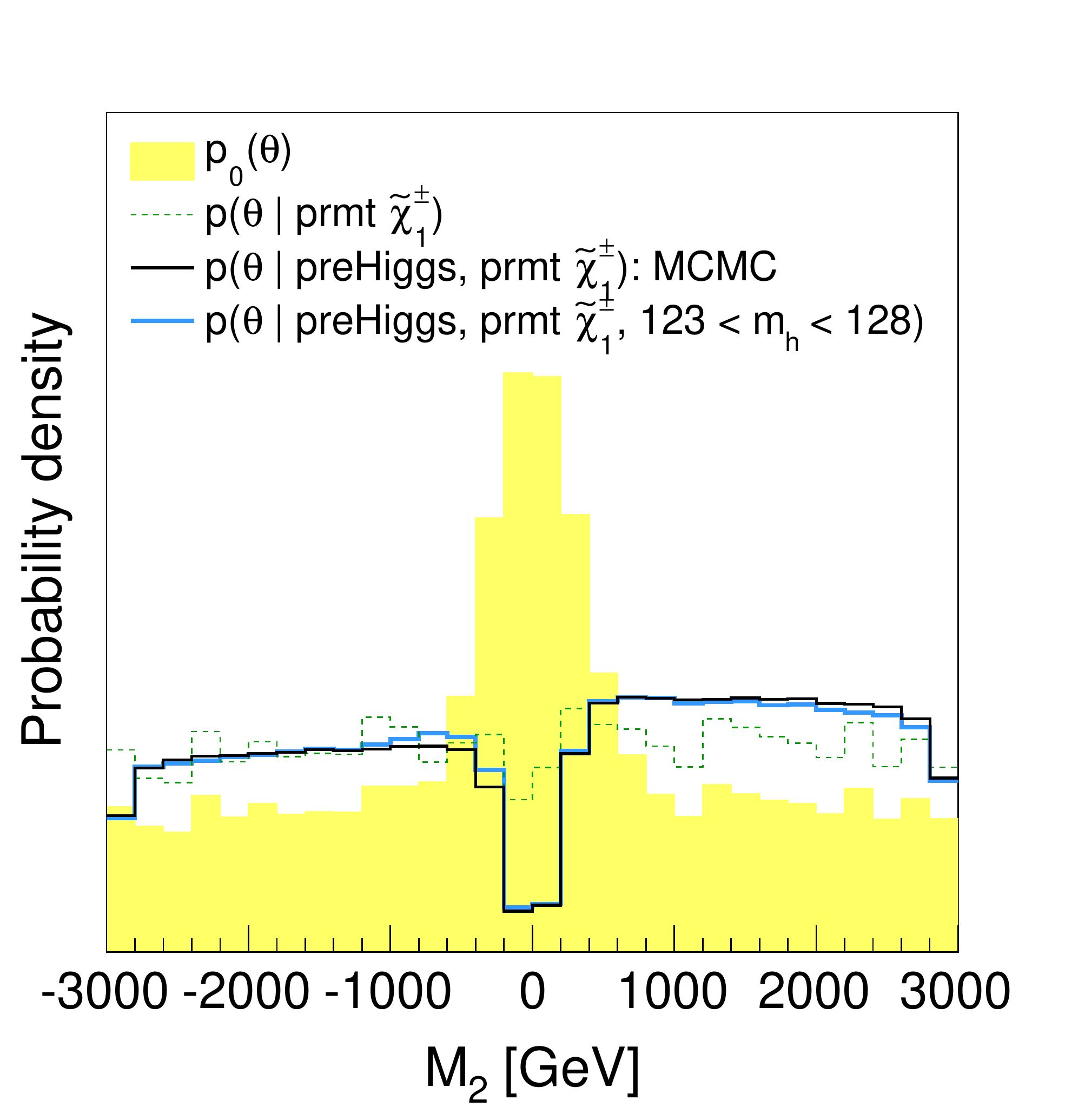}
\includegraphics[width=0.24\linewidth]{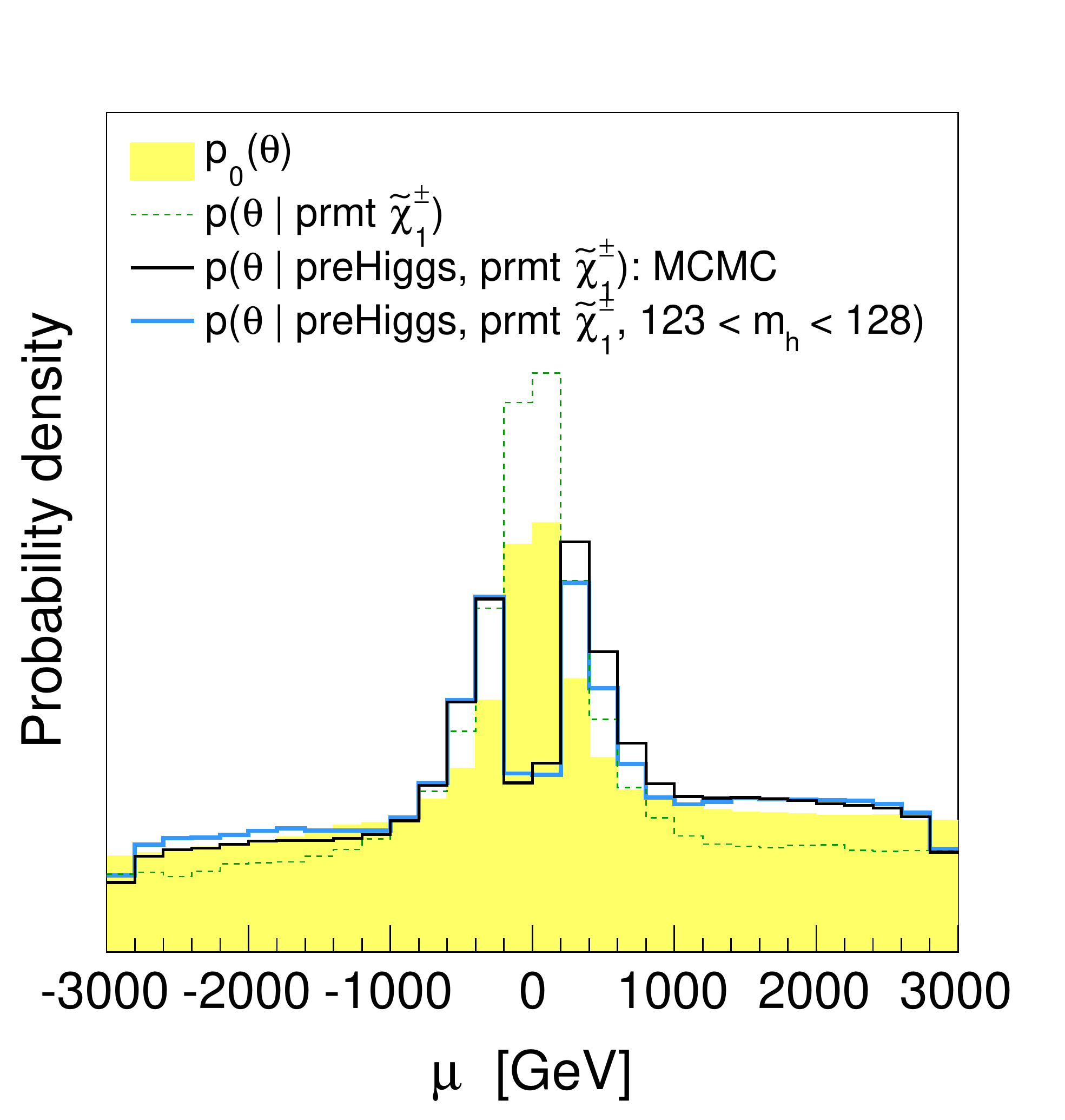} 
\includegraphics[width=0.24\linewidth]{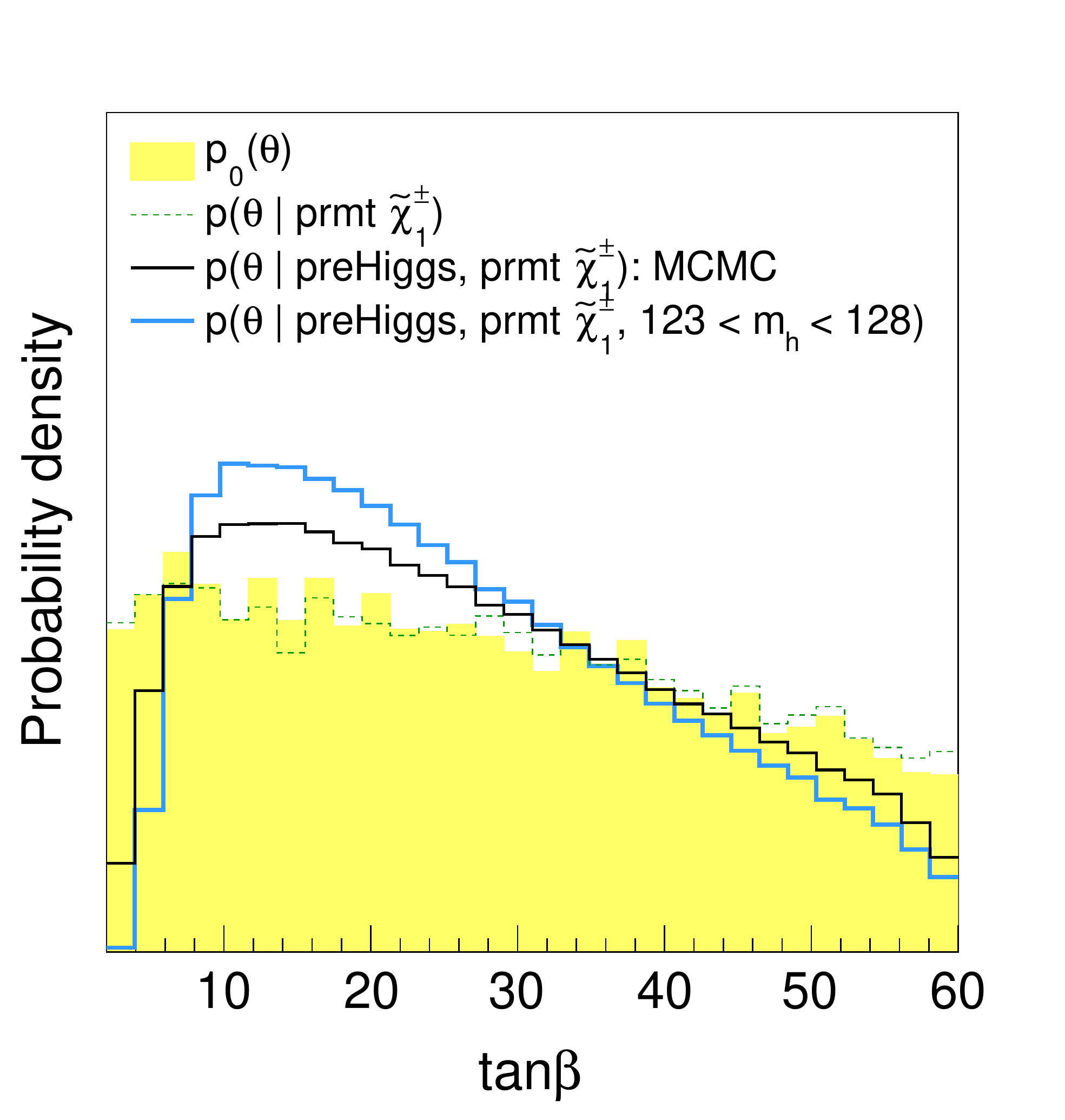}
\includegraphics[width=0.24\linewidth]{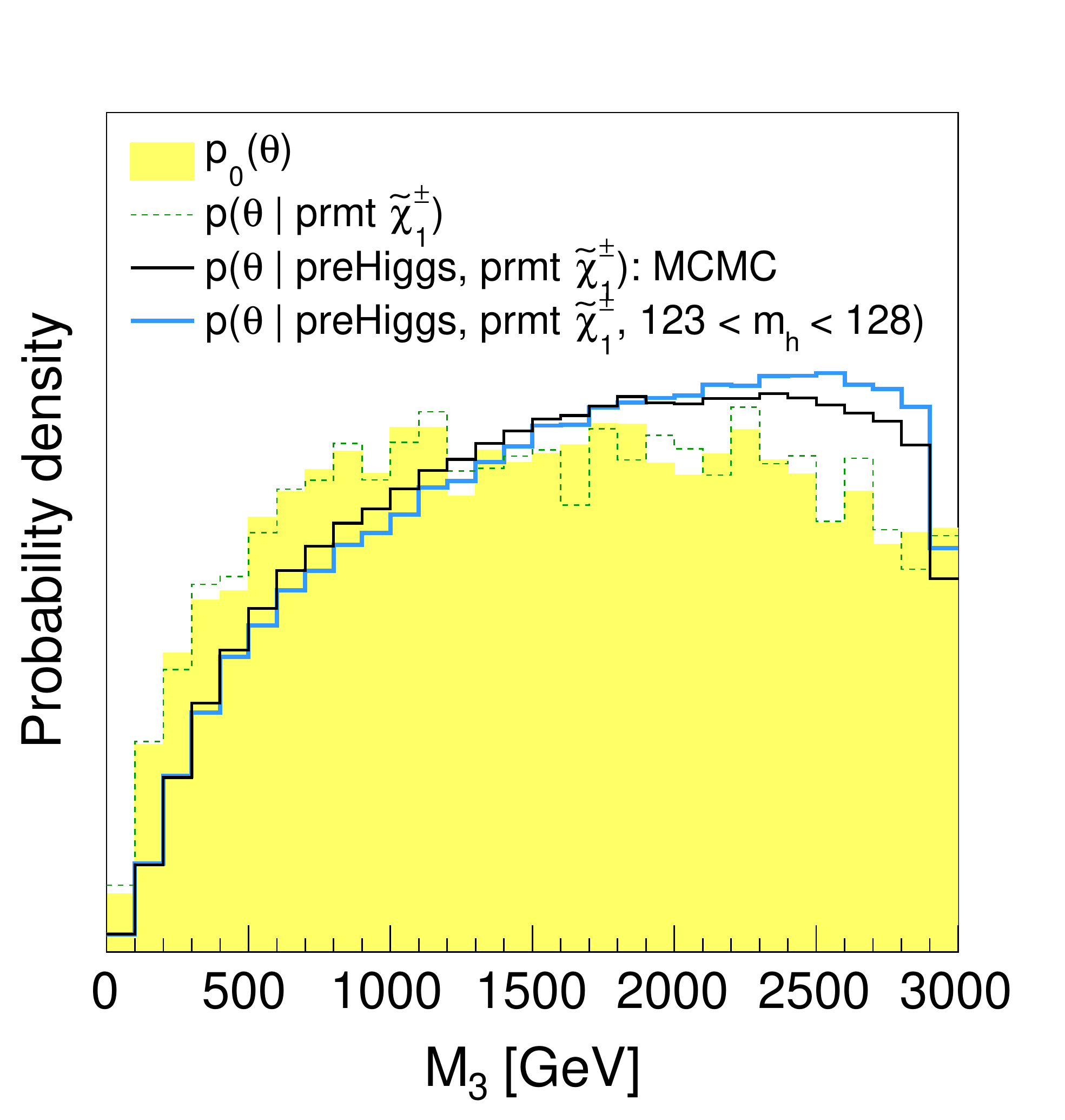}
\includegraphics[width=0.24\linewidth]{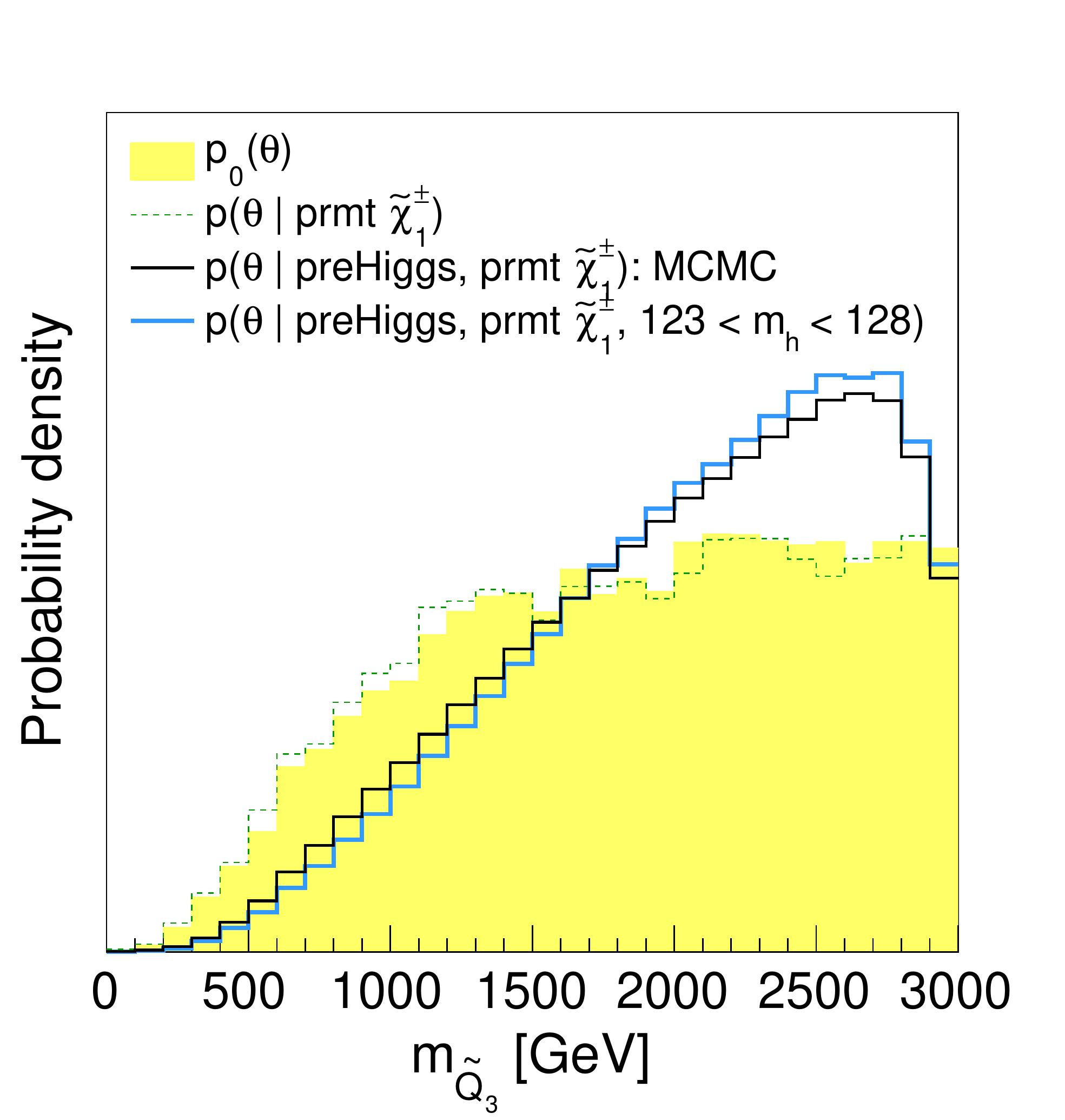} 
\includegraphics[width=0.24\linewidth]{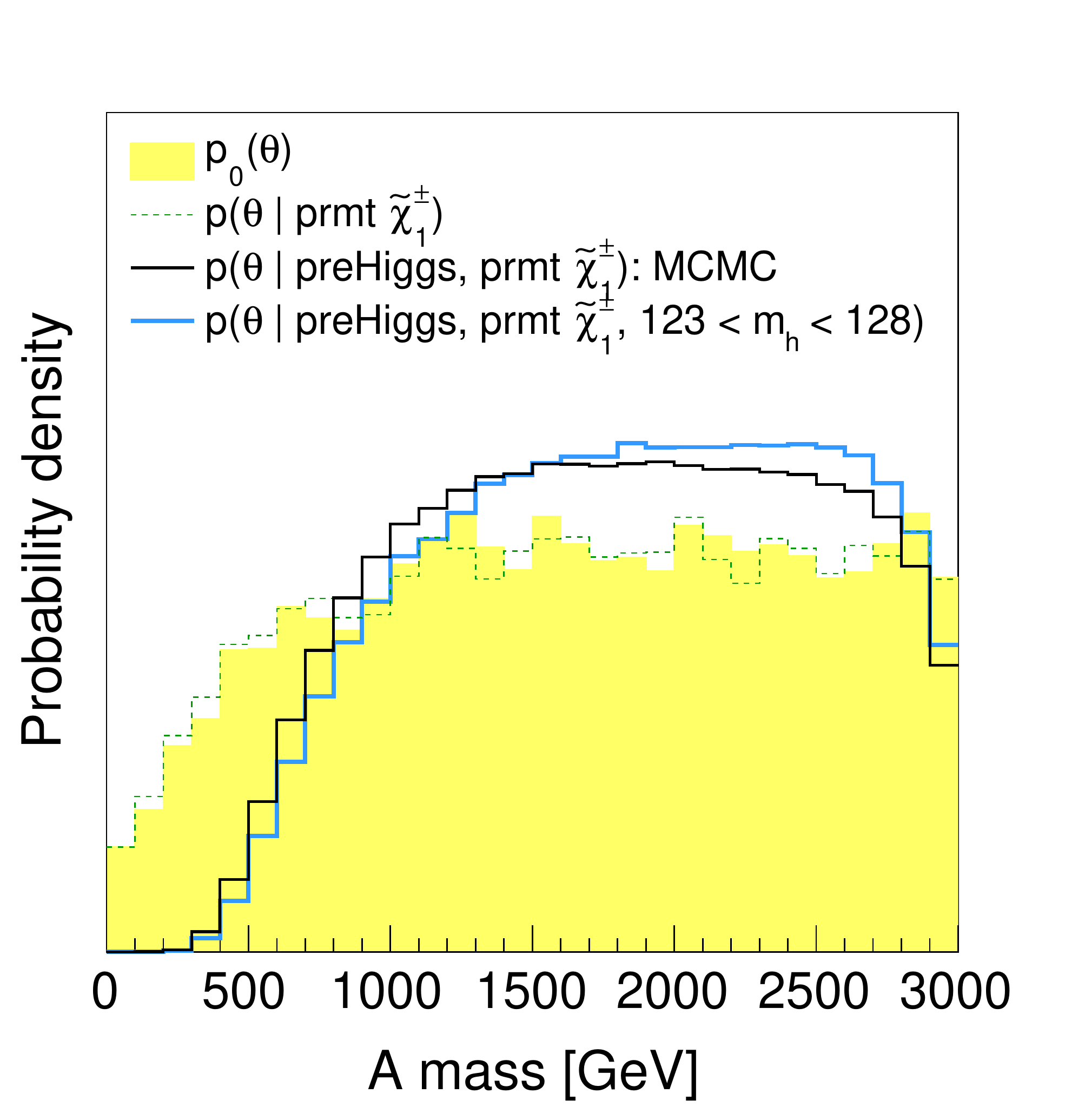} 
\includegraphics[width=0.24\linewidth]{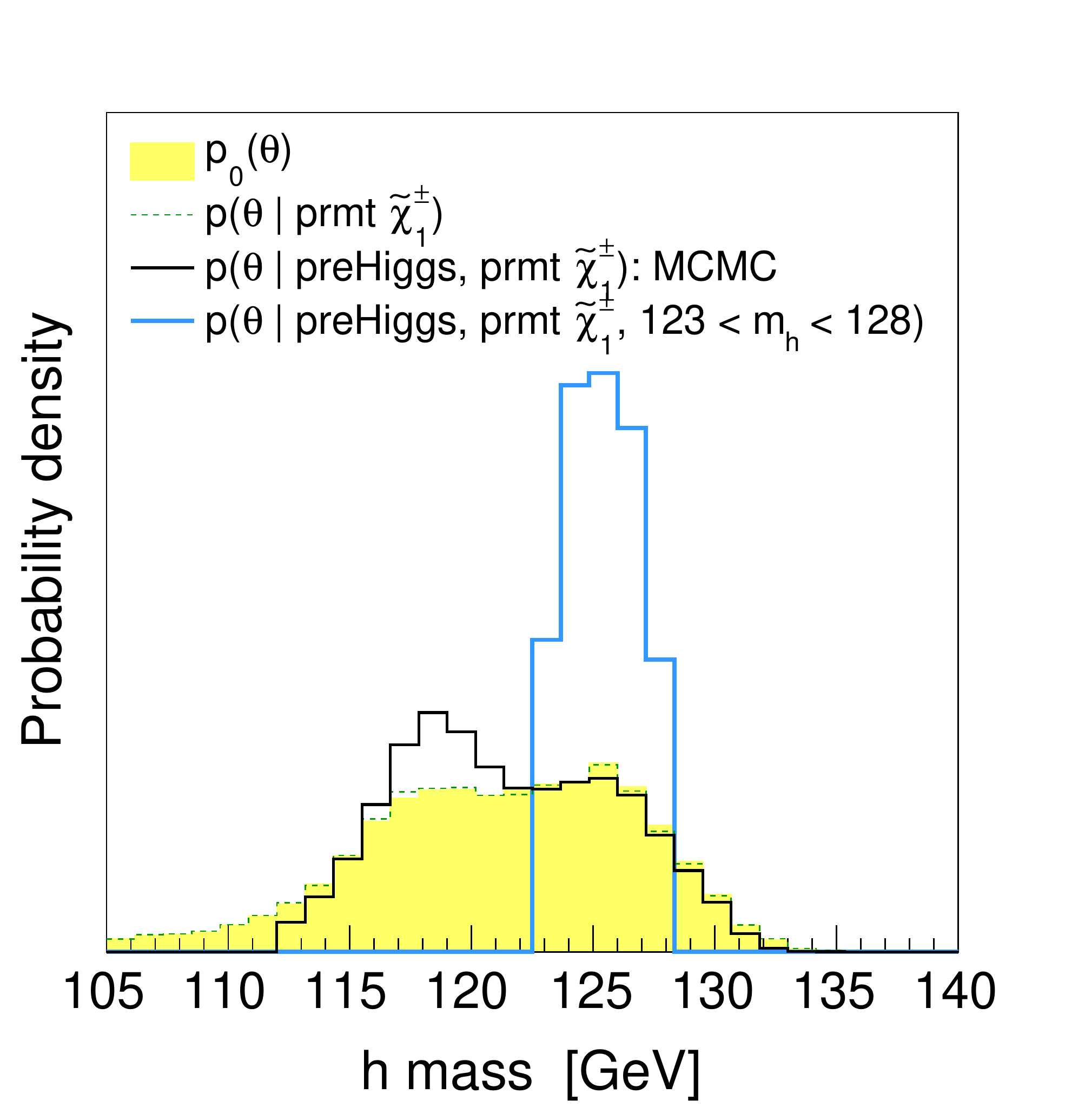}
\includegraphics[width=0.24\linewidth]{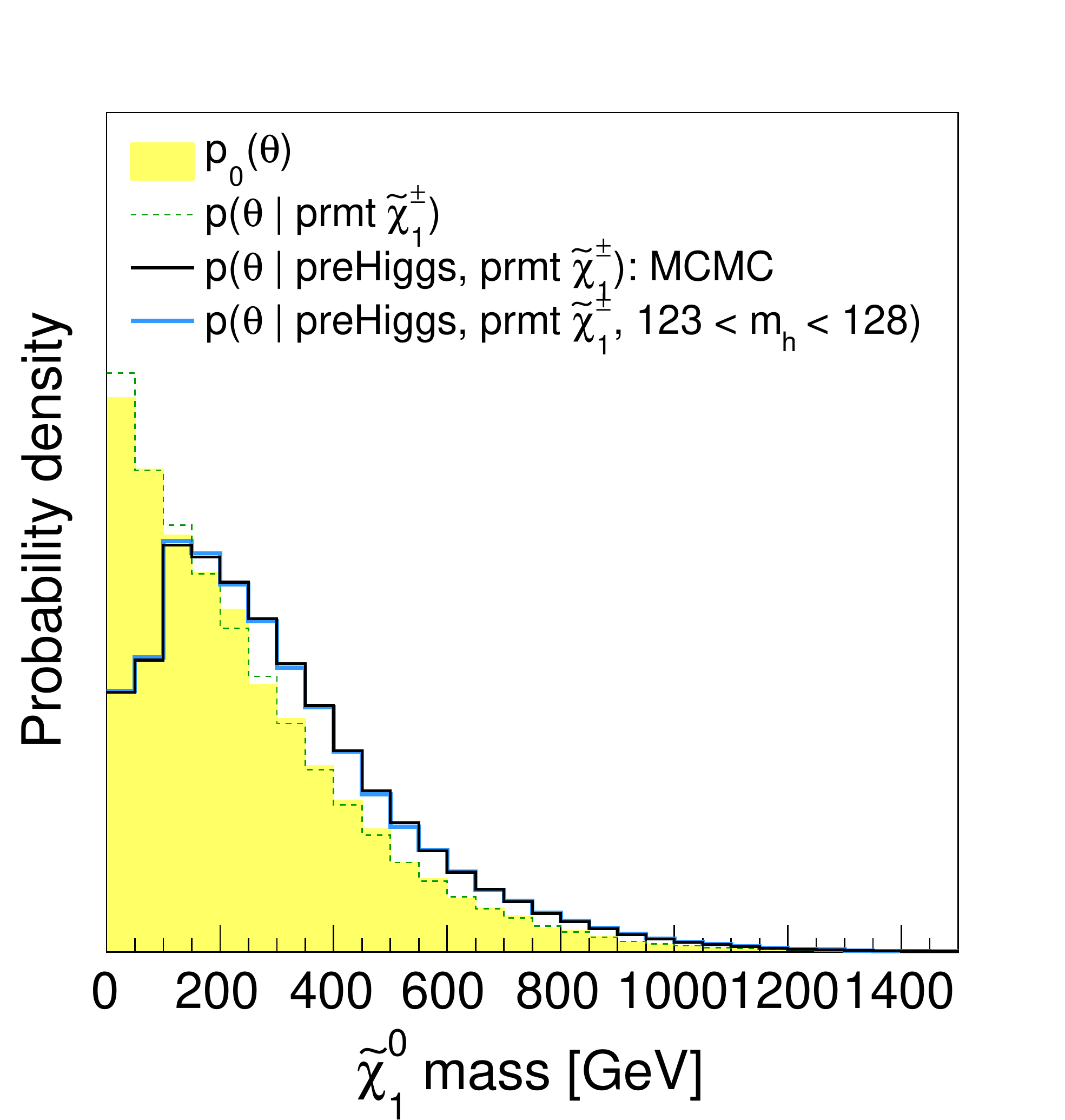}
\includegraphics[width=0.24\linewidth]{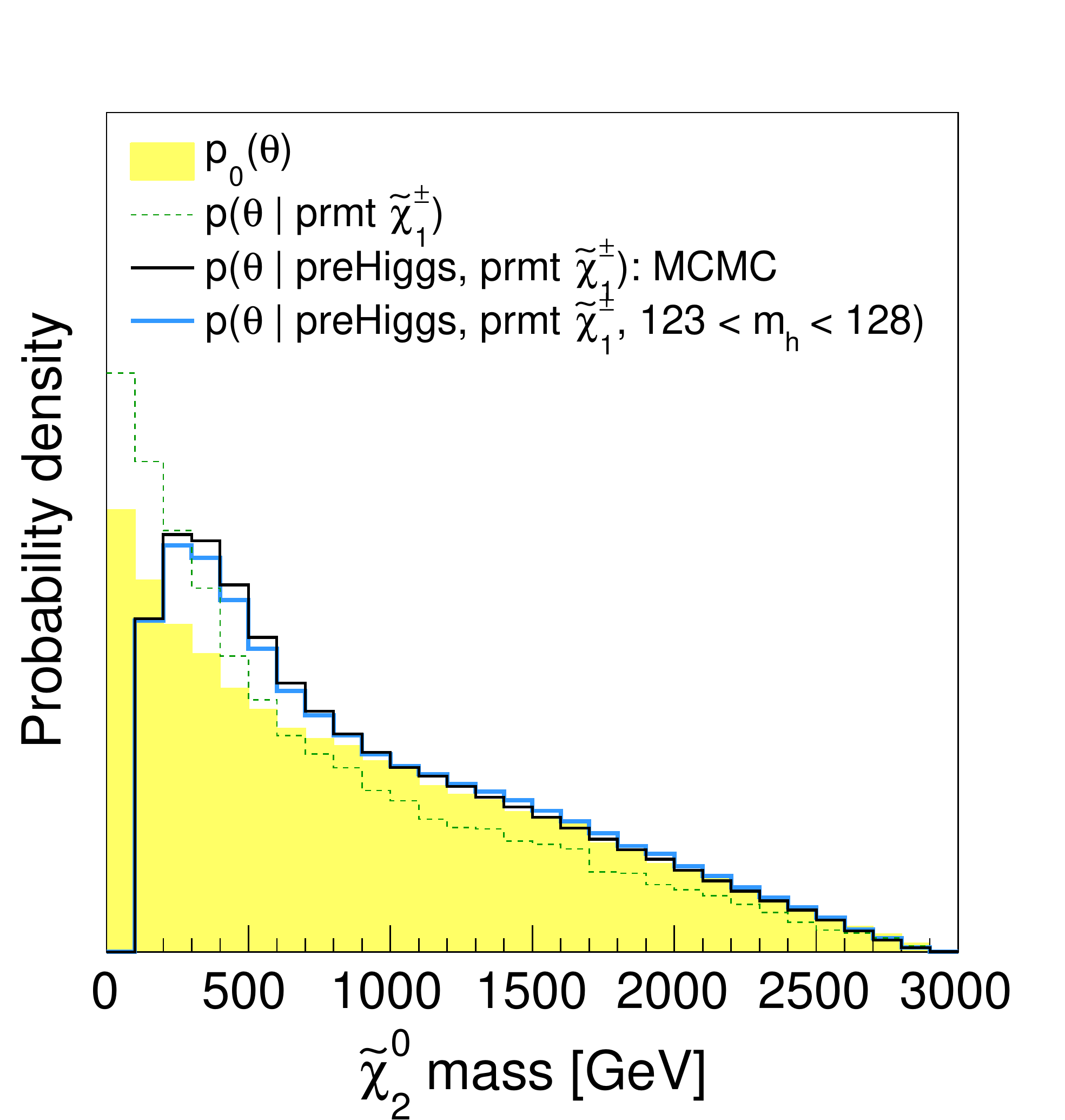} 
\includegraphics[width=0.24\linewidth]{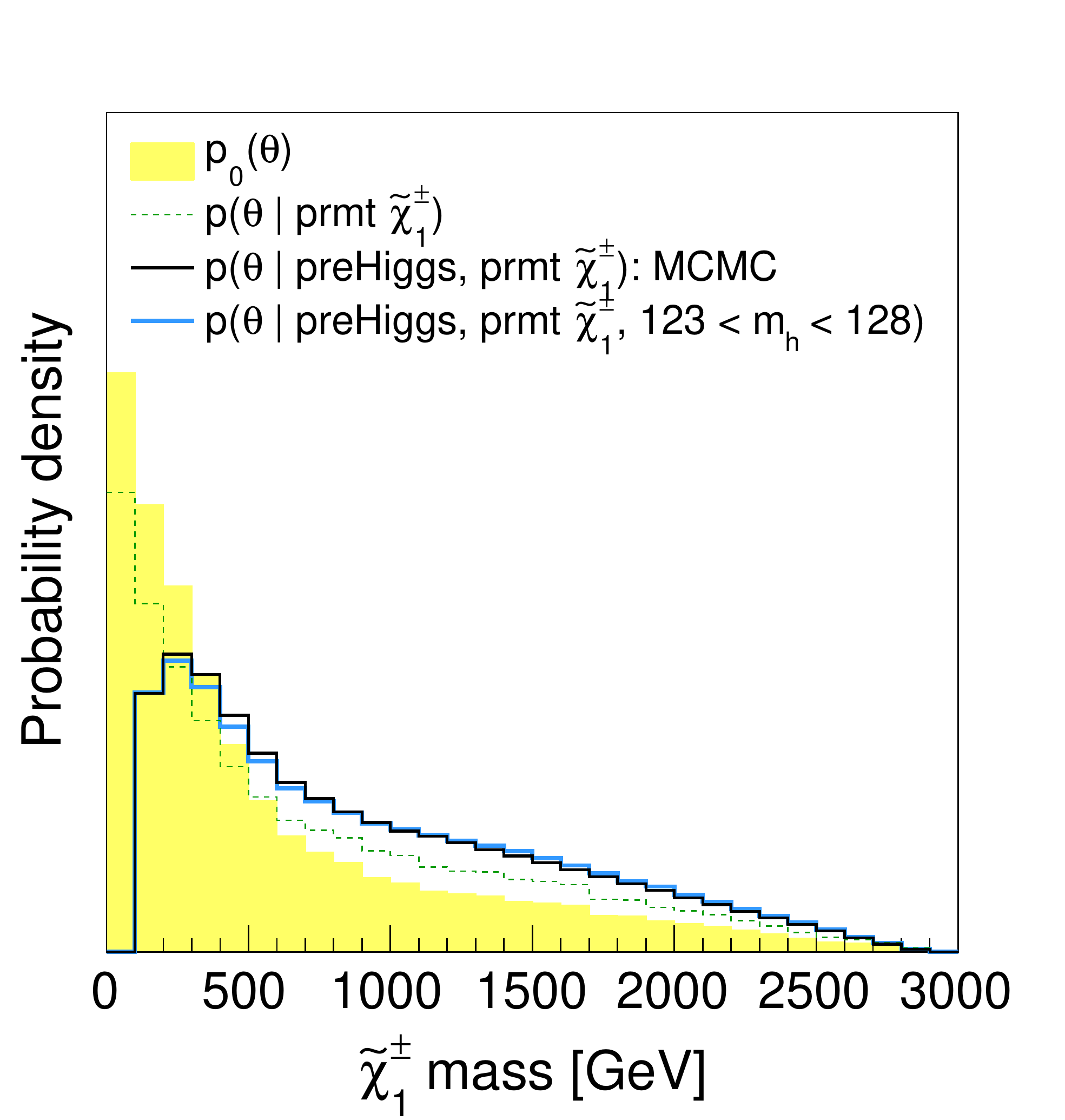}
\includegraphics[width=0.24\linewidth]{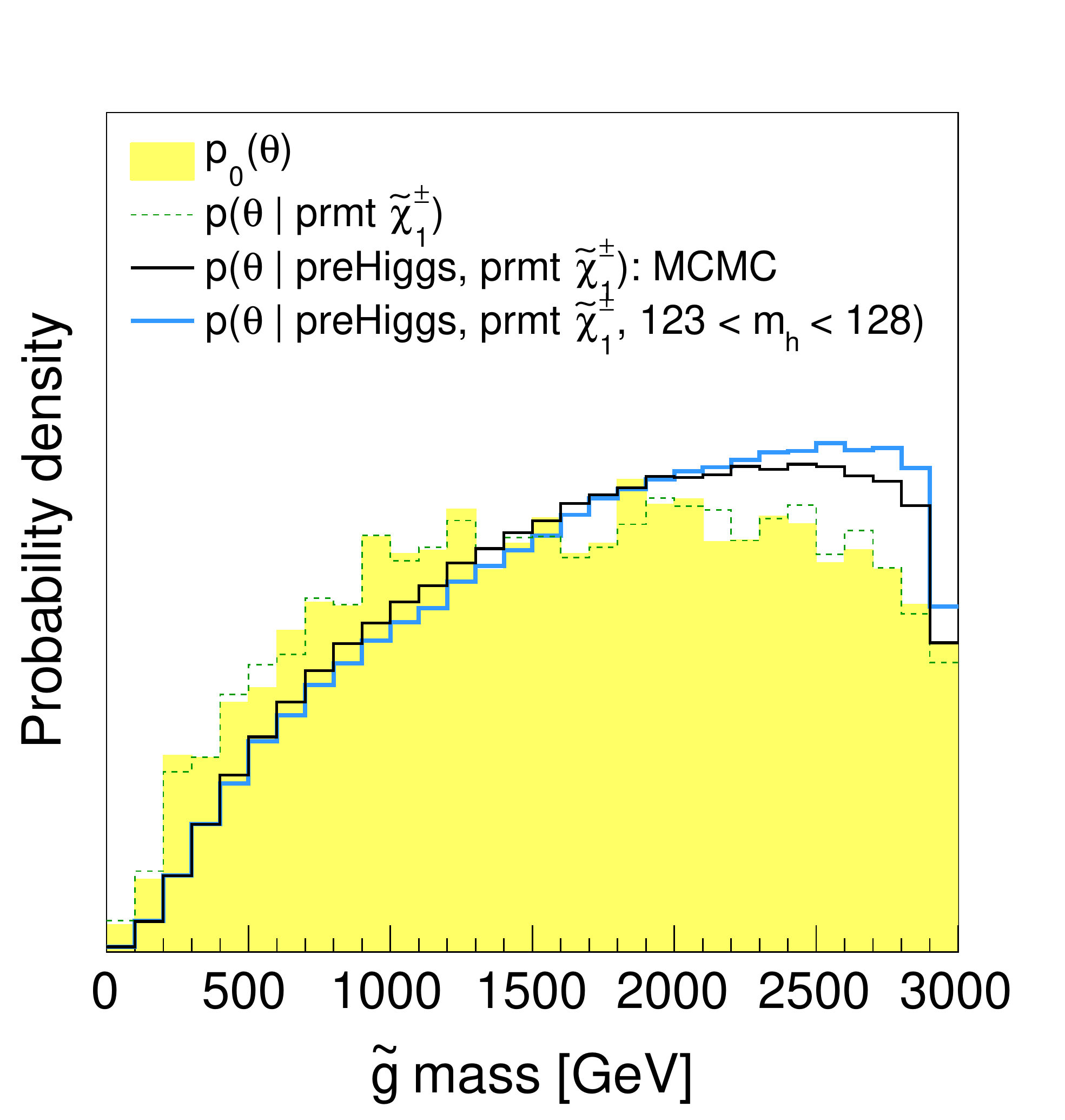}
\includegraphics[width=0.24\linewidth]{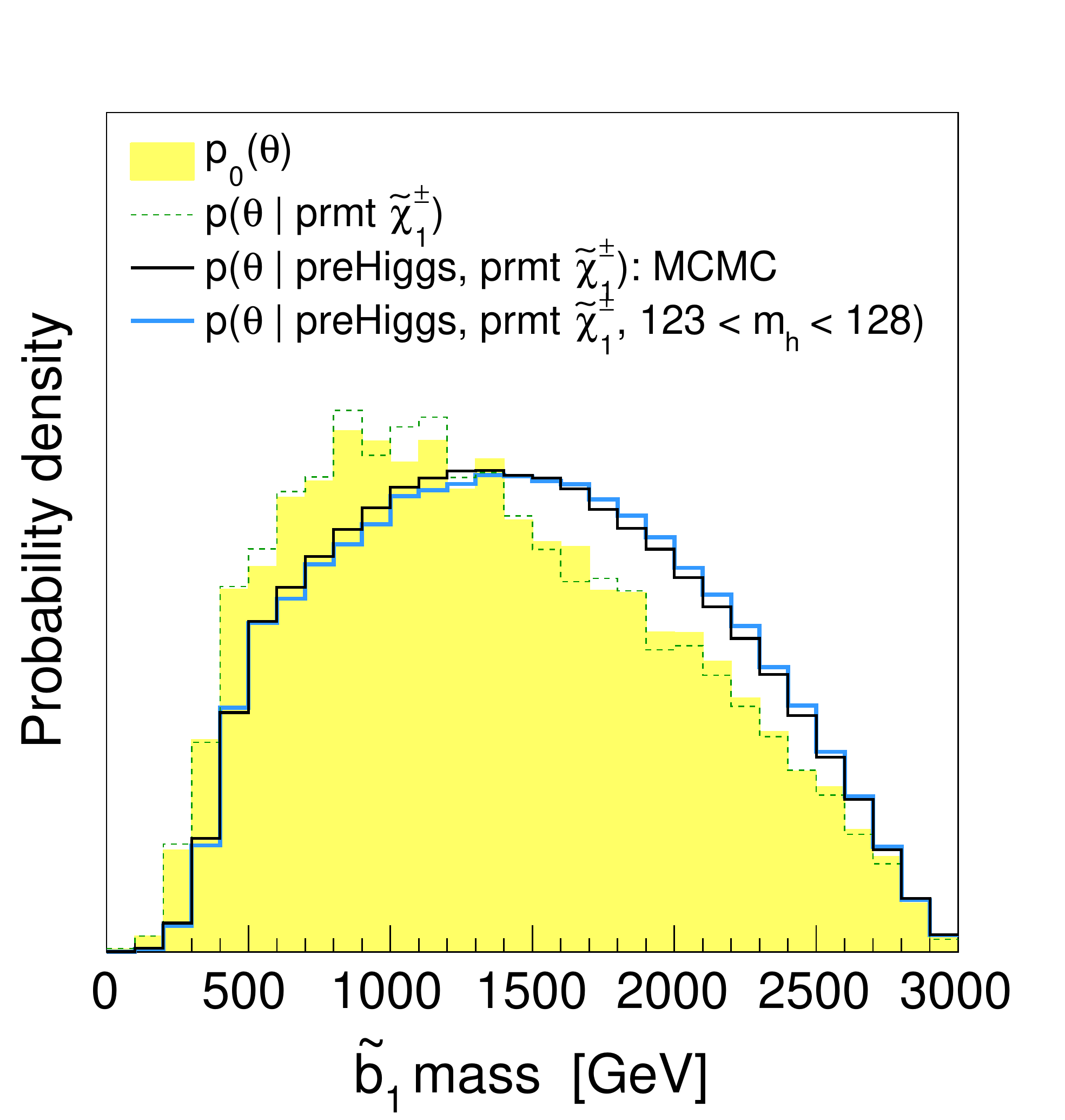} 
\includegraphics[width=0.24\linewidth]{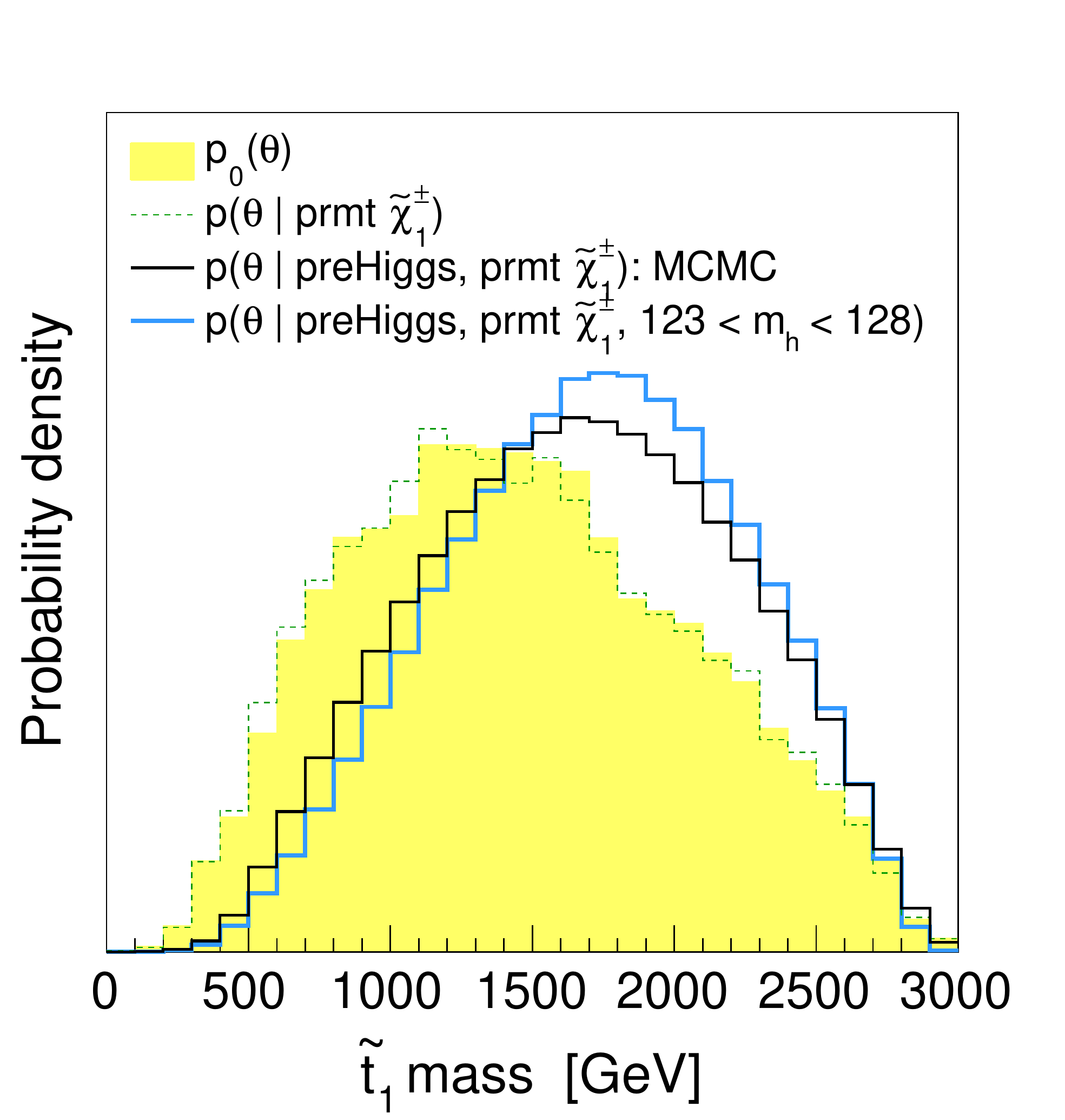}
\includegraphics[width=0.24\linewidth]{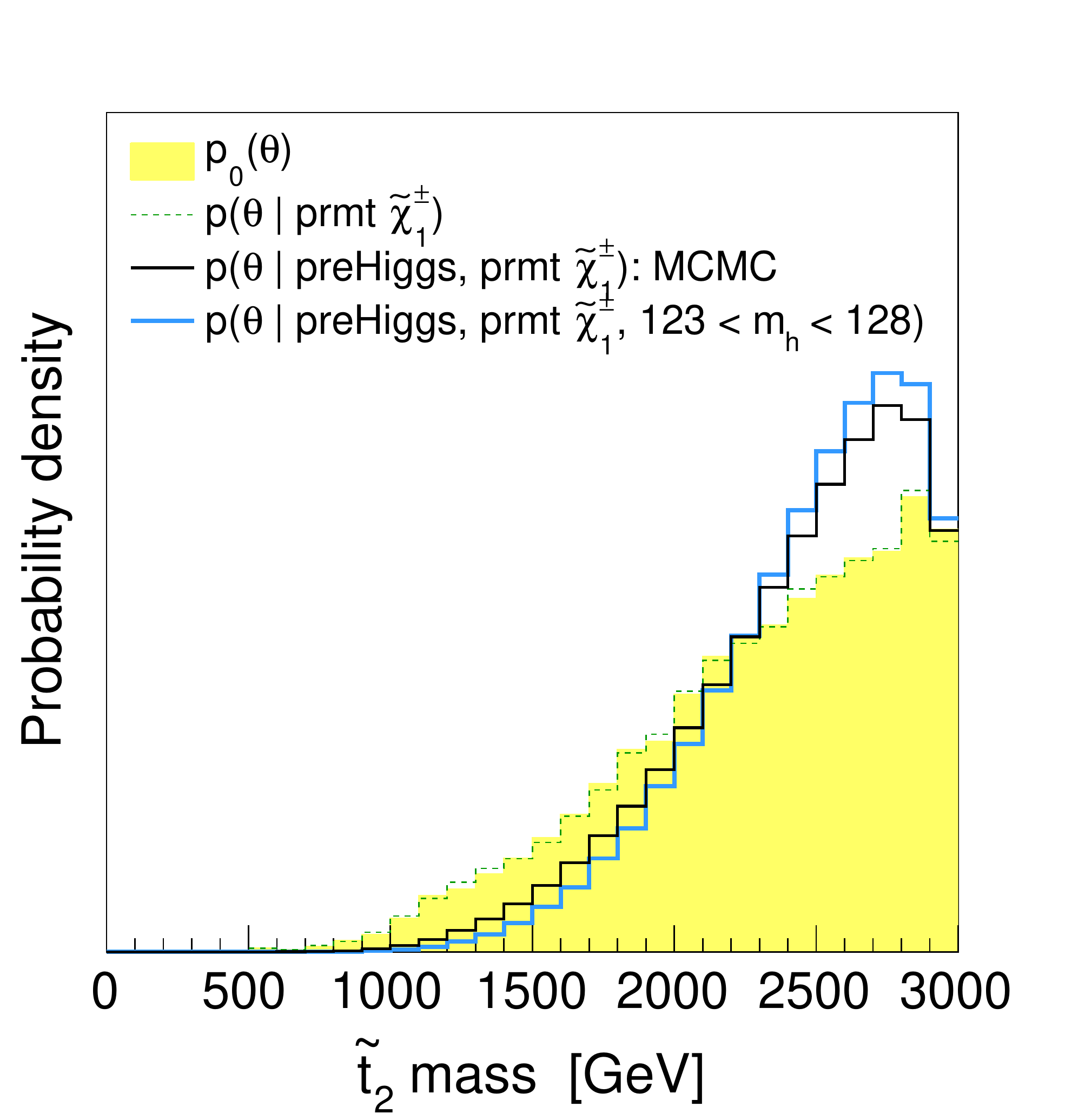}
\includegraphics[width=0.24\linewidth]{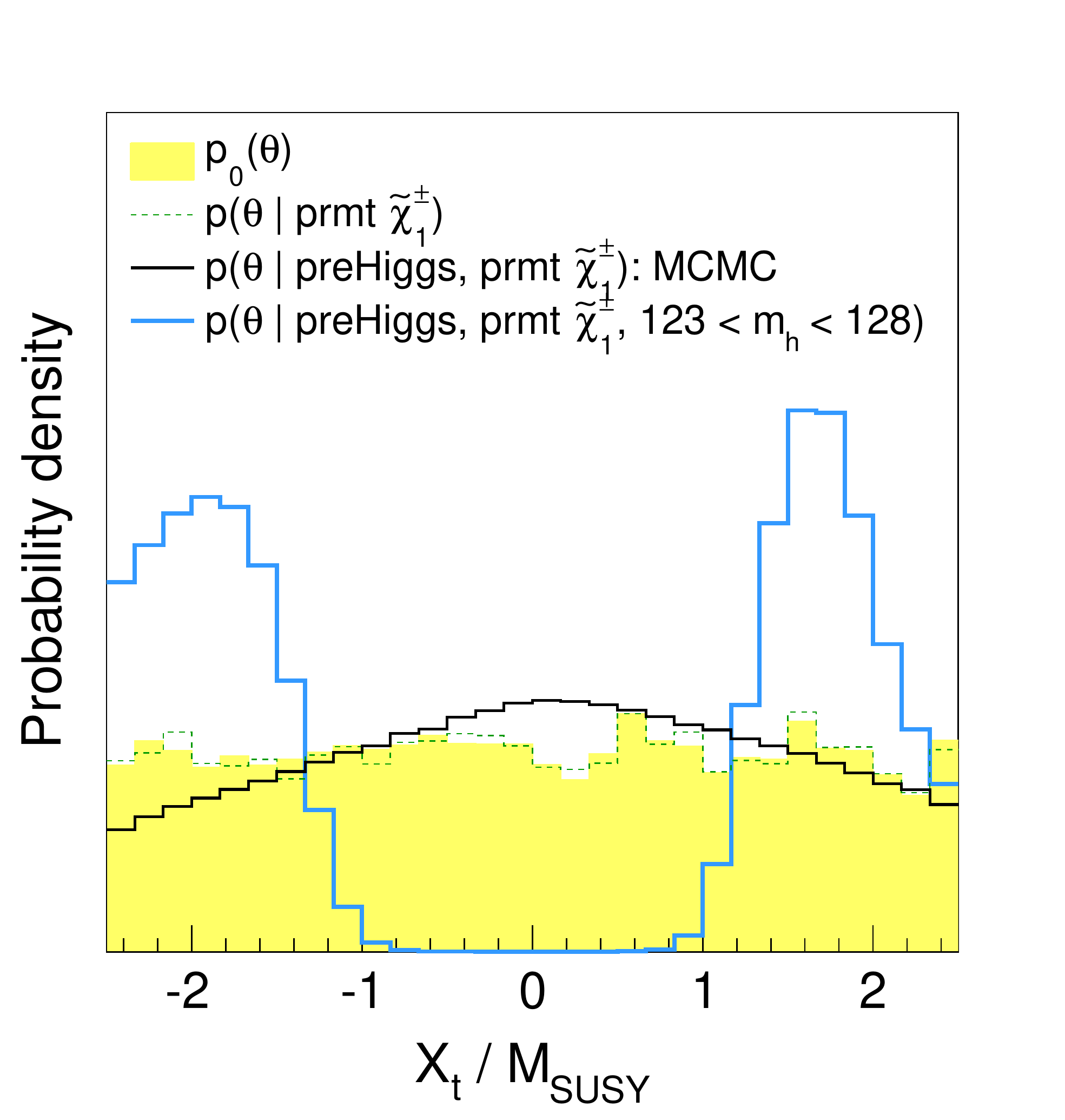} 
\caption{Marginalized 1D posterior densities for selected parameters and masses.
The yellow histograms show the sampled distributions, $p_0(\theta)$, as obtained after imposing theoretical constraints starting from a flat scan in the parameter ranges specified by Eq.~(\ref{eq:subspace}). The dashed 
green lines are the distributions  after requiring prompt charginos (prmt), the full black lines show  
the distributions  based on the ``preHiggs'' measurements of Table~\ref{tab:preHiggs}, 
and the full blue lines the ones when requiring $m_h=[123,\,128]$~GeV in addition 
to ``prmt'' and ``preHiggs''  constraints.
The  bottom right plot of $X_t/M_{\rm SUSY}$ shows that large (but not maximal) 
stop mixing is favored by the $m_h=123-128\gev$ requirement. }
\label{fig:sampling1}
\end{center}
\end{figure}

We begin our discussion by showing in Fig.~\ref{fig:sampling1} the sampled distributions of selected parameters and masses and the effect of the model prior.  All distributions except that of the pMSSM prior $p_0(\theta)$ include the prompt chargino requirement; as can be seen, this requirement substantially alters the probability distributions for 
the parameters $M_1$, $M_2$, and $\mu$ and the chargino and neutralino masses  relative to the $p_0(\theta)$ distributions, but has very little impact on the other parameters or masses. Further, in all the plots we observe that the preHiggs measurements incorporated in the MCMC influence the probability distributions relative to the simple prompt-chargino-decay distributions quite significantly, in particular shifting the neutralino, chargino, gluino, and also the stop/sbottom masses to higher values.

Also shown is the impact of requiring,  in addition,  that the mass of the light $h$ fall in the window $123~{\rm GeV} \leq m_h \leq 128~{\rm GeV}$. This Higgs mass constraint strongly affects the stop mixing parameter $X_t /M_{\rm SUSY} \equiv (A_t - \mu / \tan \beta)/\sqrt{m_{\st_1}m_{\st_2}}$, whose distribution takes on a two-peak structure emphasizing larger absolute values. More precisely, values around $|X_t /M_{\rm SUSY}|\approx 2$, \ie\ large but not maximal stop mixing is preferred. (Maximal stop mixing would mean $|X_t /M_{\rm SUSY}|=\sqrt{6}$; for a detailed discussion of the relation between $|X_t /M_{\rm SUSY}|$ and $m_h$ see, \eg,  \cite{Djouadi:2005gj,Brummer:2012ns}). 
It is interesting to note here that, in view of naturalness, the optimal stop mixing is indeed somewhat shy of maximal~\cite{Wymant:2012zp}. The optimal value is actually quite close to that which has the highest probability in the pMSSM context, despite the fact that no measure of naturalness is input into the pMSSM likelihood analyses.    
The Higgs mass window requirement also results in a shift of the $\tilde t_1$ mass distribution to slightly larger values; however, compared to the impact of the preHiggs constraints the effect is quite small.  Aside from an increased preference for values of $\tan\beta\approx 10-20$, 
the other parameters and masses are hardly affected by the Higgs mass window. 

\begin{figure}[t!]
\begin{center}
\includegraphics[width=0.3\linewidth]{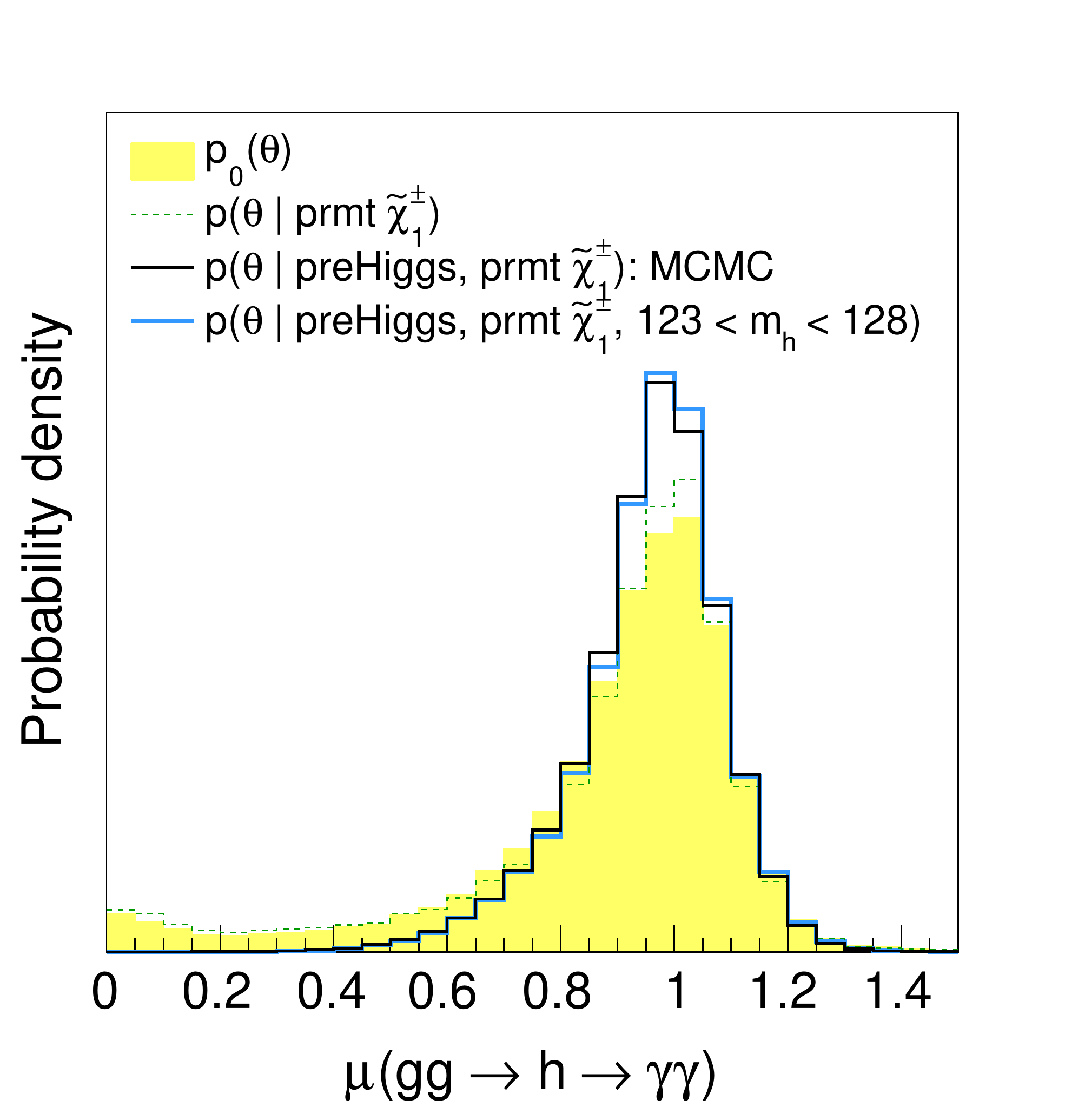}
\includegraphics[width=0.3\linewidth]{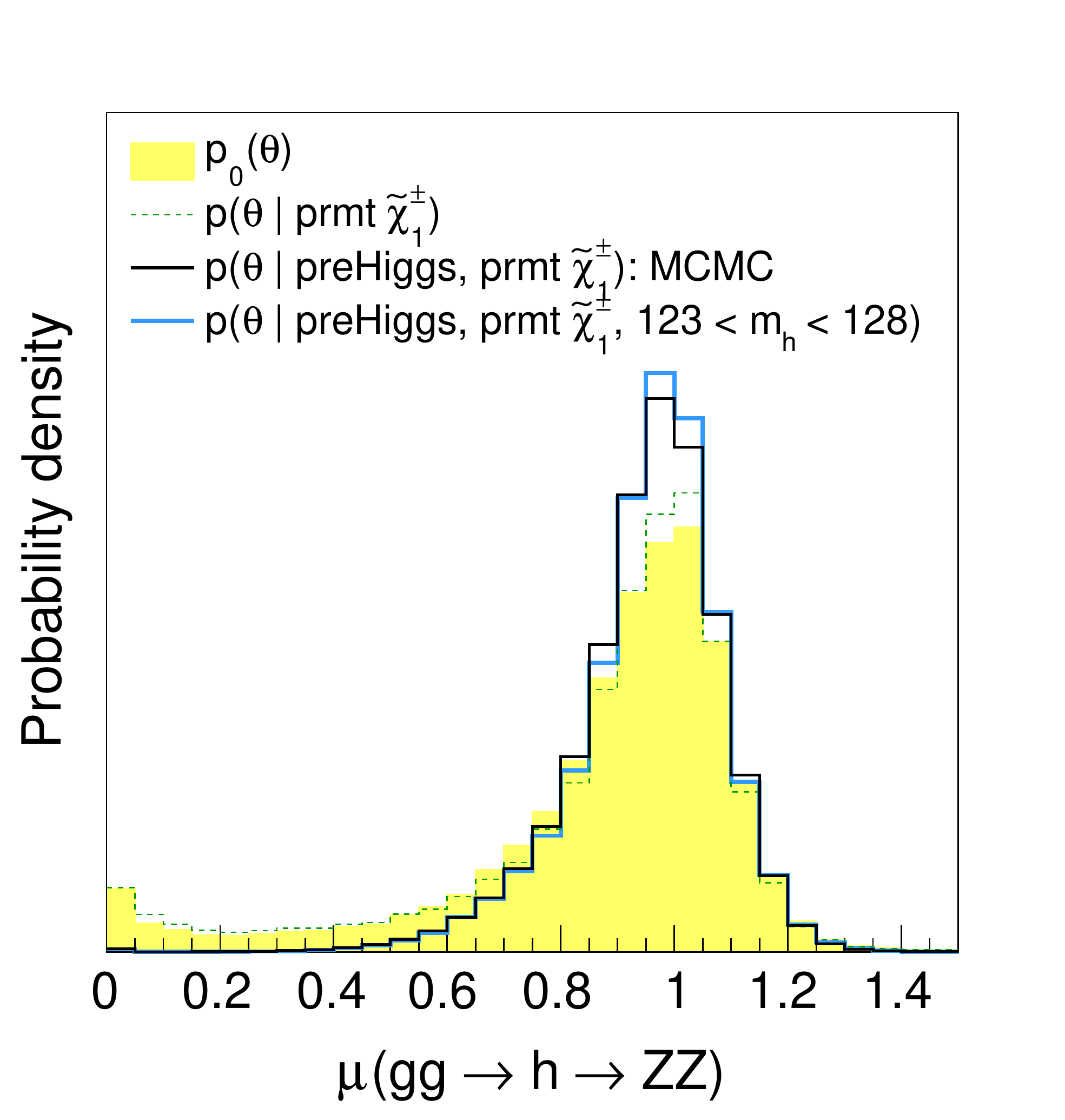}
\includegraphics[width=0.3\linewidth]{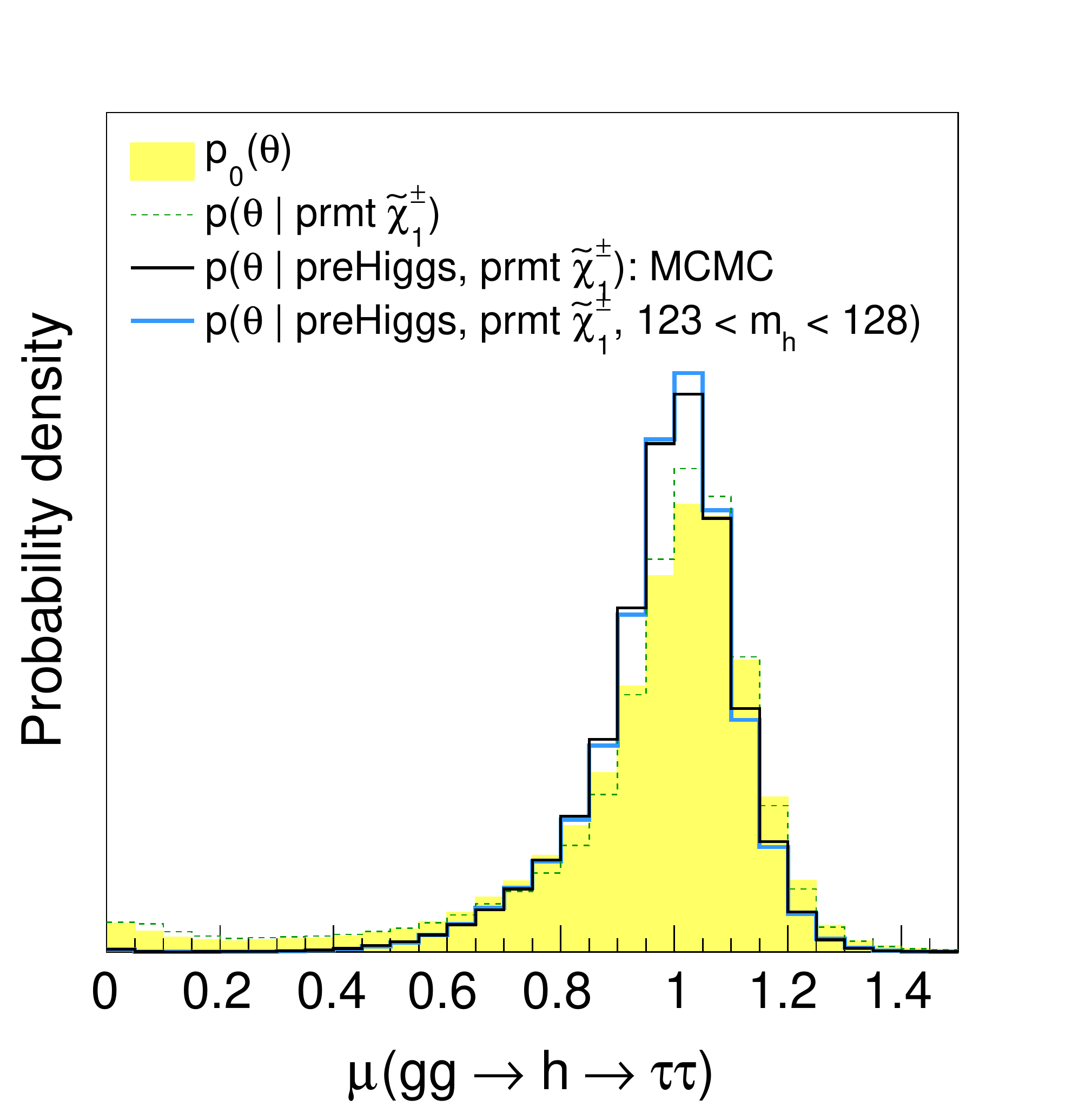}
\includegraphics[width=0.3\linewidth]{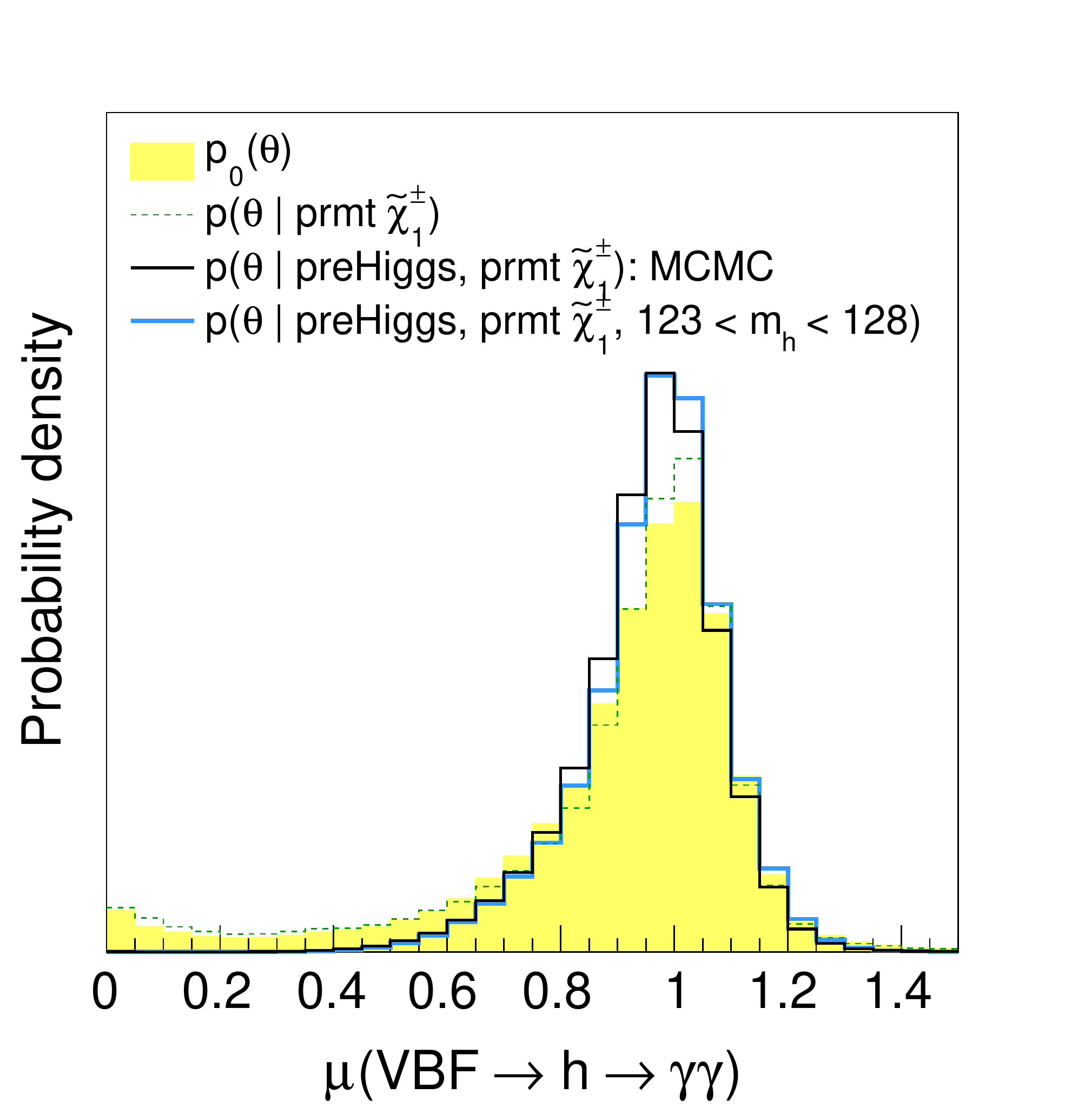}
\includegraphics[width=0.3\linewidth]{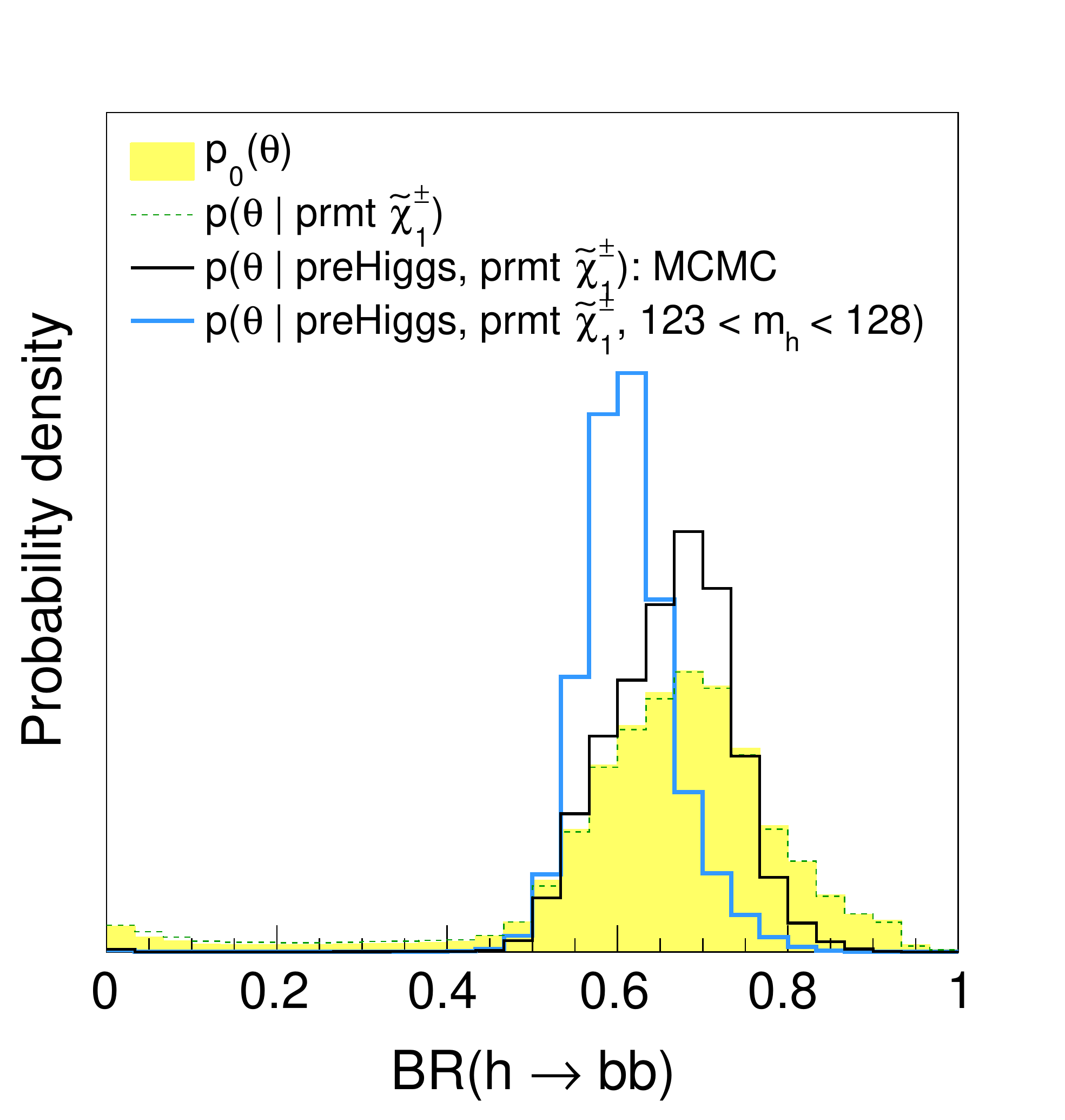}
\includegraphics[width=0.3\linewidth]{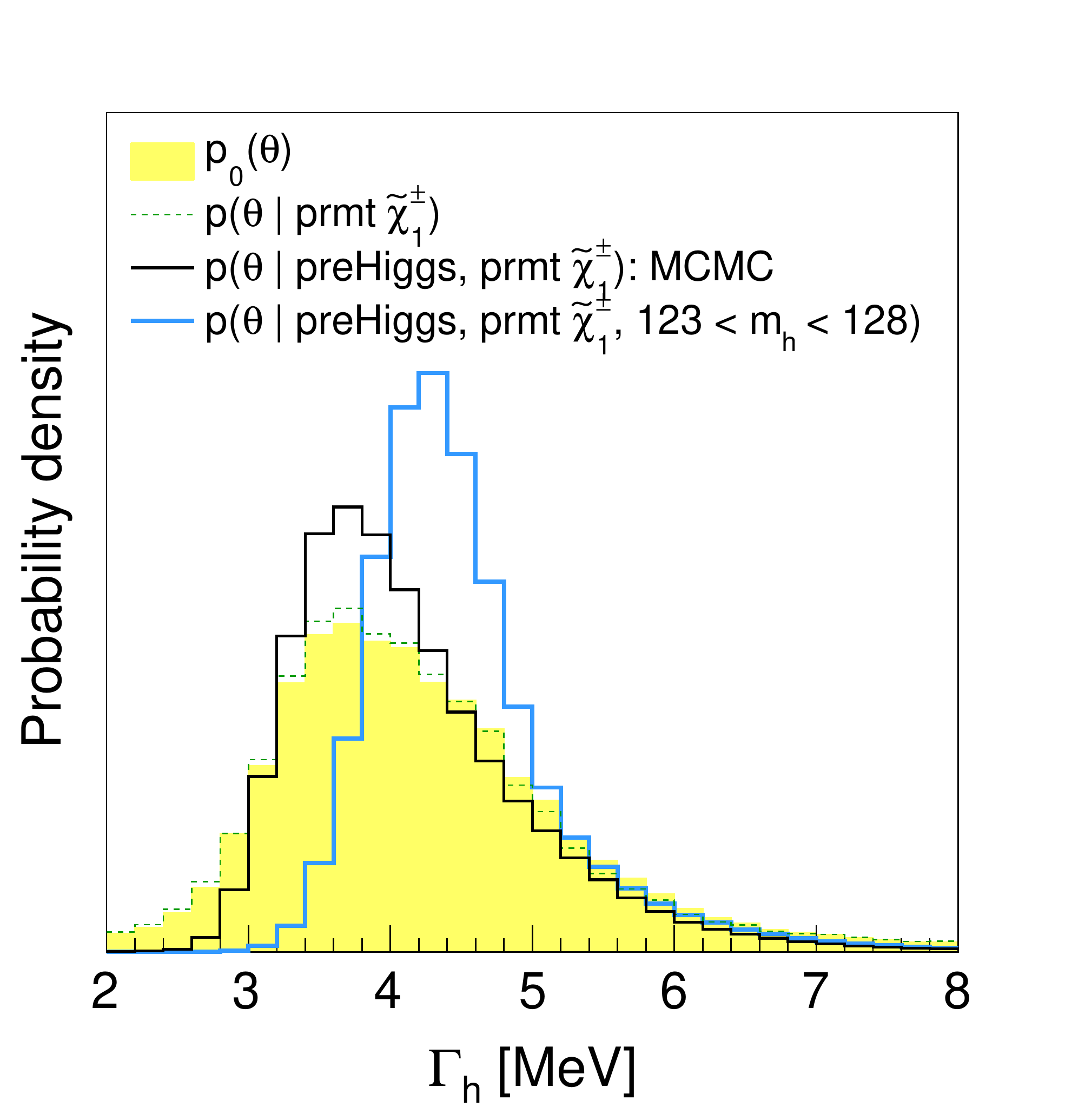}
\caption{Same as Fig.~\ref{fig:sampling1} but for selected $h$ signal strengths, $\br(h\to b\bar b)$ and the total decay width $\Gamma_h$. The VBF distributions look practically the same as the ggF distributions, as exemplified for the 
${\rm VBF}\to h\to \gamma\gamma$ case, though they show a slightly larger effect from requiring $m_h=123-128$~GeV than the ggF distributions.  
%Also shown are the  $\gam\gam$, $b\anti b$ and $gg$ partial widths relative to the SM values (as computed in the decoupling limit --- see text).
}
\label{fig:sampling3}
\end{center}
\end{figure}

It is also interesting to consider the $h$ signal at this level. Some relevant distributions are shown in Fig.~\ref{fig:sampling3}.  While generically the $h$ signal strength can go down to zero in the MSSM, already the ``preHiggs'' constraints eliminate very small values below $\mu\approx 0.6$ and narrow the signal strength distributions to a range of $\mu\approx 1\pm0.4$. This is coming from two different effects. First, in the low-$m_A$ region the heavier scalar $H$ can be more SM-like than $h$. Second, in the region where the LSP is light ($m_{\tilde \chi^0_1} \lesssim 65$~GeV) a large increase of the total width, resulting in  reduced signal strengths, is possible through $h \to \tilde \chi^0_1 \tilde \chi^0_1$.
The low-$m_A$ region is mostly disfavored from flavor constraints, while a light neutralino---if mainly wino or higgsino---is excluded by the LEP bound on charginos. In both cases, requiring $m_h=123-128$~GeV only has a very small additional effect.

One might expect that the influence of the Higgs mass is larger in the ggF channels than in the VBF channels (because of the negative loop contribution from maximally mixed stops affecting the former) but, in fact, the effect is very small and goes in the opposite direction,  as can be seen by comparing the top-left and the bottom-left plots in Fig.~\ref{fig:sampling3}. 
The observables which are really influenced by the Higgs mass are  the branching ratio into $b\bar b$, which becomes  centered around $\br(h\to b\bar b)\approx 0.6$, and the $h$ total width, for which the most likely value is shifted a bit upwards to $\Gamma_h\approx4$--5~MeV. However, this is not really a SUSY effect: the same happens for the SM Higgs when going from $m_H\lesssim 120$~GeV to $m_H\approx 125$~GeV.

%----------------------------------------------------------------------------------------------------------------------------------
\subsection{Impact of Higgs signal strengths}\label{sec:BluePlots}
%----------------------------------------------------------------------------------------------------------------------------------

\begin{figure}[htbp]
\begin{center}
\includegraphics[width=0.24\linewidth]{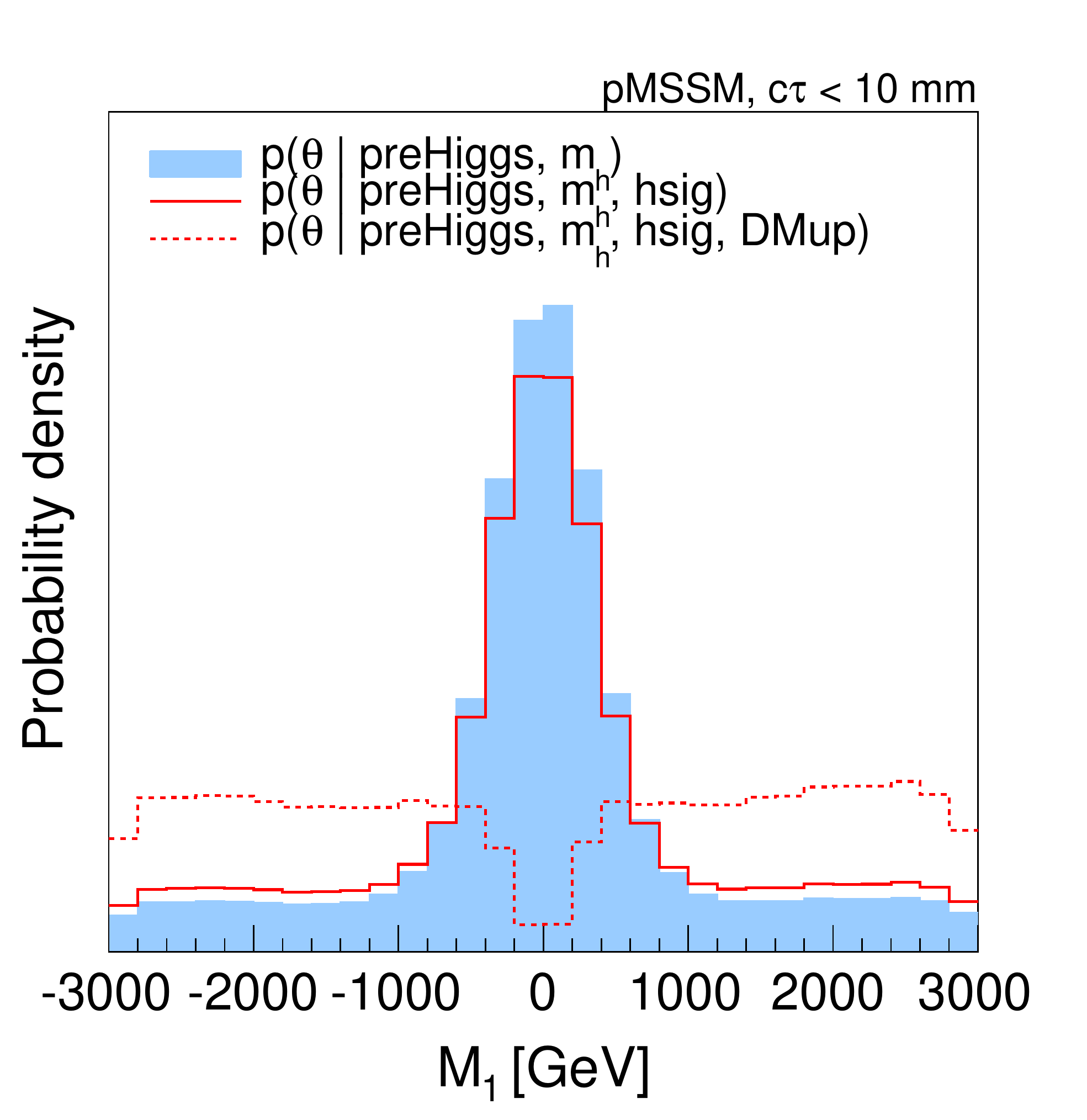}
\includegraphics[width=0.24\linewidth]{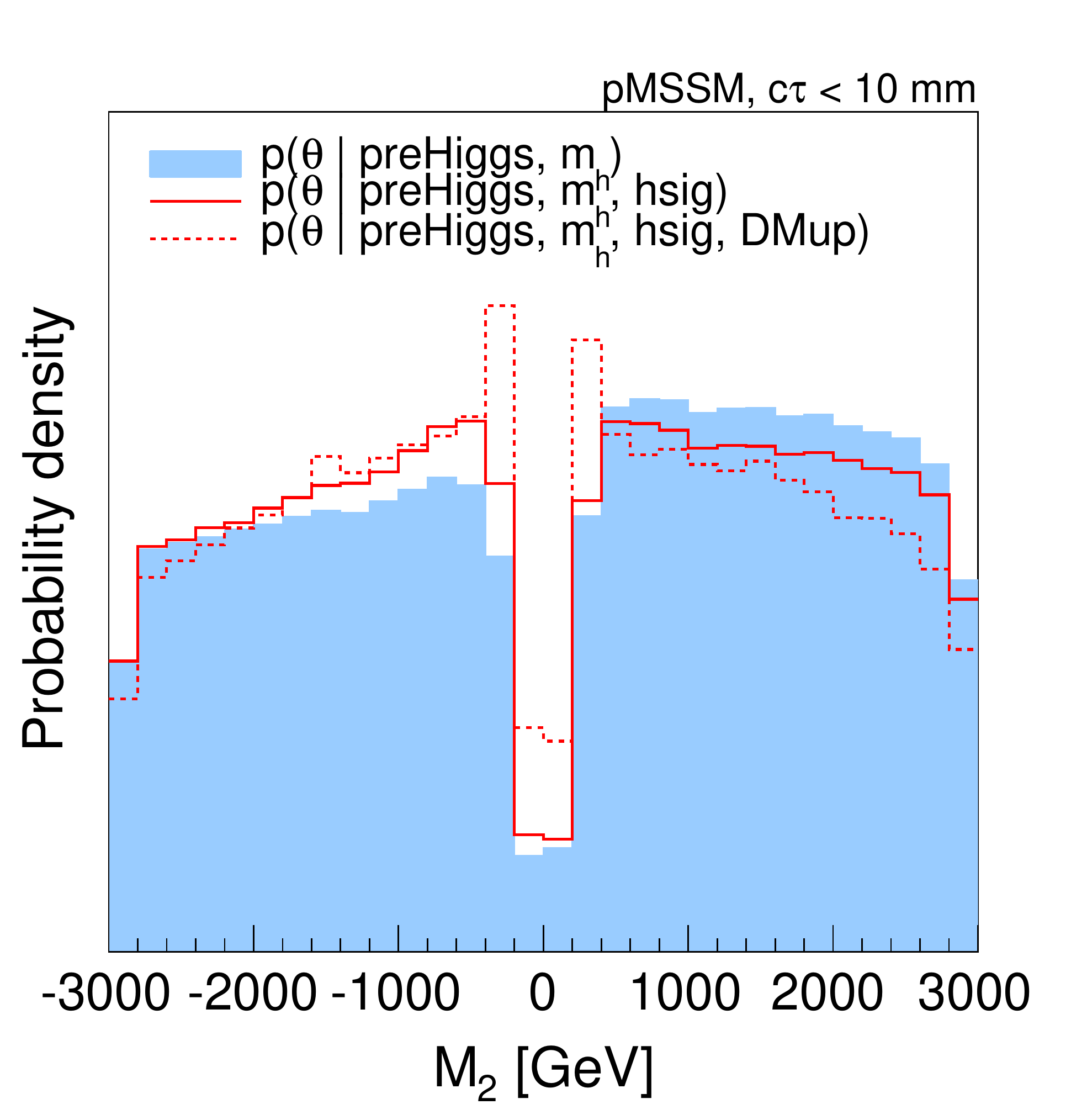}
\includegraphics[width=0.24\linewidth]{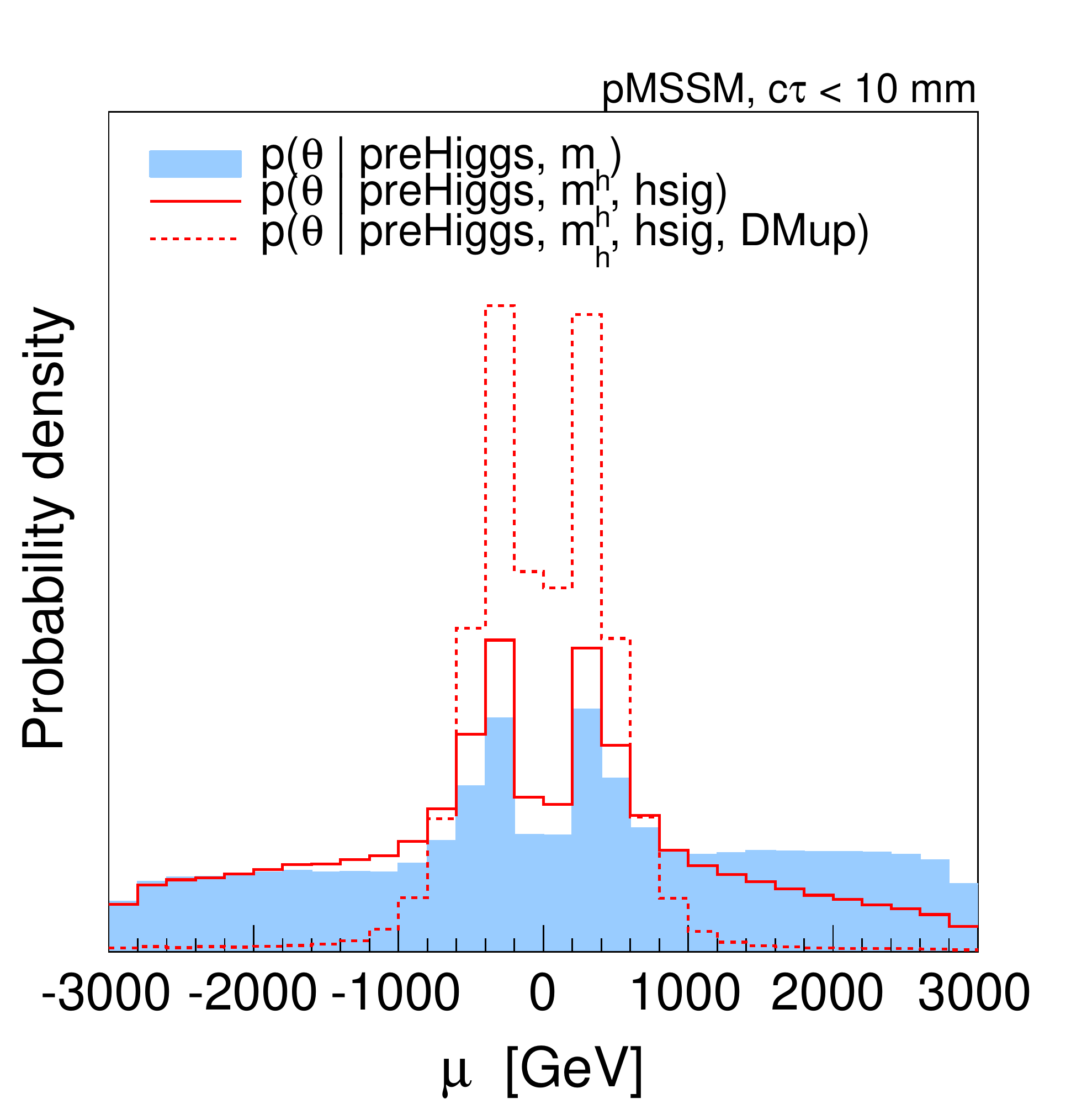} 
\includegraphics[width=0.24\linewidth]{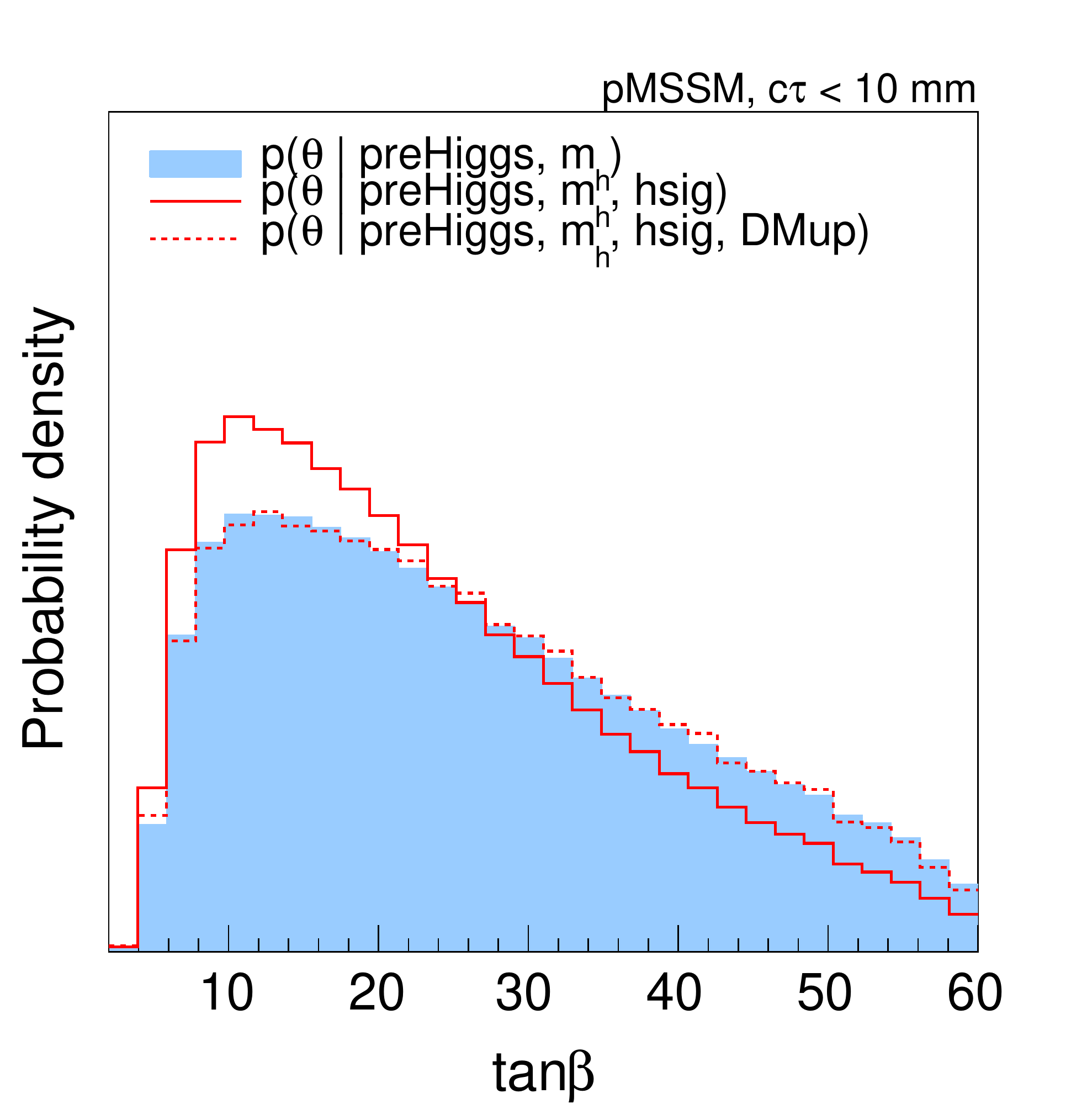}
\includegraphics[width=0.24\linewidth]{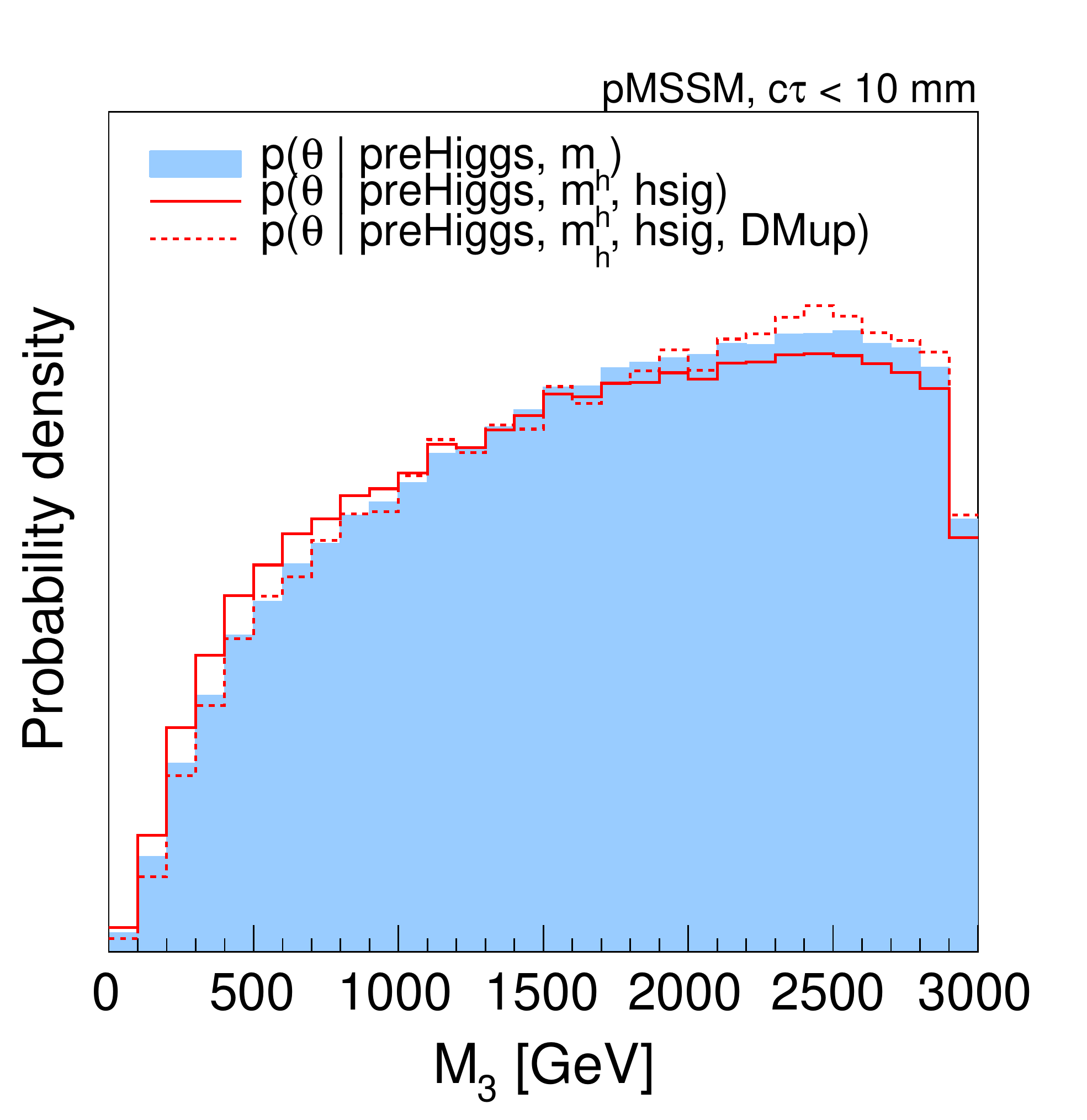}
\includegraphics[width=0.24\linewidth]{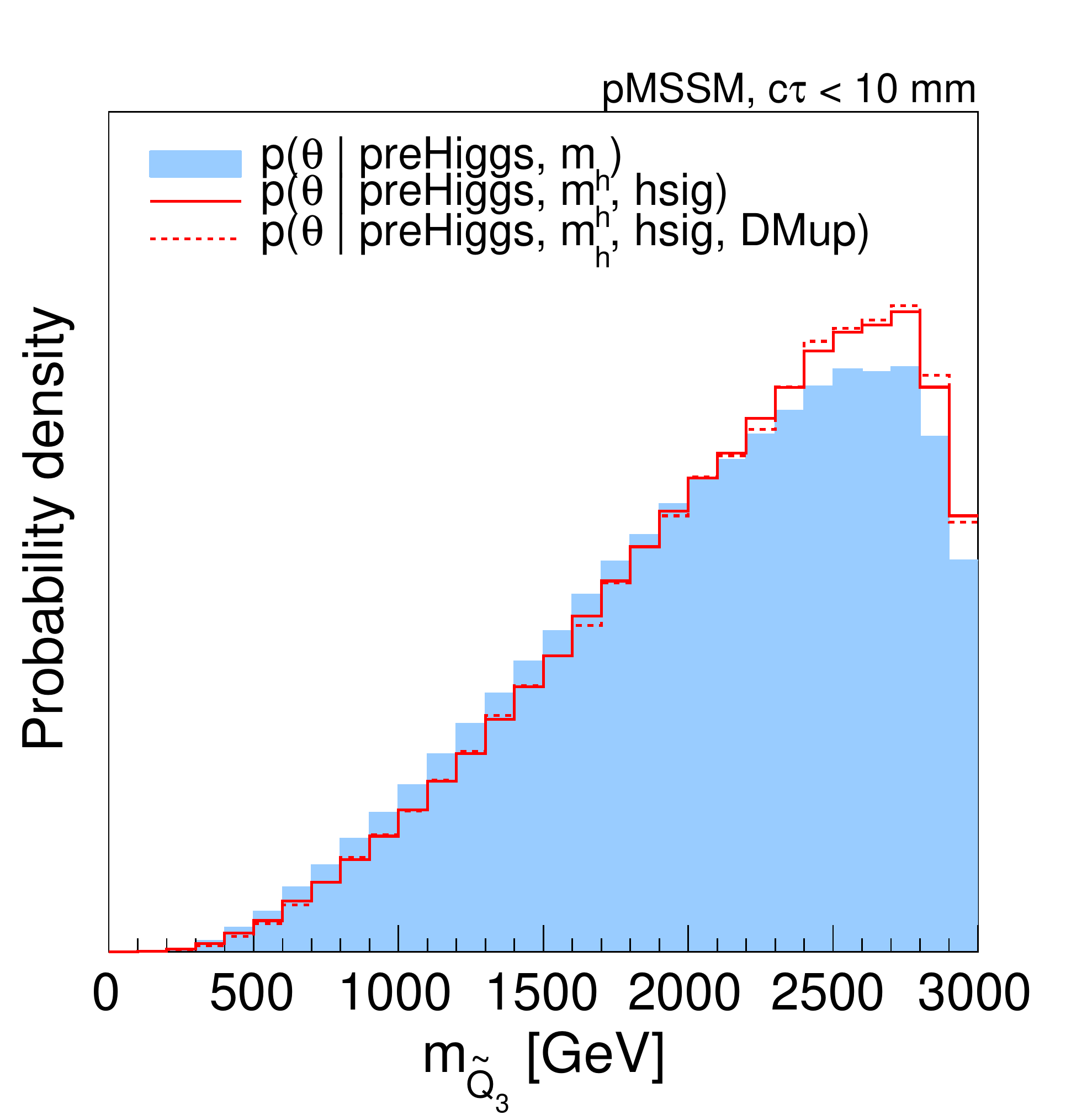}
\includegraphics[width=0.24\linewidth]{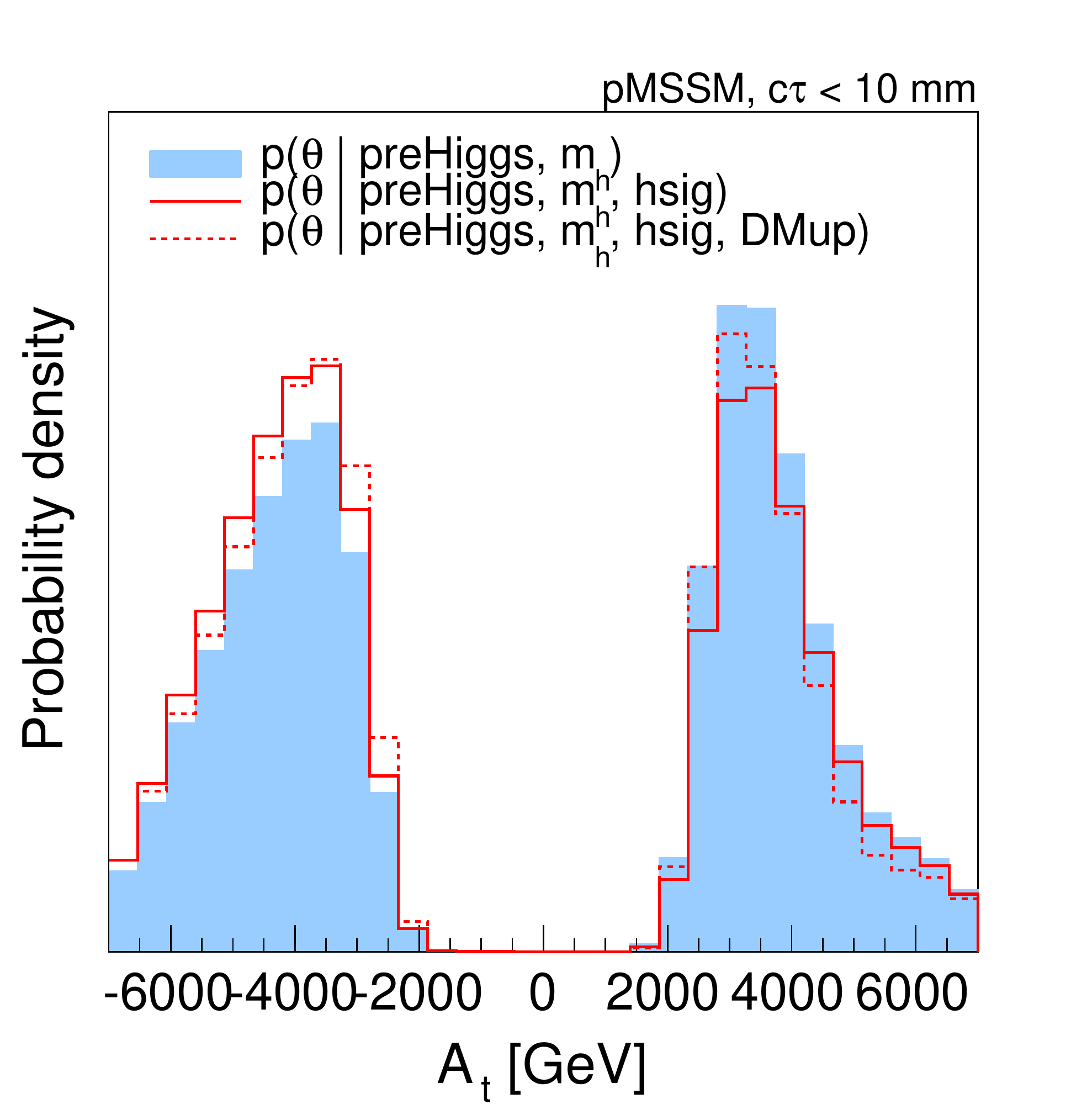}
\includegraphics[width=0.25\linewidth]{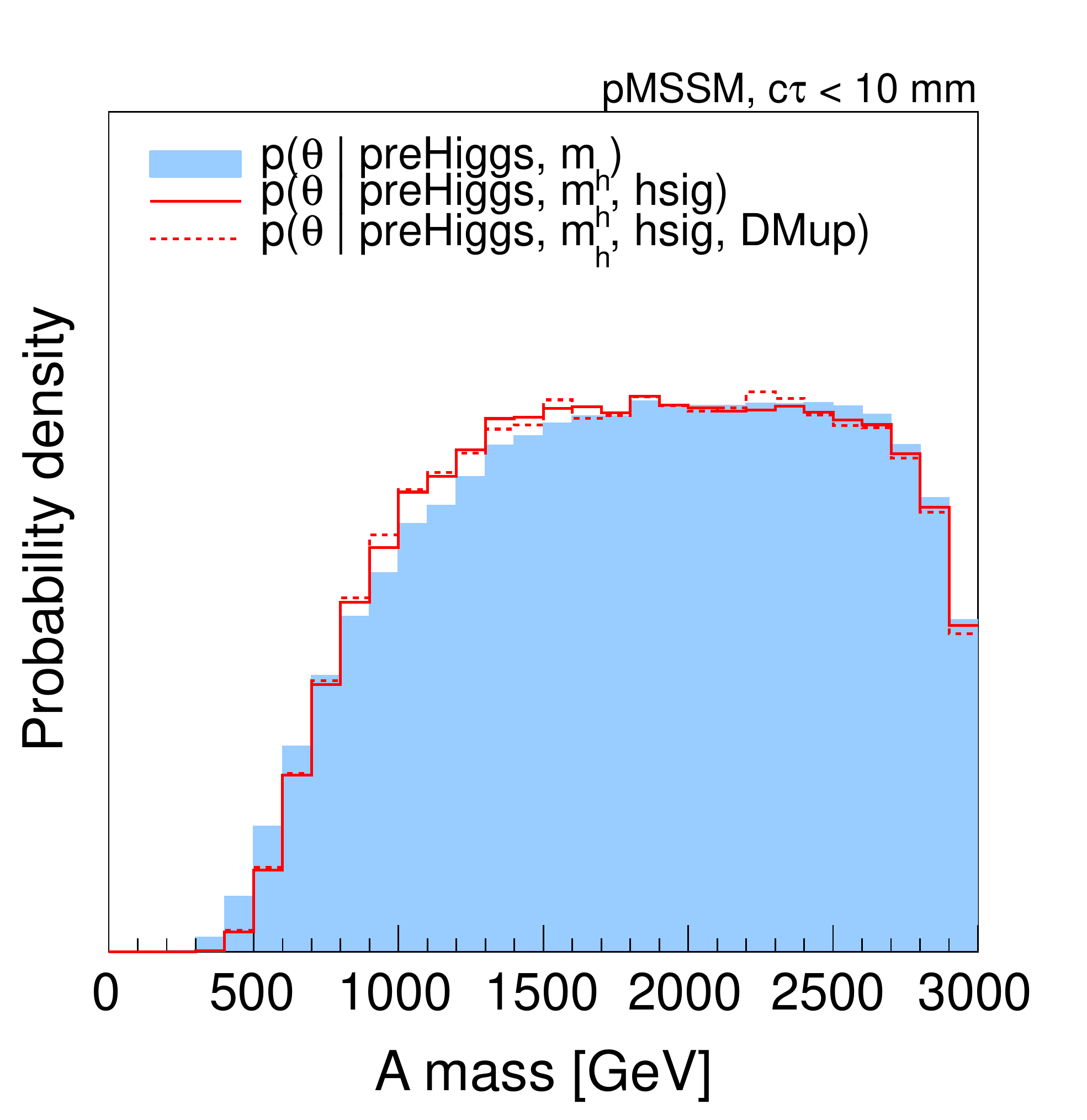}
\includegraphics[width=0.24\linewidth]{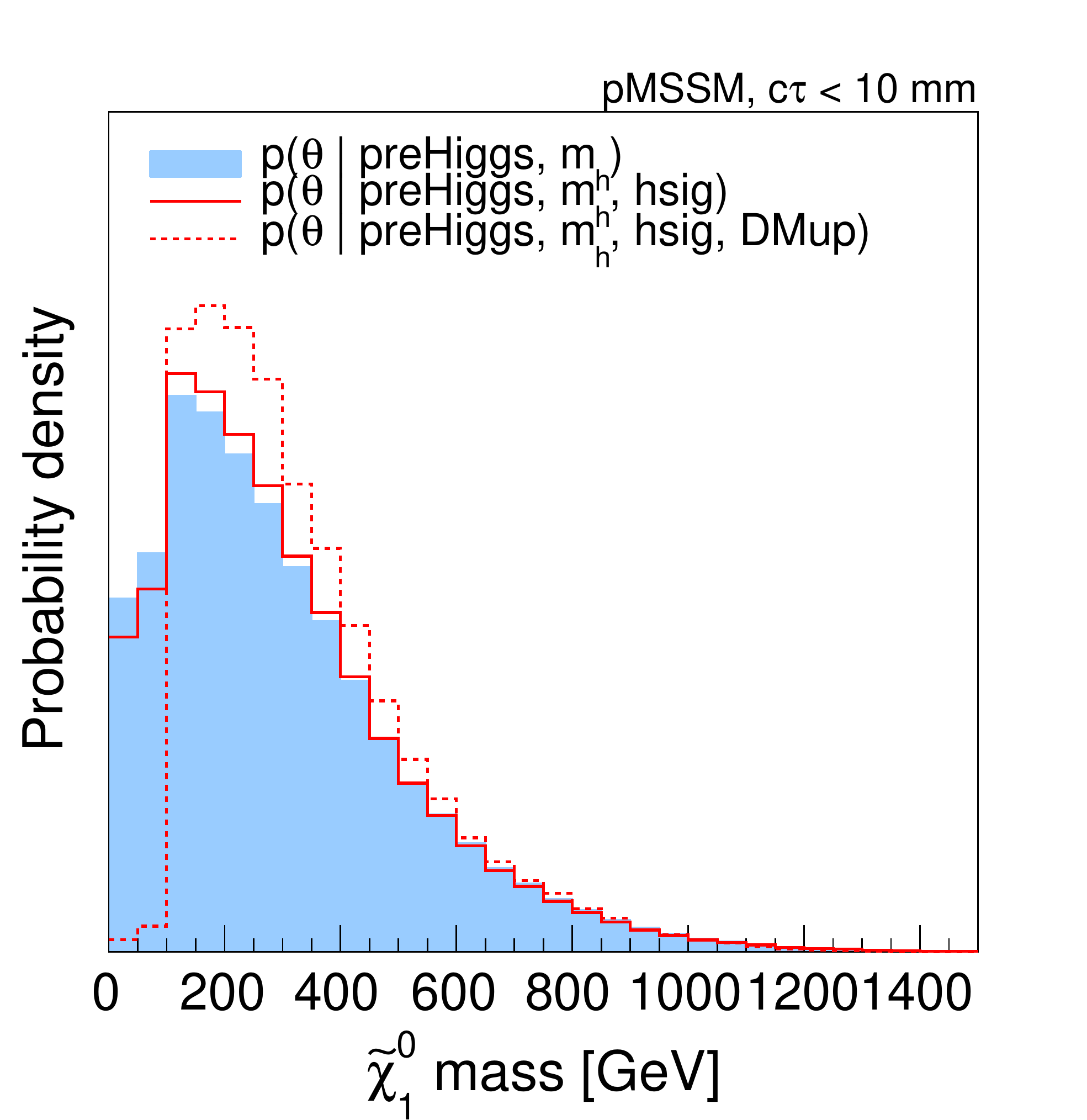}
\includegraphics[width=0.24\linewidth]{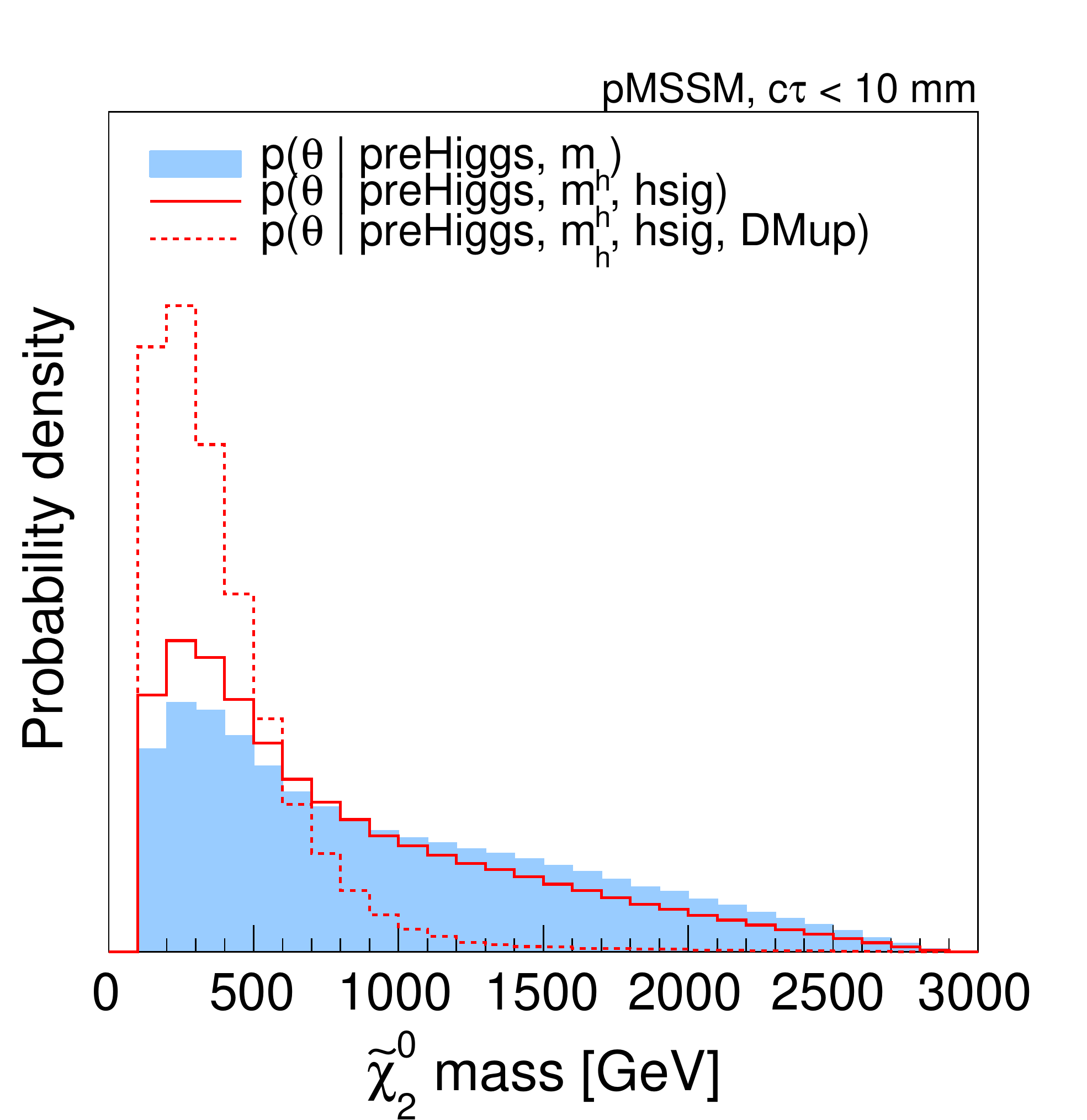}
\includegraphics[width=0.24\linewidth]{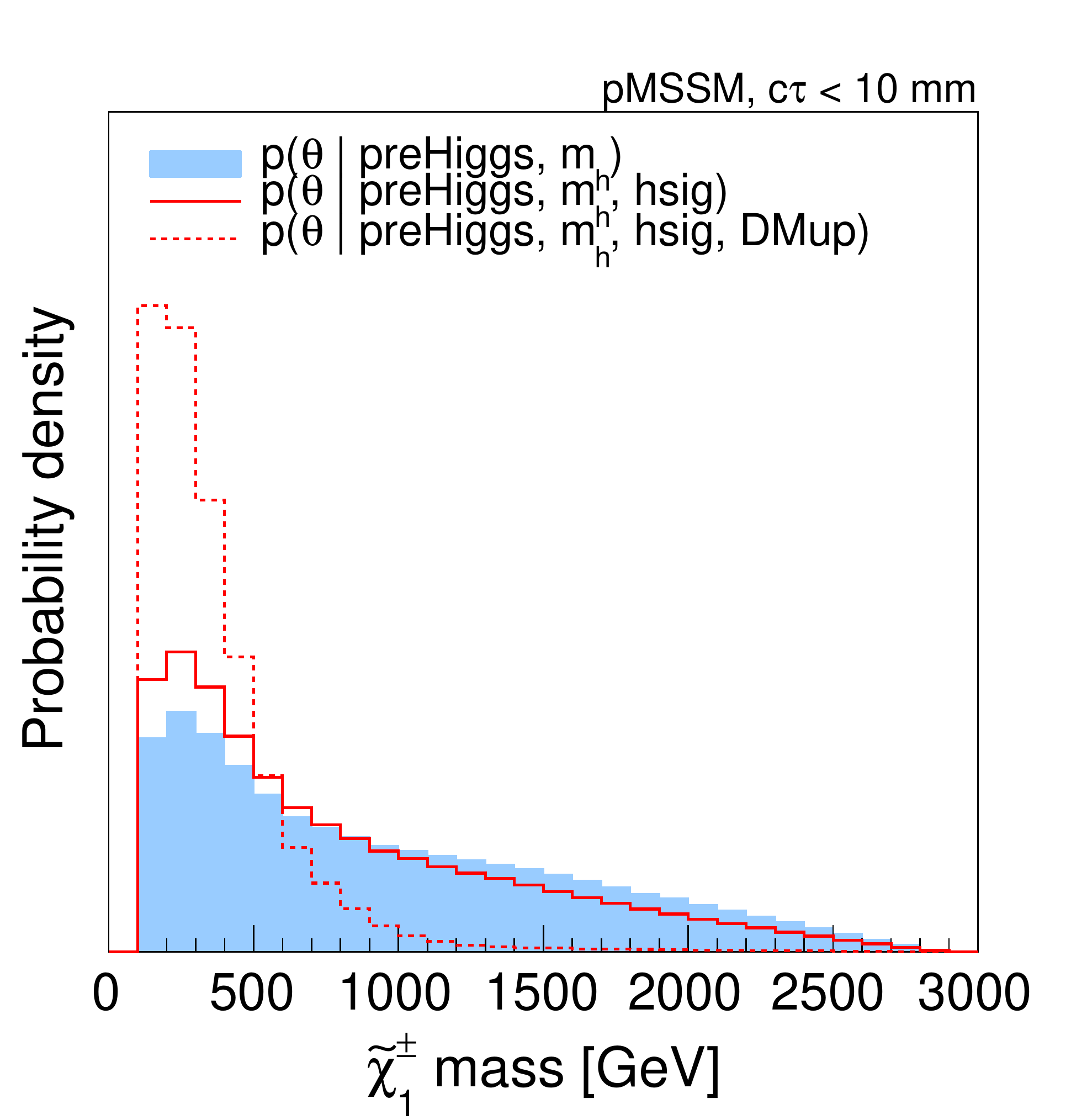}
\includegraphics[width=0.24\linewidth]{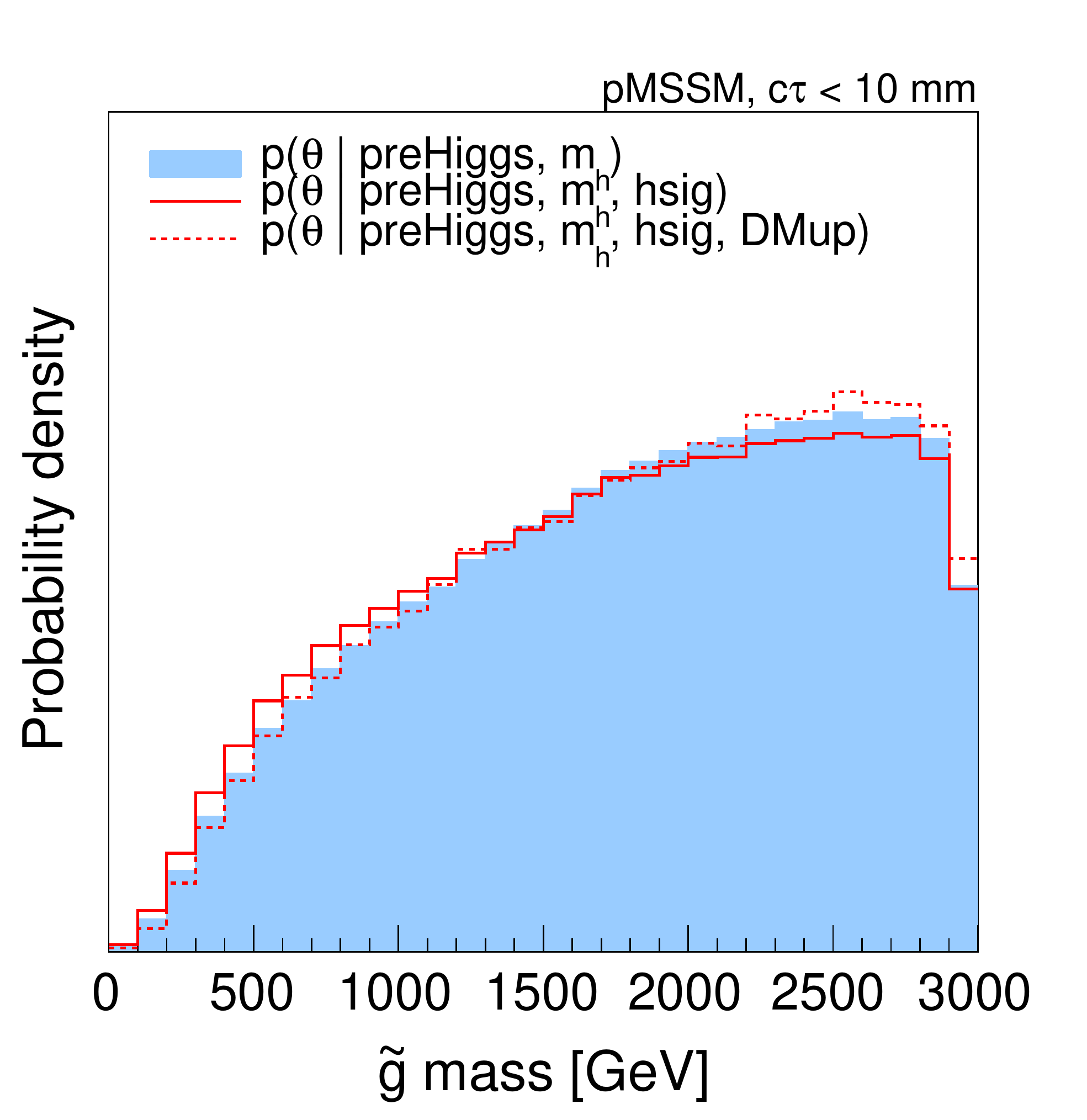}
\includegraphics[width=0.24\linewidth]{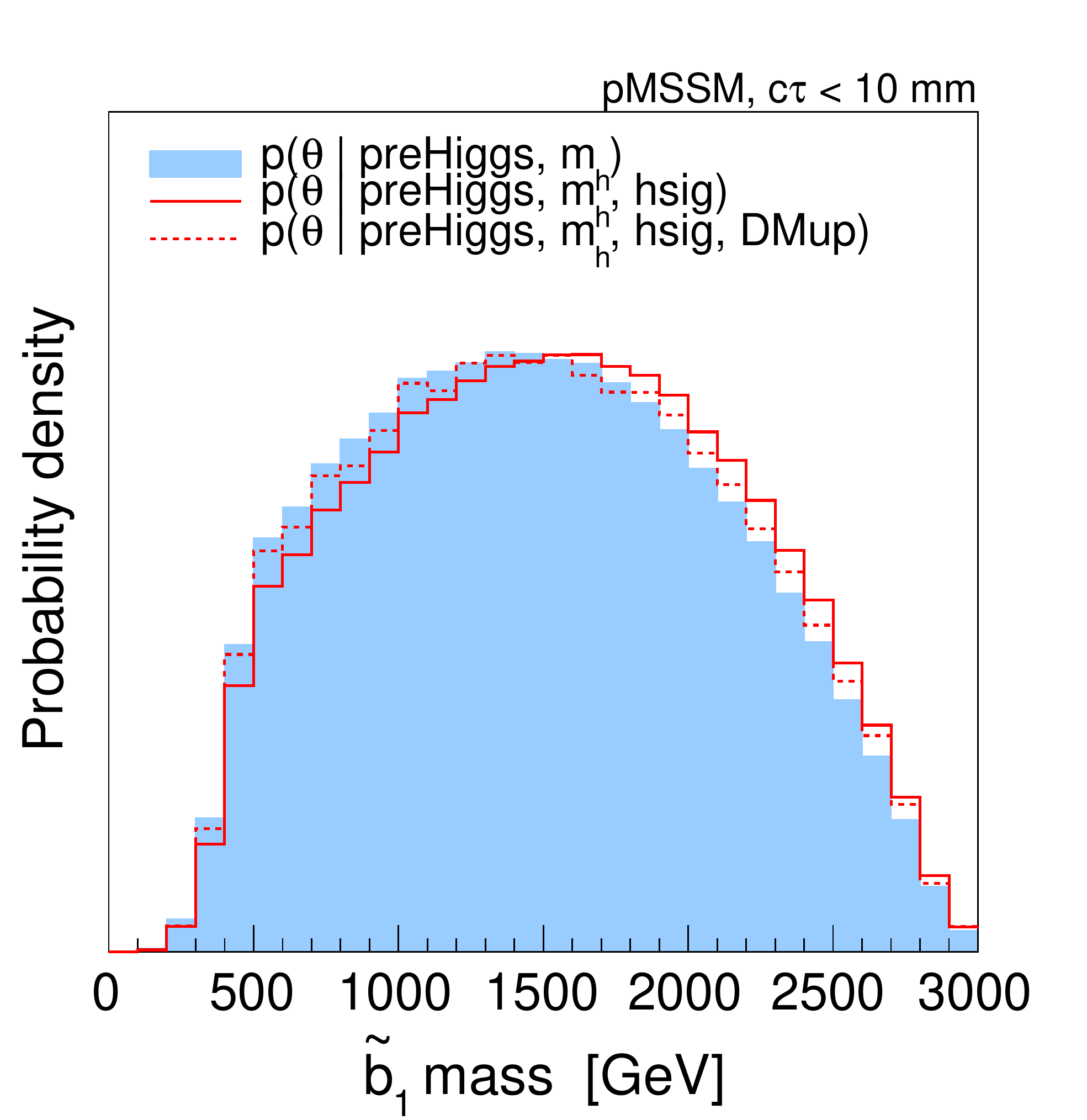}
\includegraphics[width=0.24\linewidth]{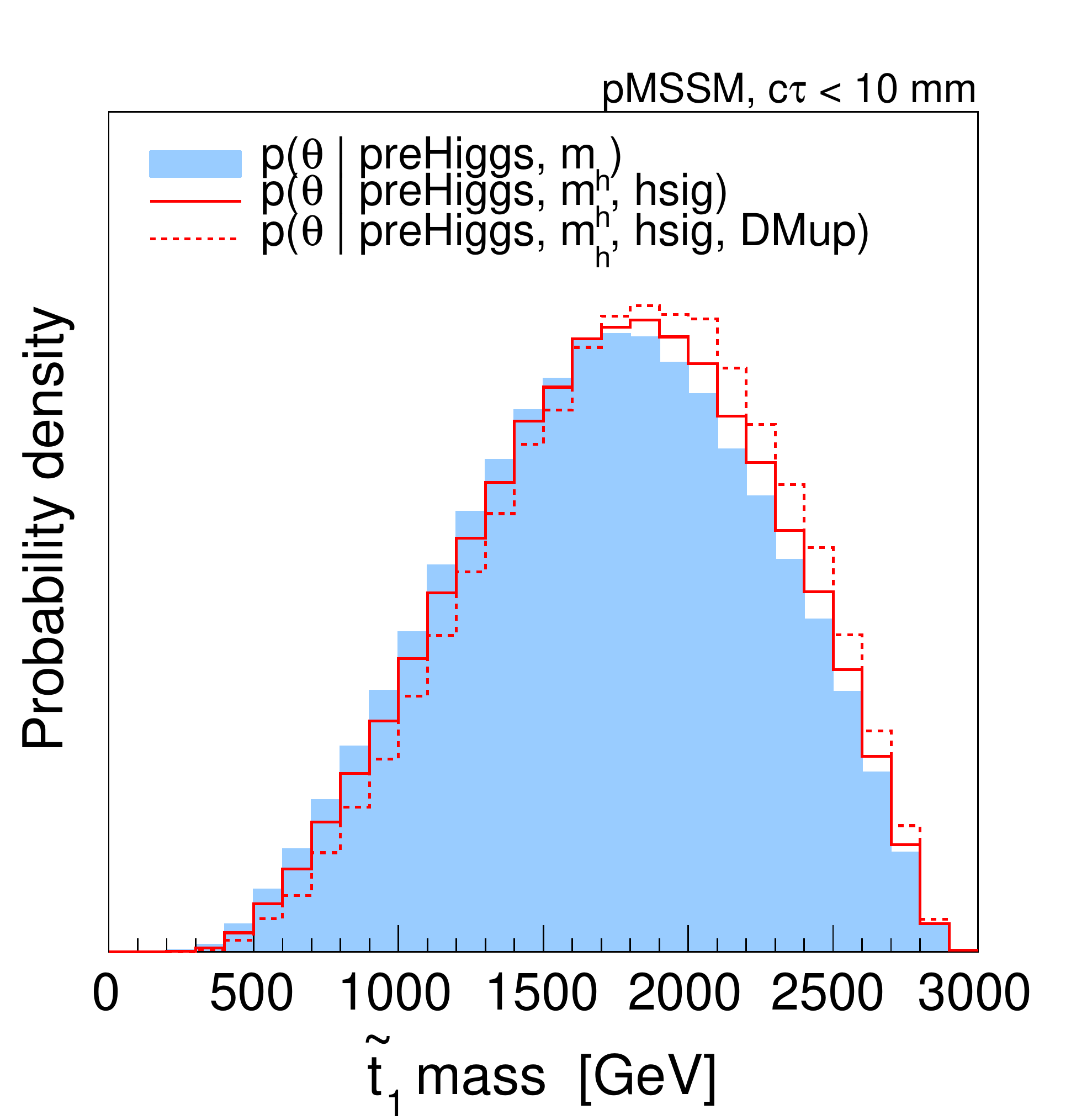}
\includegraphics[width=0.24\linewidth]{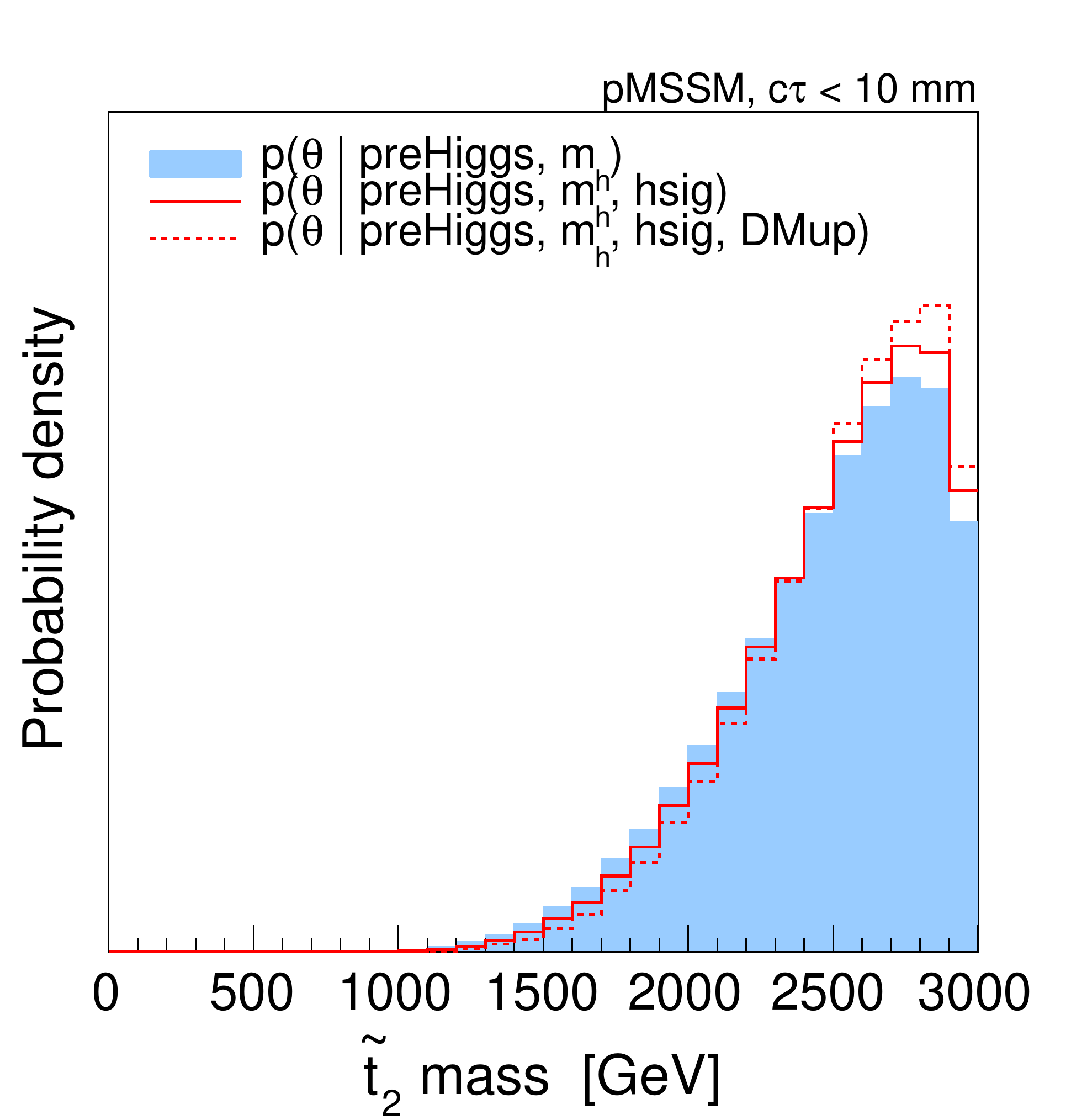}
\includegraphics[width=0.24\linewidth]{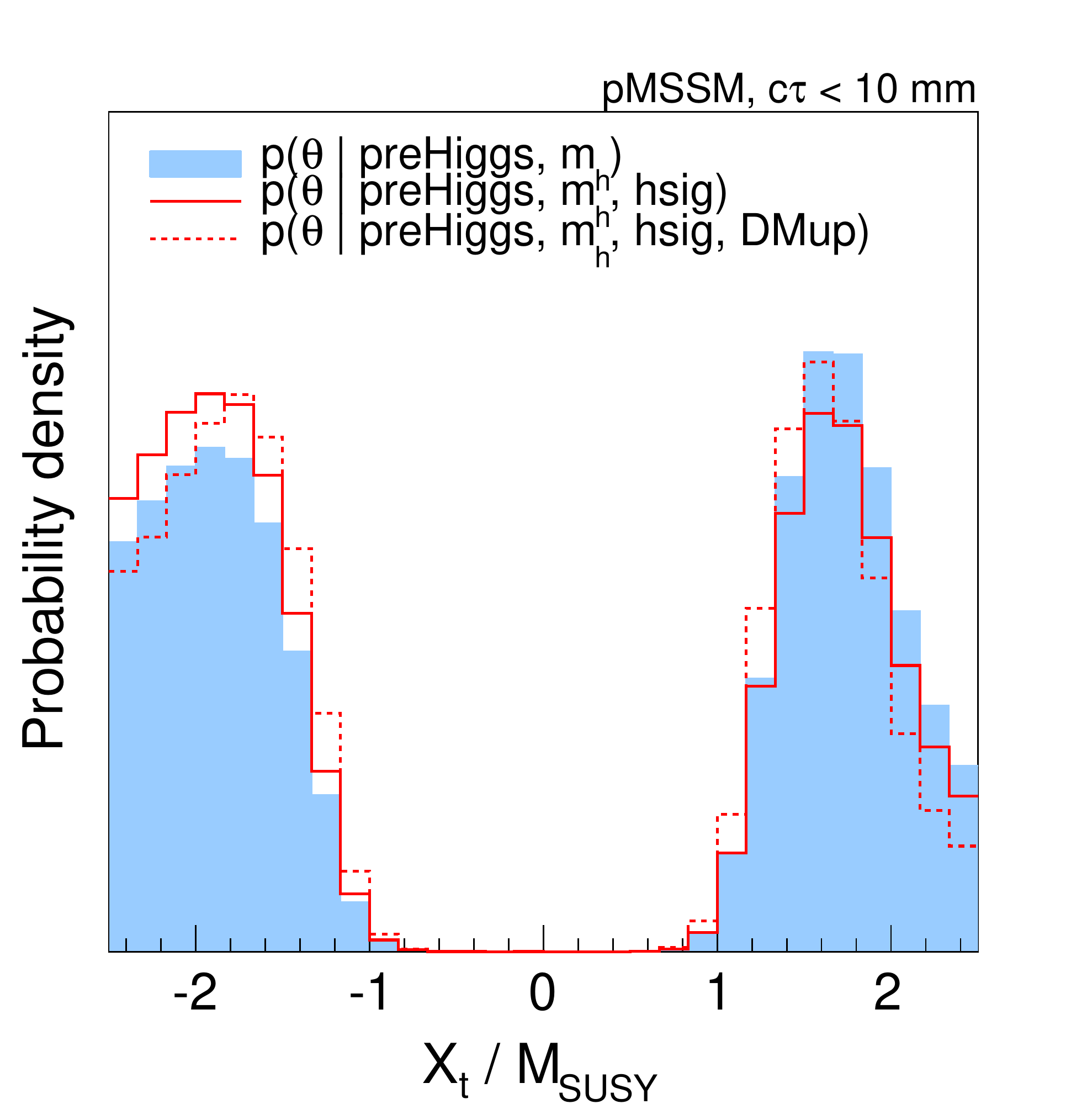}  
\caption{Marginalized 1D posterior densities for selected parameters and masses, showing the effect of the 
Higgs signal strength measurements. 
The light blue histograms show the distributions  based on the ``preHiggs'' measurements 
of Table~\ref{tab:preHiggs} plus requiring in addition $m_h\in [123,\,128]$~GeV.
The solid red lines, labelled ``hsig'', are the distributions when moreover taking into account 
the measured Higgs signal strengths in the various channels. The limits from searches for the heavy Higgses ($H$ and $A$) 
are also included in the red line-histograms, but have a totally negligible effect. 
The dashed red lines, labelled ``DMup'', include in addition an upper limit on the 
neutralino relic density and the recent direct DM detection limit from LUX as explained in the text.}
\label{fig:likehiggs1}
\end{center}
\end{figure}

\begin{figure}[t!]
\includegraphics[width=0.33\linewidth]{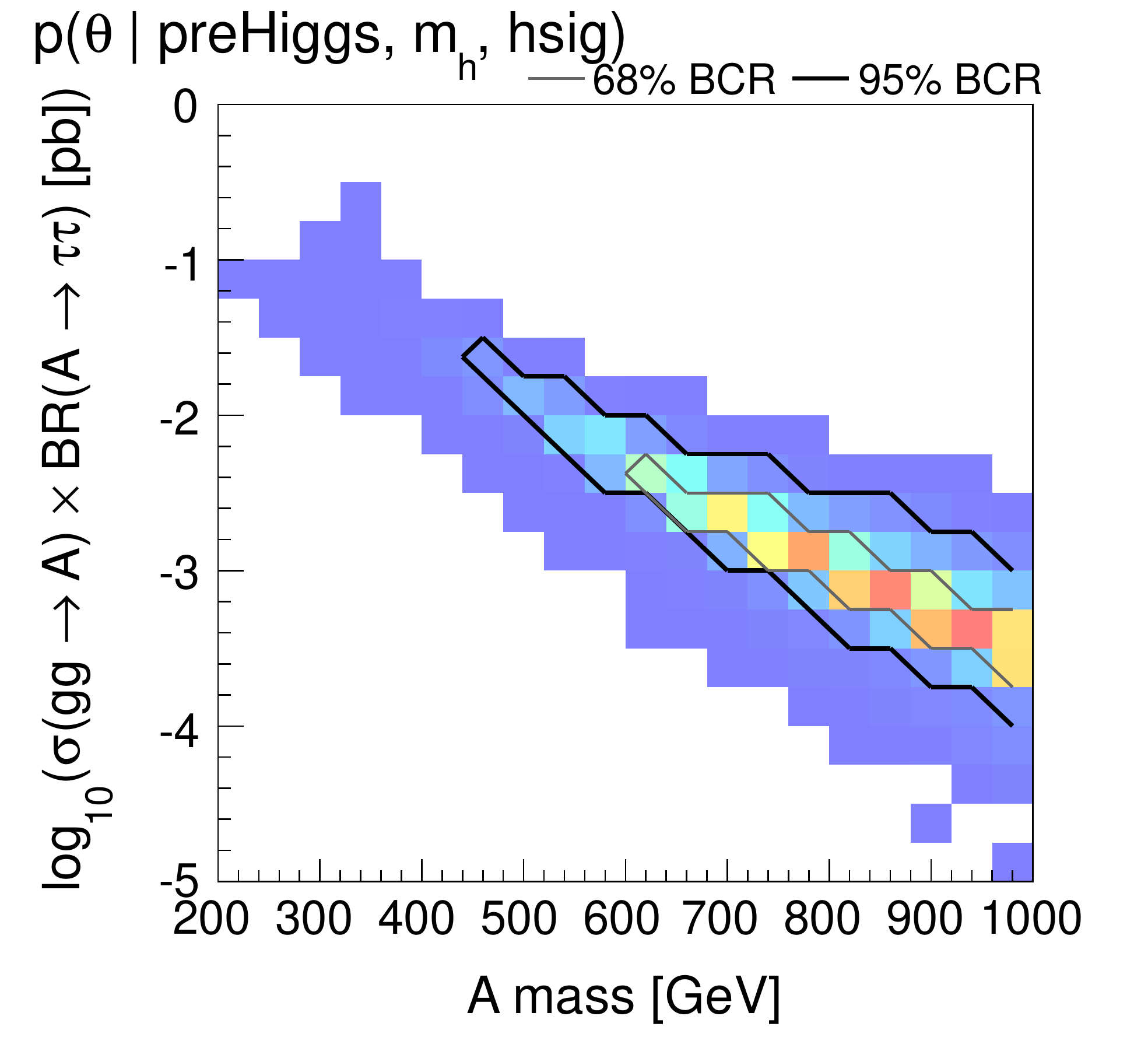}\includegraphics[width=0.33\linewidth]{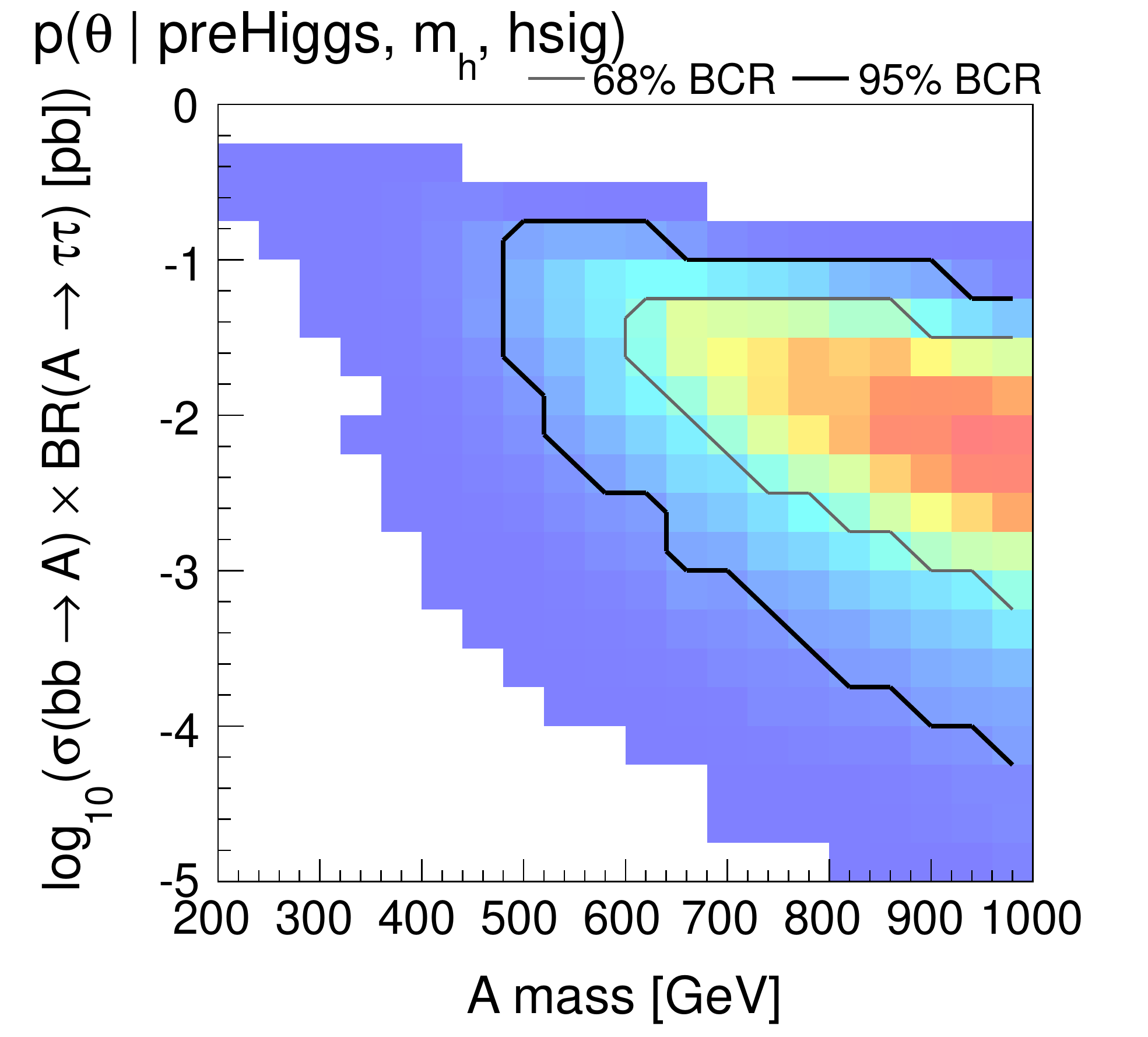}\includegraphics[width=0.33\linewidth]{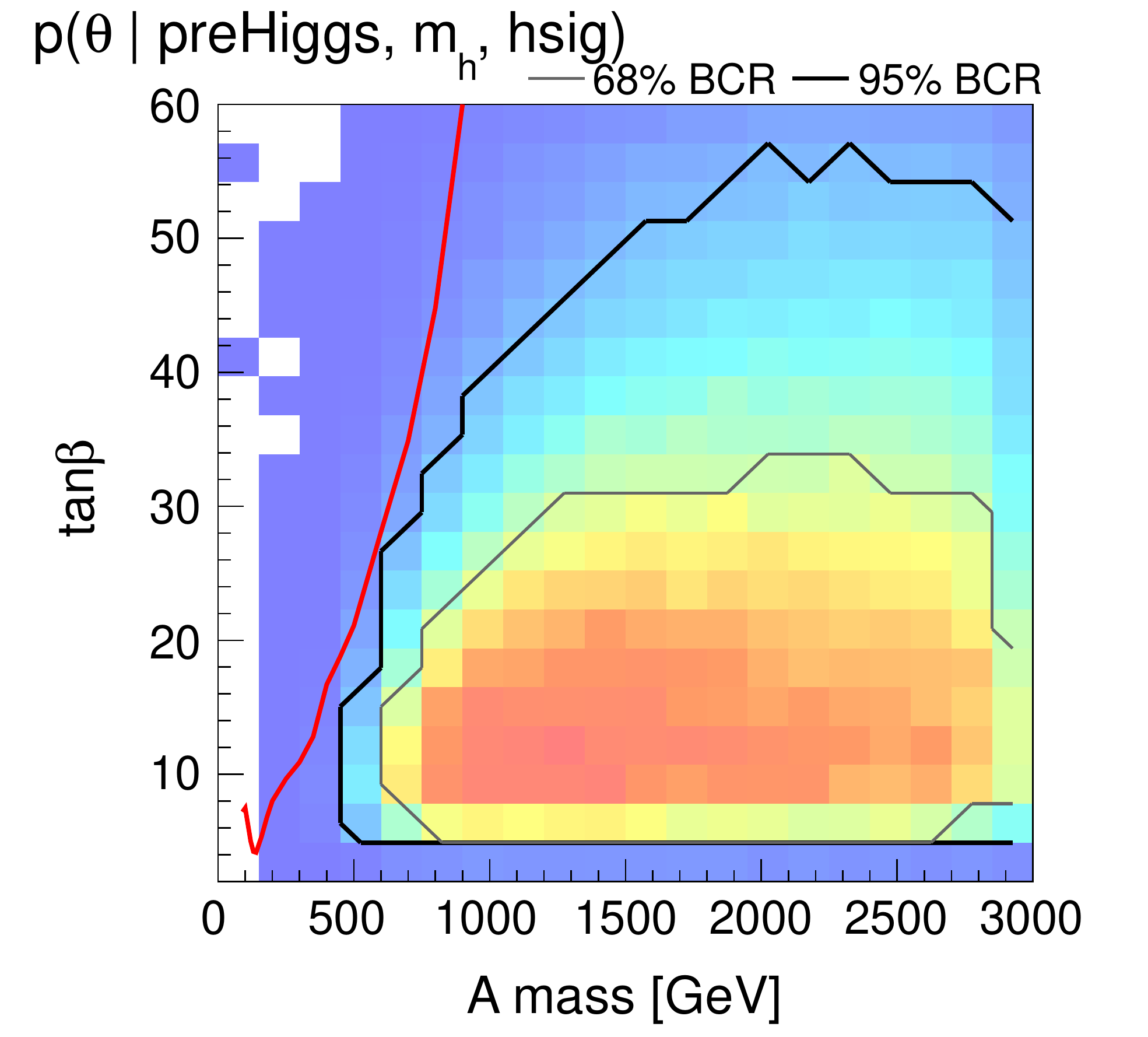}\\\includegraphics[width=0.33\linewidth]{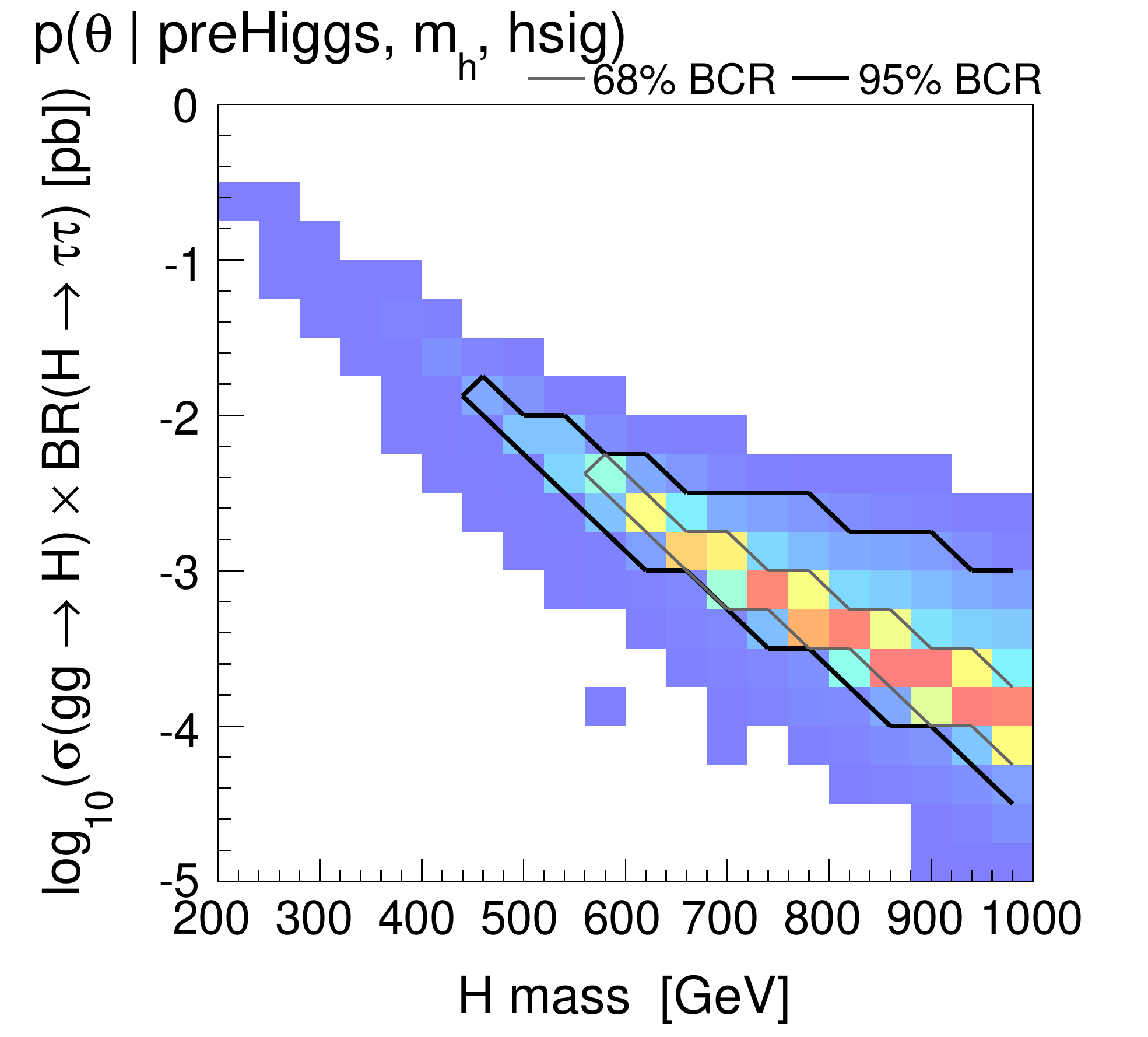}\includegraphics[width=0.33\linewidth]{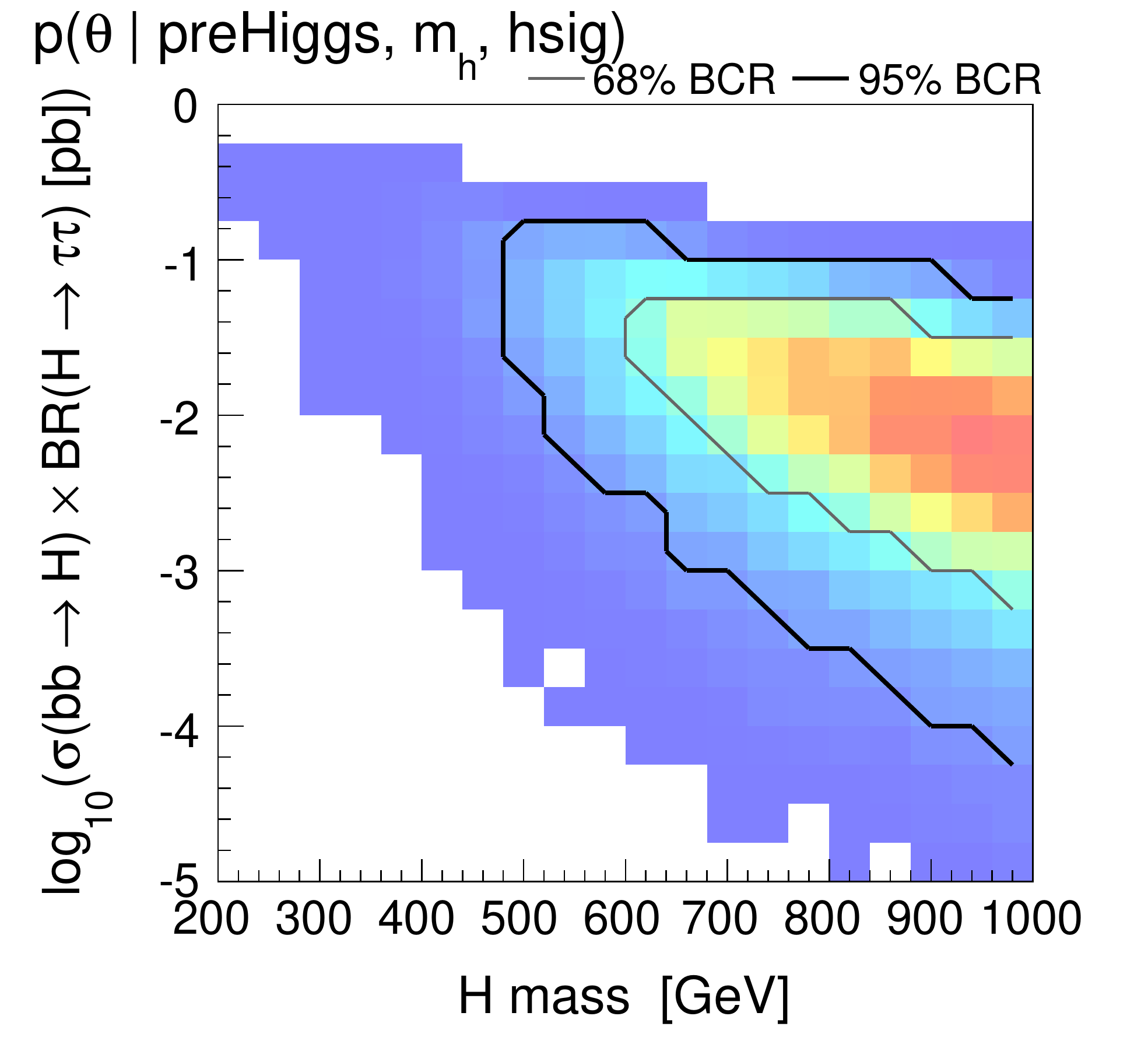}\includegraphics[width=0.33\linewidth]{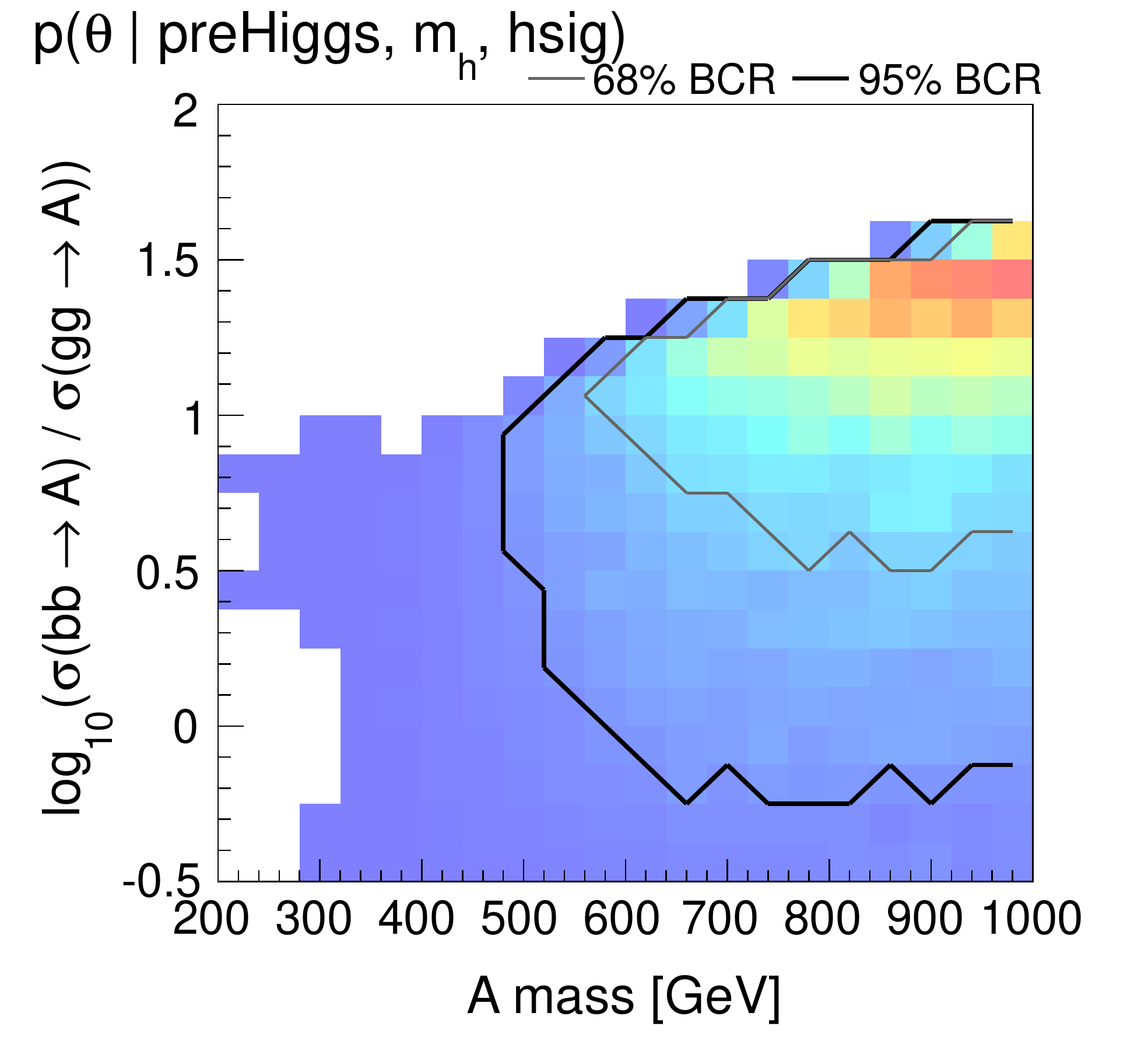}
\caption{Marginalized posterior densities in 2D for the heavy MSSM Higgses $A$ and $H$. The plots on the left and in the middle  show $\sigma \times \br$ in the $\tau\tau$ final state, from $bb$ and $gg$ production at $\sqrt{s} = 14$~TeV, versus the $A$ or $H$ mass. 
The top-right plot shows the posterior density in the $\tan\beta$ versus $m_A$ plane with the latest 95\%~CL from the CMS search for MSSM $H,A\to\tau\tau$~\cite{CMS-PAS-HIG-13-021} superimposed. 
The bottom-right plot compares $bb$ to $gg$ production as function of $m_A$.
In all plots, 
the probability density is represented by color shading, ranging from low values in blue to high values in red. The grey and black lines are contours of 68\% and 95\% Bayesian Credibility, respectively.}
\label{fig:heavierHiggses2D}
\end{figure}

As the next step, we include in addition the detailed properties of the $h$ signal in the computation of the likelihood as outlined in Section~\ref{sec:higgslikeli}. 
The effects of the Higgs observations on the pMSSM parameters and on the particle masses are shown in Fig.~\ref{fig:likehiggs1}.  
In these plots, the light blue histograms show the distributions  based on the ``preHiggs'' measurements 
of Table~\ref{tab:preHiggs} plus requiring in addition $m_h\in [123,\,128]$~GeV, \ie\ they correspond 
to the blue line-histograms of Fig.~\ref{fig:sampling1}.
The solid red lines are the distributions when moreover taking into account the measured Higgs signal strengths 
in the various channels as outlined in Section~2. Note that the limits from the MSSM $H,A\to \tau\tau$ searches,  
which are also included in the red line-histograms, have a negligible effect. 
(For completeness, a plot of the $\tan\beta$ versus $m_A$ plane is given in Fig.~\ref{fig:heavierHiggses2D}.)
Finally, the dashed red lines also take into account upper limits from the DM relic density 
and direct DM searches, as explained in Section~\ref{sec:dmconst}.

Let us first discuss the effect of the Higgs measurements, \ie\ consider the solid red lines only. 
We observe a significant preference for small or negative 
$\mu$ and smaller $\tan\beta$ values when including the Higgs signal strength likelihood. 
The main reason is the $\mu\tan\beta$ correction to the bottom Yukawa coupling~\cite{Carena:1999py,Eberl:1999he}, which for large $\tan\beta$ and large positive (negative) $\mu$ enhances (reduces) $\Gamma(h\to b\bar b)$ and the total $h$ width, hence reducing (increasing) all signal strengths except $\mu(Vh \to b\bar{b})$.
The preference for positive $\mu$ comes from the slight excess in the VBF and VH channels of $\gamma\gamma$ (mainly seen by ATLAS). In Ref.~\cite{Belanger:2013xza}, $\mu({\rm VBF+VH}, \gamma\gamma) = 1.72 \pm 0.59$ is found, while other combined signal strengths are fully compatible with 1 at 68\% CL. An overall excess (negative $\mu$) is therefore preferred over a general deficit (positive $\mu$). To a good approximation, the correction to the bottom Yukawa coupling is given by
\begin{eqnarray}
   \Delta_b \equiv \frac{\Delta m_b}{m_b} \simeq 
   \left[ \frac{2\alpha_s}{3\pi} \mu m_{\tilde{g}}\, I(m_{\tilde{g}}^2, m_{\tilde{b}_1}^2, m_{\tilde{b}_2}^2) +
            \frac{\lambda_t^2}{16\pi^2} A_t \mu \, I(\mu^2,m_{\tilde{t}_1}^2, m_{\tilde{t}_2}^2) \right] \tan\beta \,, \label{Deltamb} 
\end{eqnarray}
where $I(x,y,z)$ is of order $1/{\rm max}(x,y,z)$~\cite{Djouadi:2005gj}. The shifts to higher values of all four stops and sbottoms masses and to lower values for the gluino mass also come from $\Delta_b$. In addition, negative values of $A_t$ are more likely after taking into account the Higgs likelihood. This comes from the second term of Eq.~(\ref{Deltamb}): in order to compensate the first, dominant term, ${\rm sgn}(A_t \mu) = -{\rm sgn}(\mu)$ is required, hence a negative $A_t$.
The tree-level coupling $hbb$ also has an effect. It is given by
\beq
   g_{hbb} \simeq 1 - \frac{M_Z^2} {2 m_A^2} \sin 4\beta \tan \beta \,,
\eeq
for $m_A \gg M_Z$~\cite{Djouadi:2005gj}, and disfavors relatively light $A$ and $H$, with masses below about $700$~GeV (the effect from imposing the CMS $H,A \to \tau\tau$ limit is subdominant). Finally, $M_2$ shows a slight preference towards negative values. This is a direct consequence of the asymmetry in the distribution of $\mu$, since ${\rm sgn}(\mu M_2) > 0$ is required for $\Delta a^{\rm SUSY}_{\mu} > 0$ as suggested by the data. 

The DM constraints, on the other hand, have a dramatic effect on the bino and higgsino mass parameters and in turn on the chargino and neutralino masses. Since a mostly bino $\tilde \chi_1^0$  generically leads to a large $\Omega_{\tilde{\chi}^0_1}h^2$, low values of $M_1$ are strongly disfavored. 
The preferred solutions have a relevant higgsino or wino fraction of the LSP; 
therefore $\tilde\chi^\pm_1$ and $\tilde\chi^0_2$ masses  below about 1~TeV are strongly favored.  
At the same time, very light LSP masses below about 100~GeV are severely limited because of the LEP bound on the chargino mass. 
The preferred value of $\tan\beta$ is also affected; in fact, the preference for lower $\tanb$ coming from the Higgs signal strengths is removed by the DM constraints. The reason for this is an enhancement of $A$-funnel annihilation to comply with the upper limit on $\omhsq$.

%%% signal strengths and hbb and total width %%%

\begin{figure}[t!]
\includegraphics[width=0.24\linewidth]{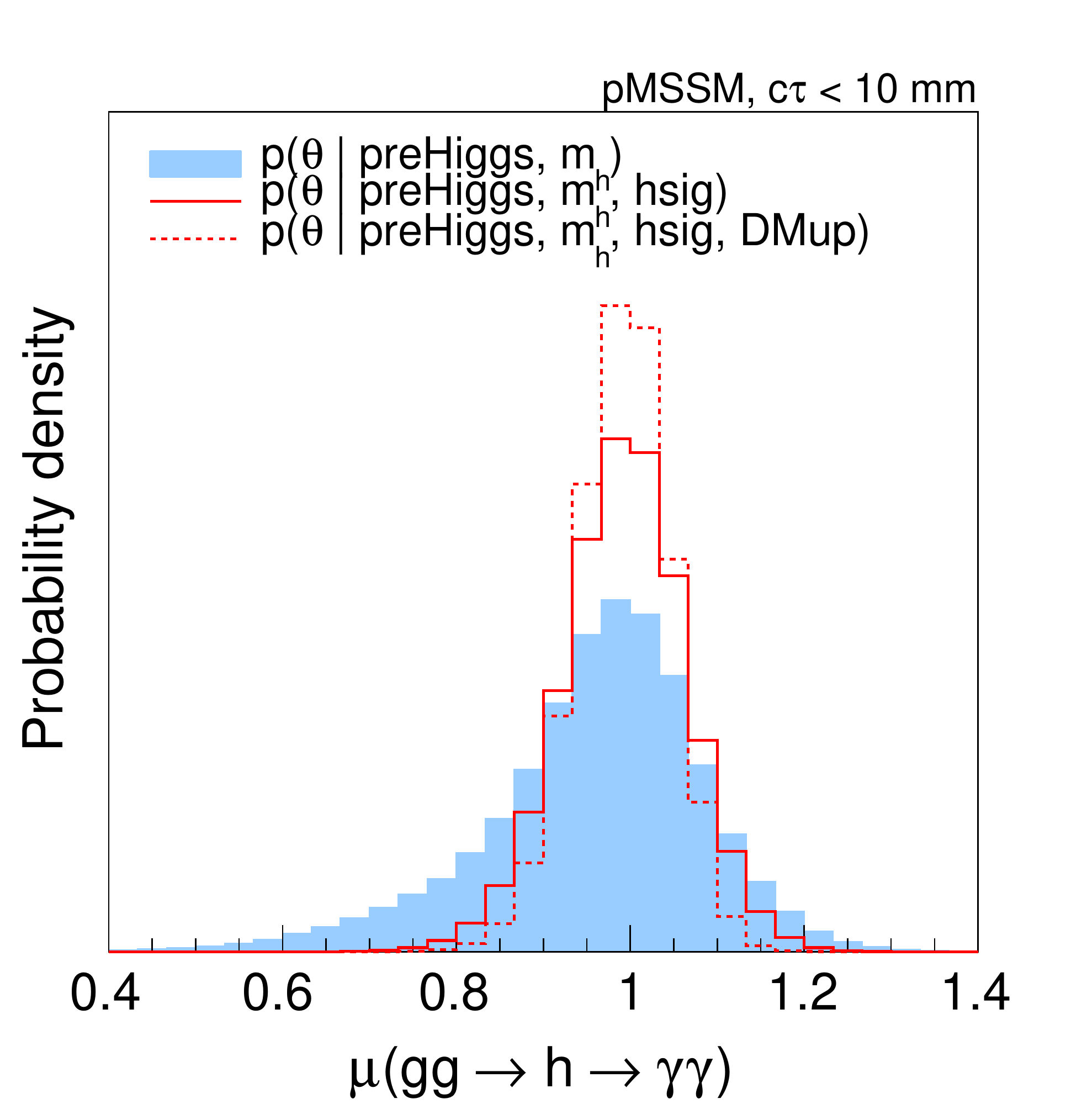}
\includegraphics[width=0.24\linewidth]{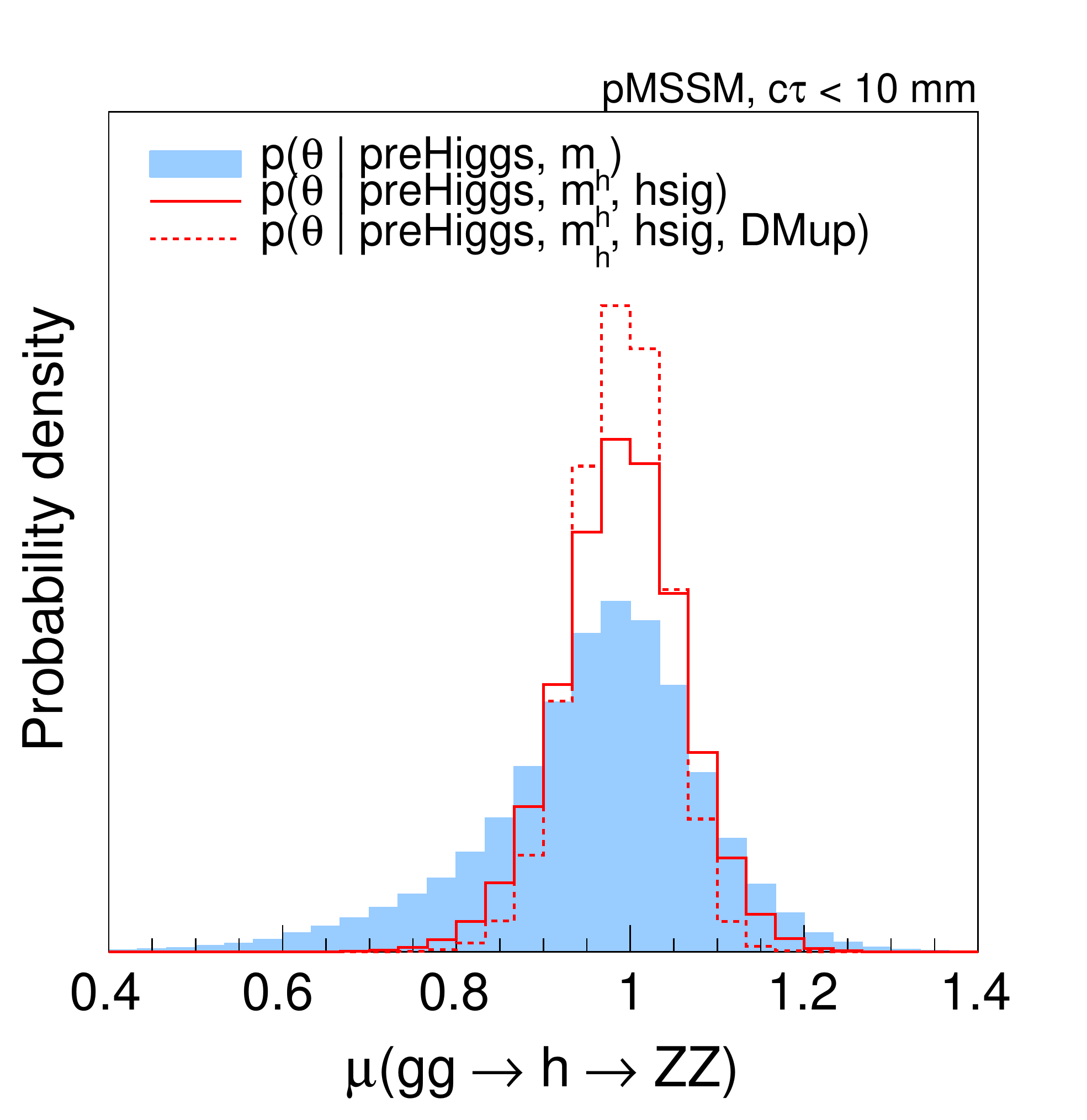}
\includegraphics[width=0.24\linewidth]{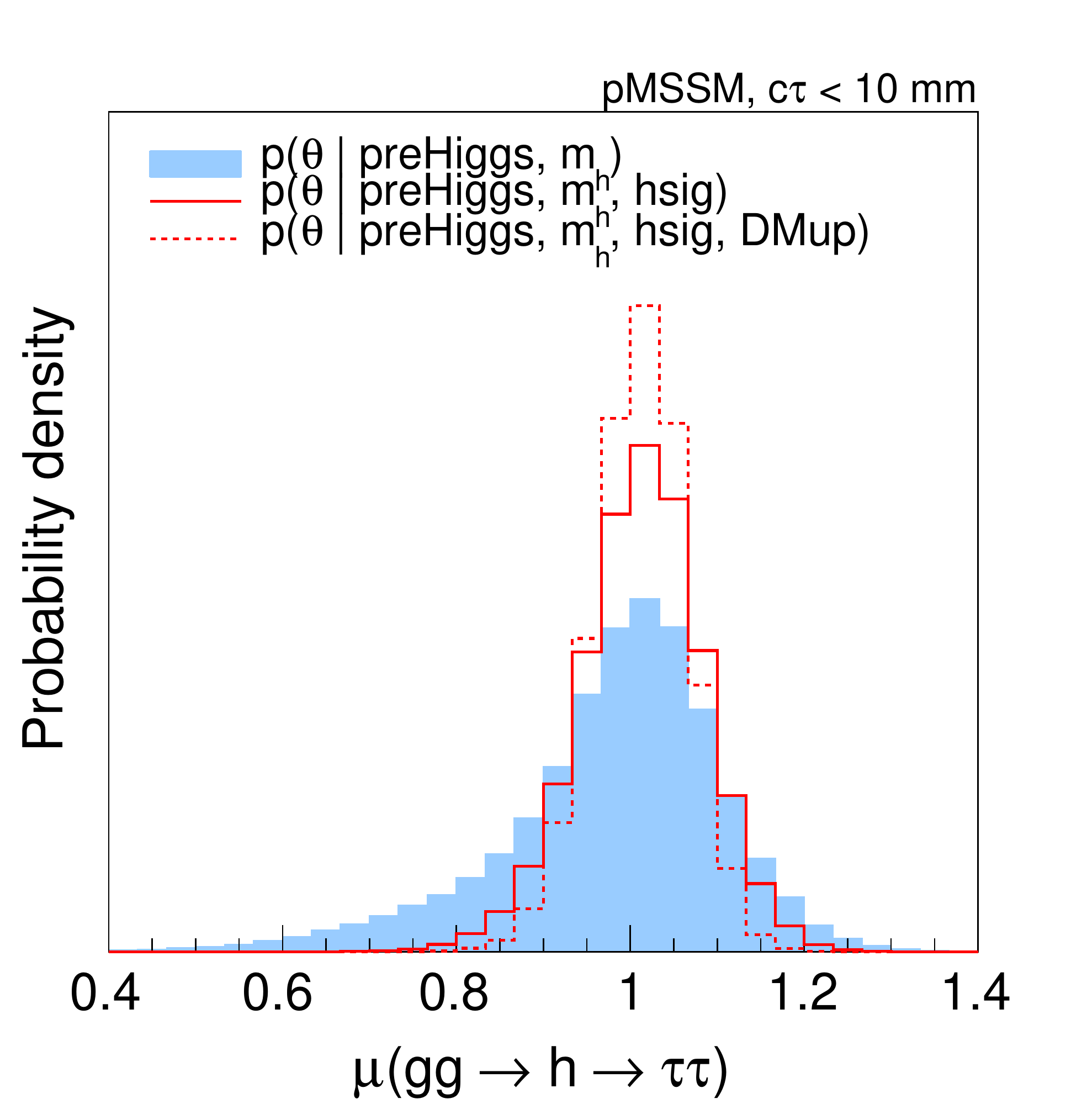}\\
\includegraphics[width=0.24\linewidth]{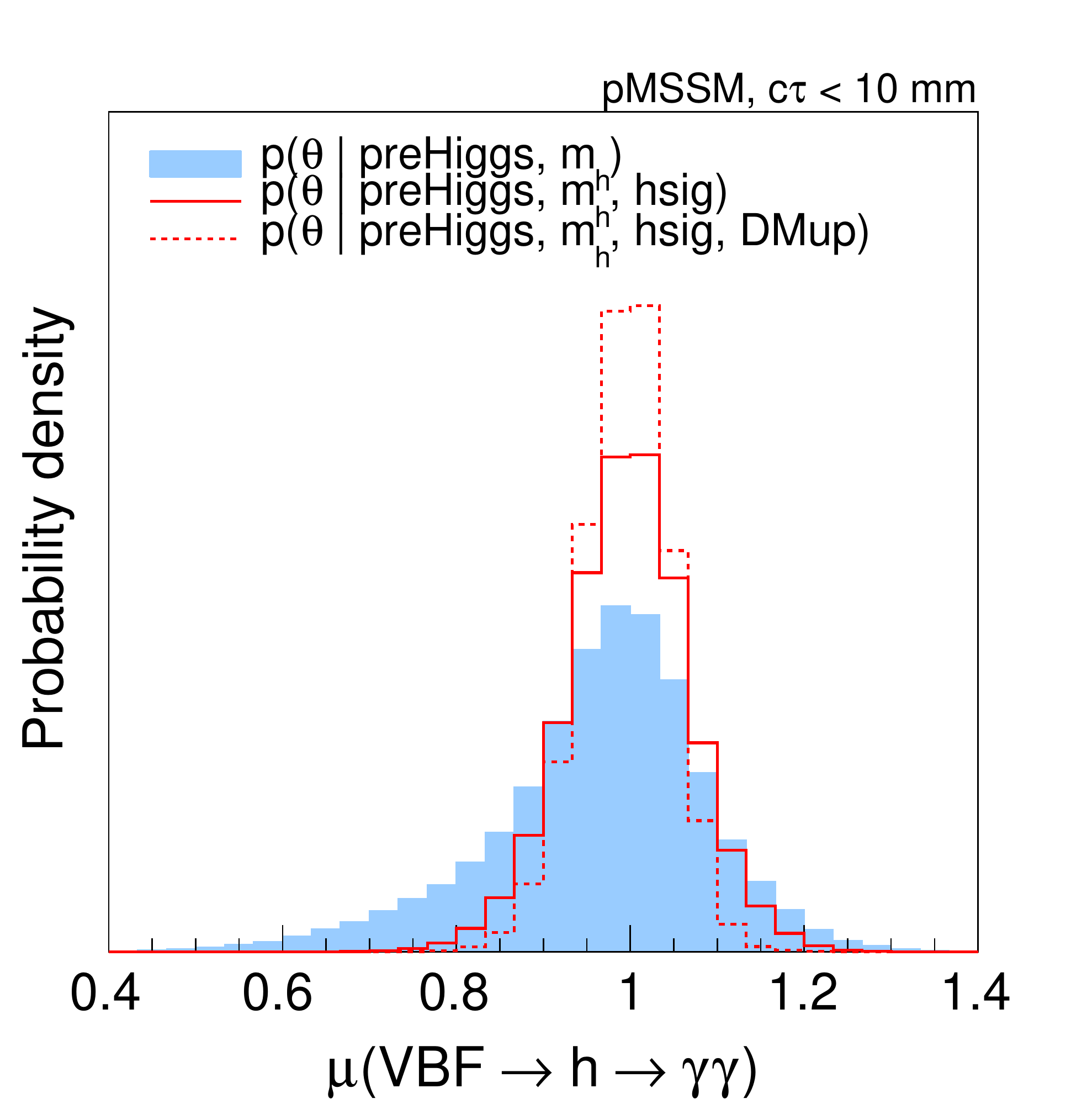}
\includegraphics[width=0.24\linewidth]{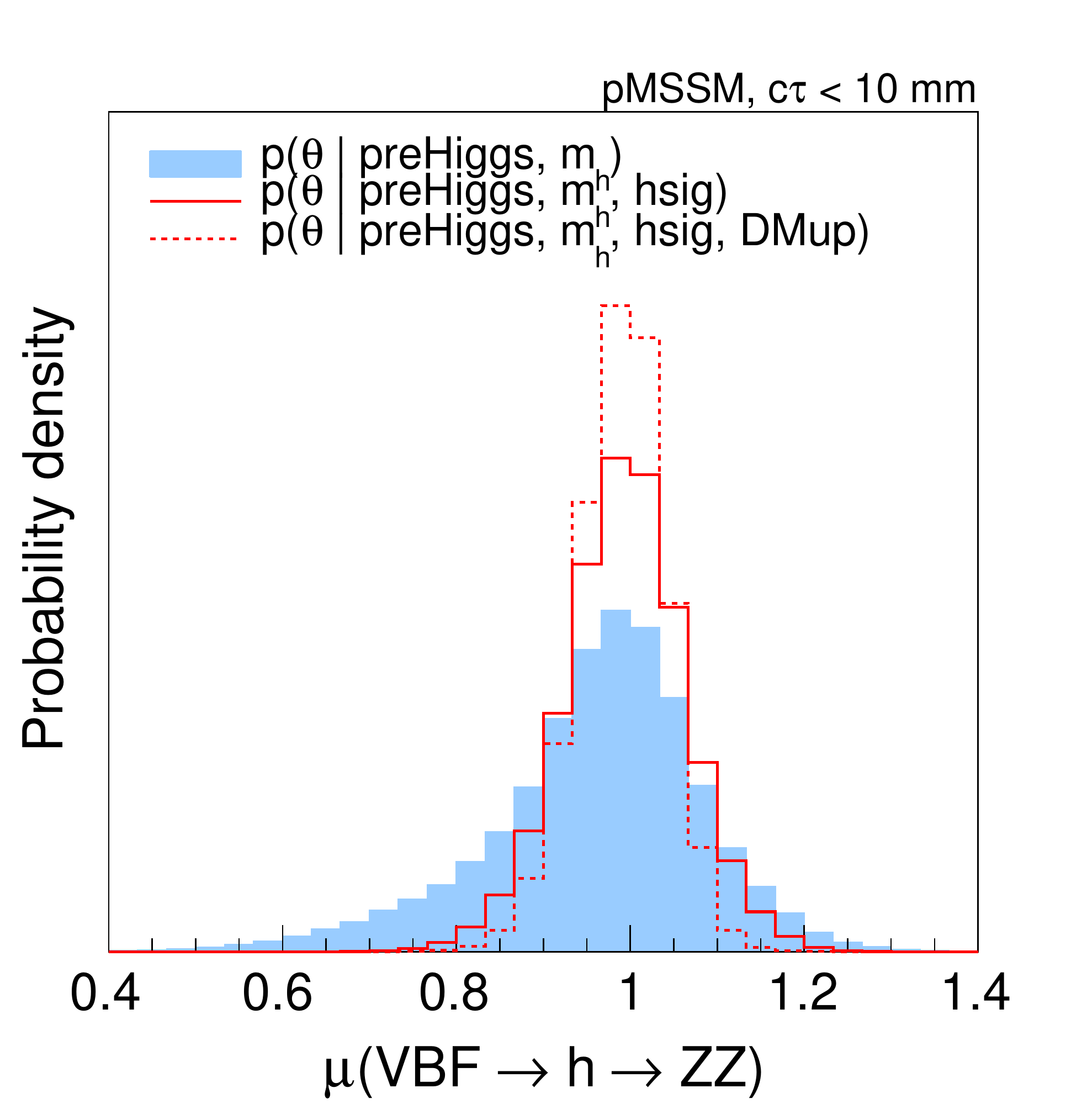}
\includegraphics[width=0.24\linewidth]{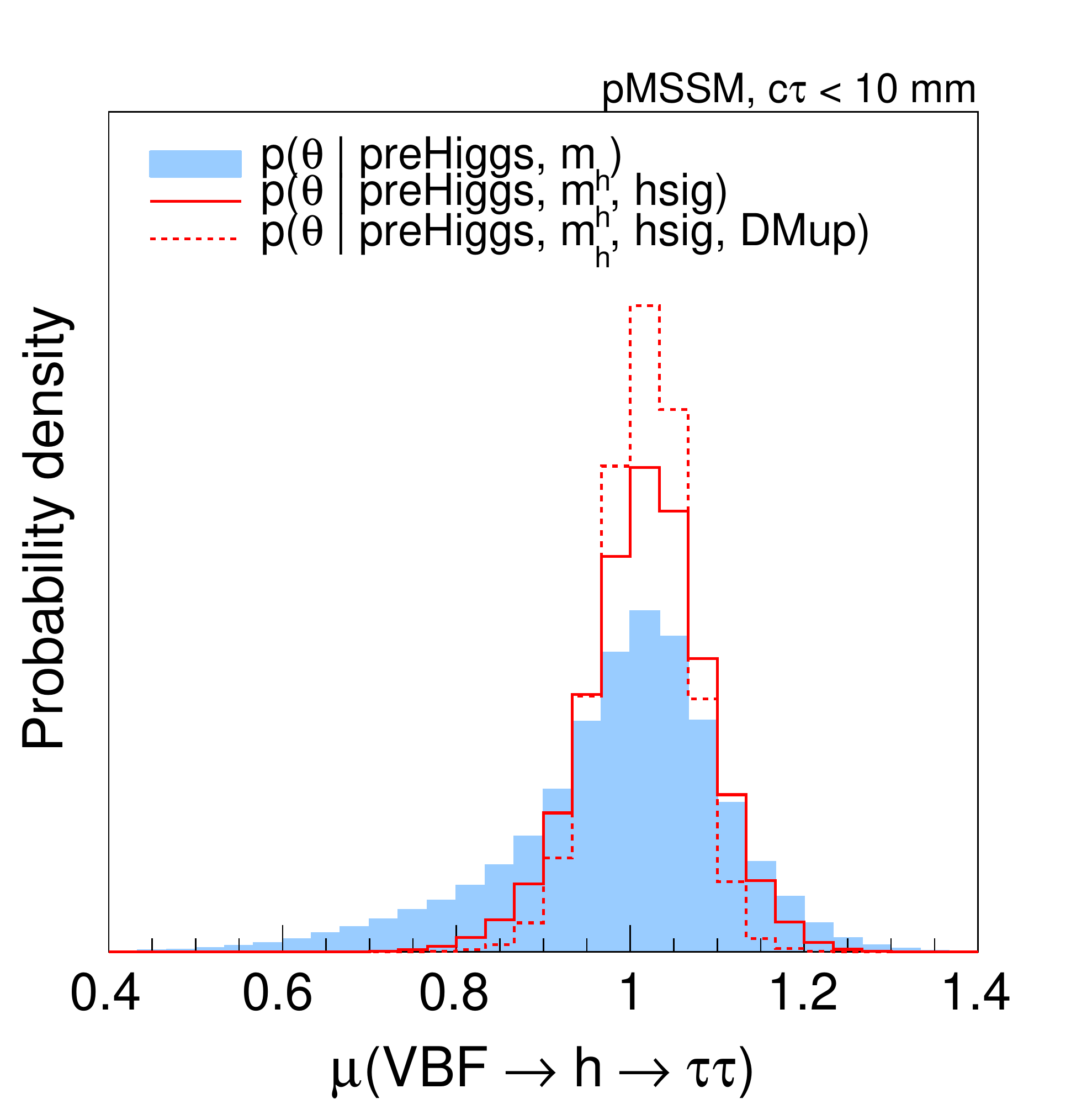}
\includegraphics[width=0.24\linewidth]{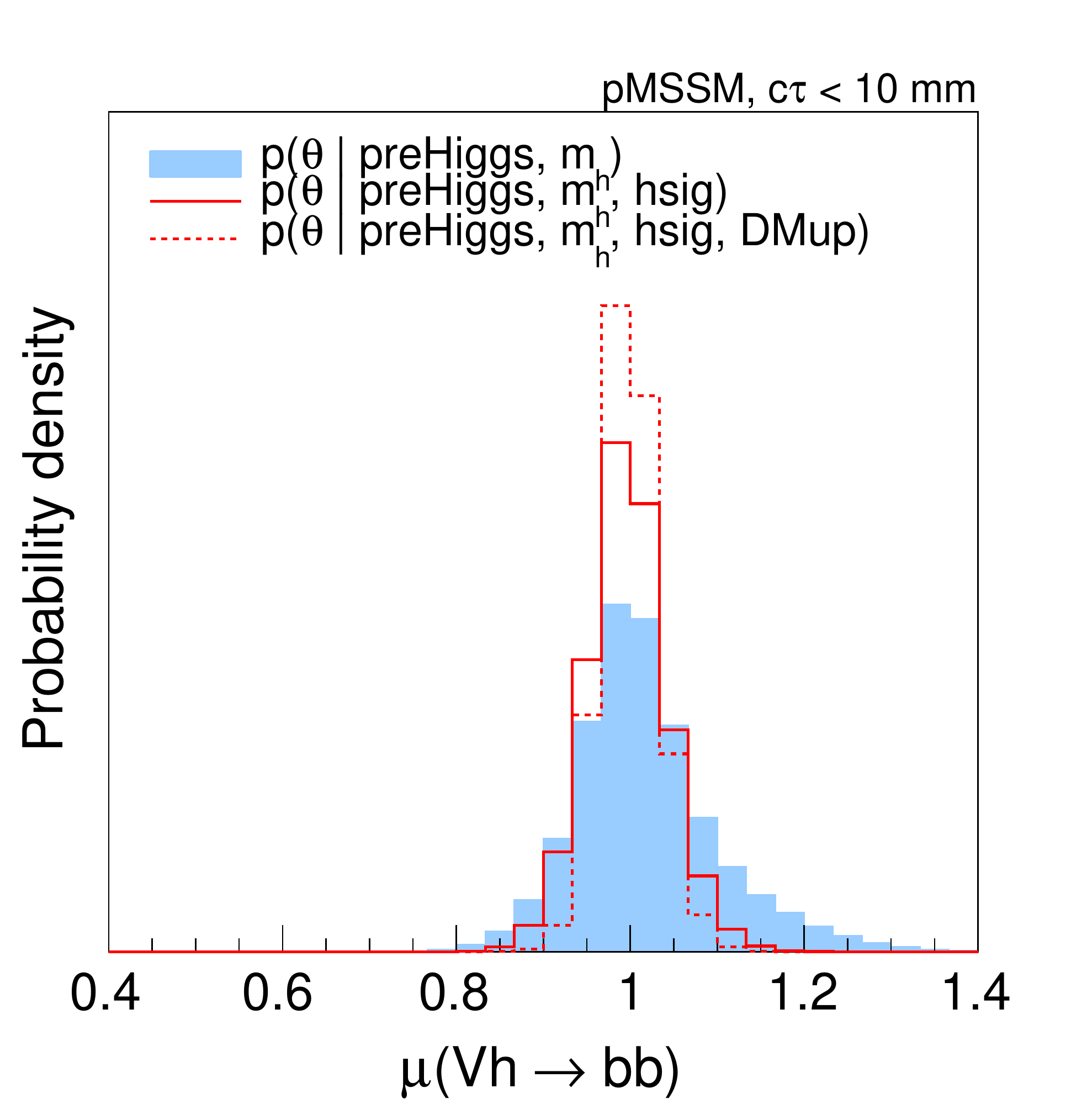}
\caption{Same as Fig.~\ref{fig:likehiggs1} but for the relevant $h$ signal strengths.}
\label{fig:likehiggs3}
\end{figure}

\begin{figure}[t!]
\begin{center}
\includegraphics[width=0.30\linewidth]{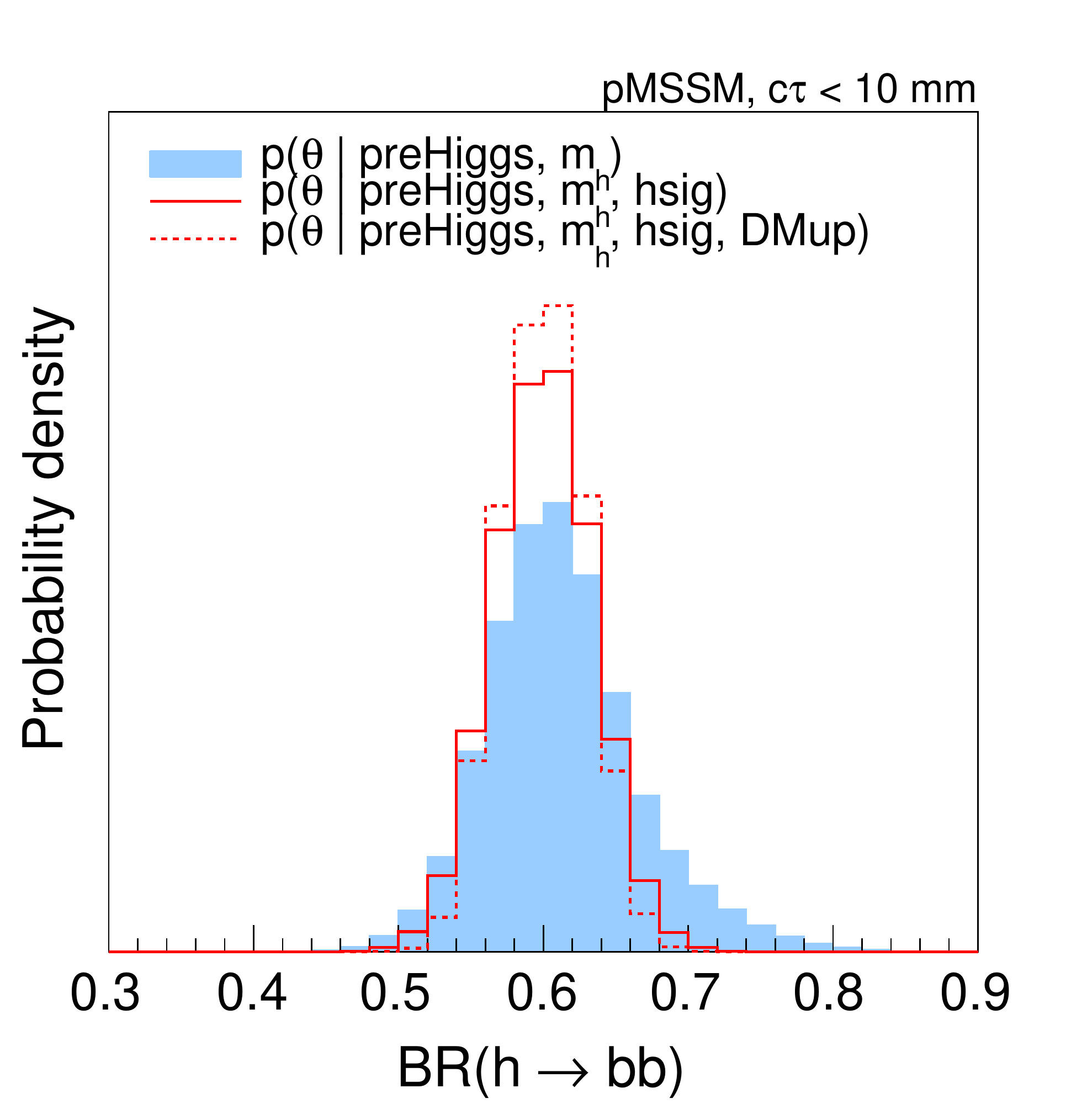}
\includegraphics[width=0.30\linewidth]{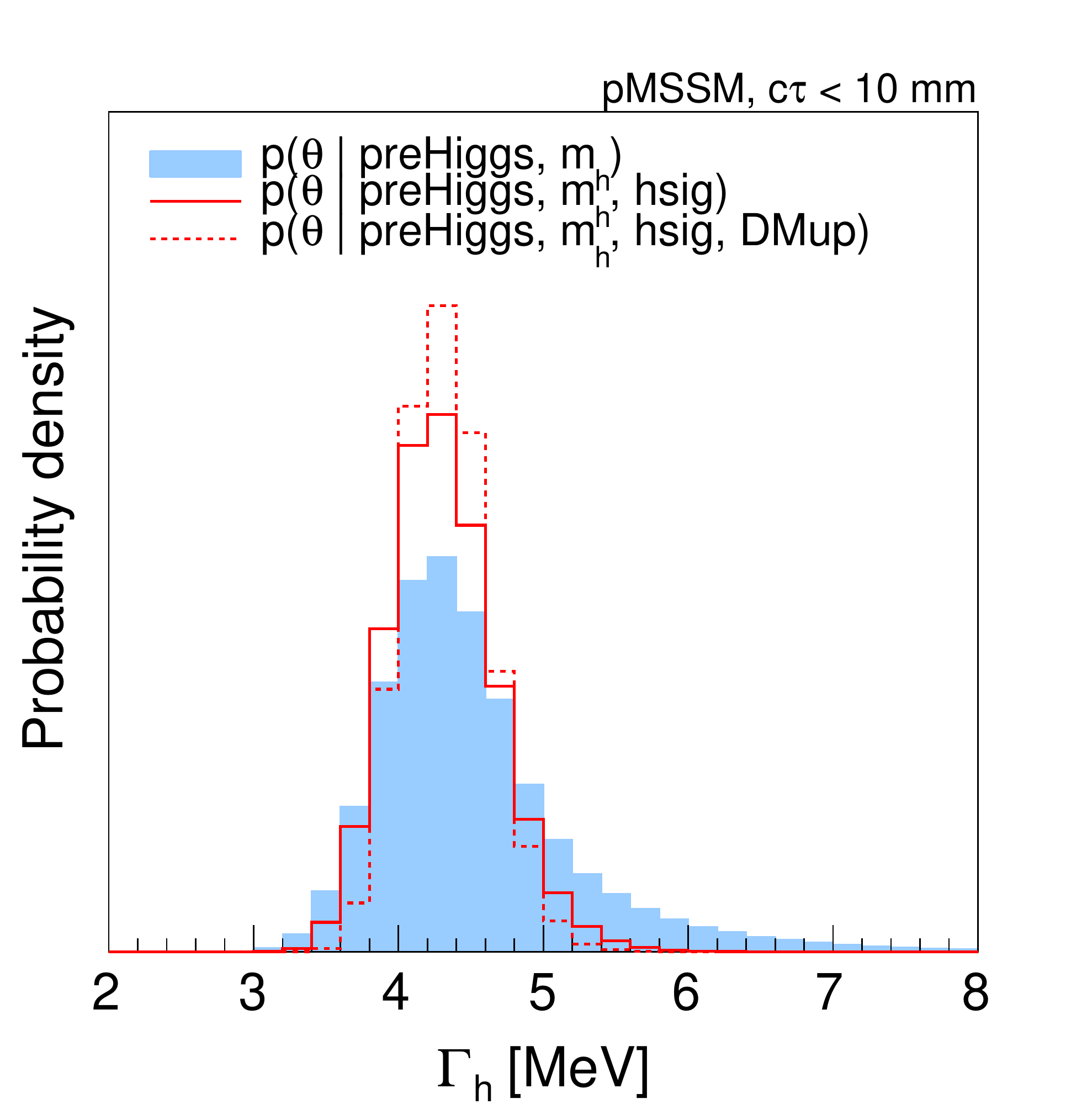}\\
\includegraphics[width=0.30\linewidth]{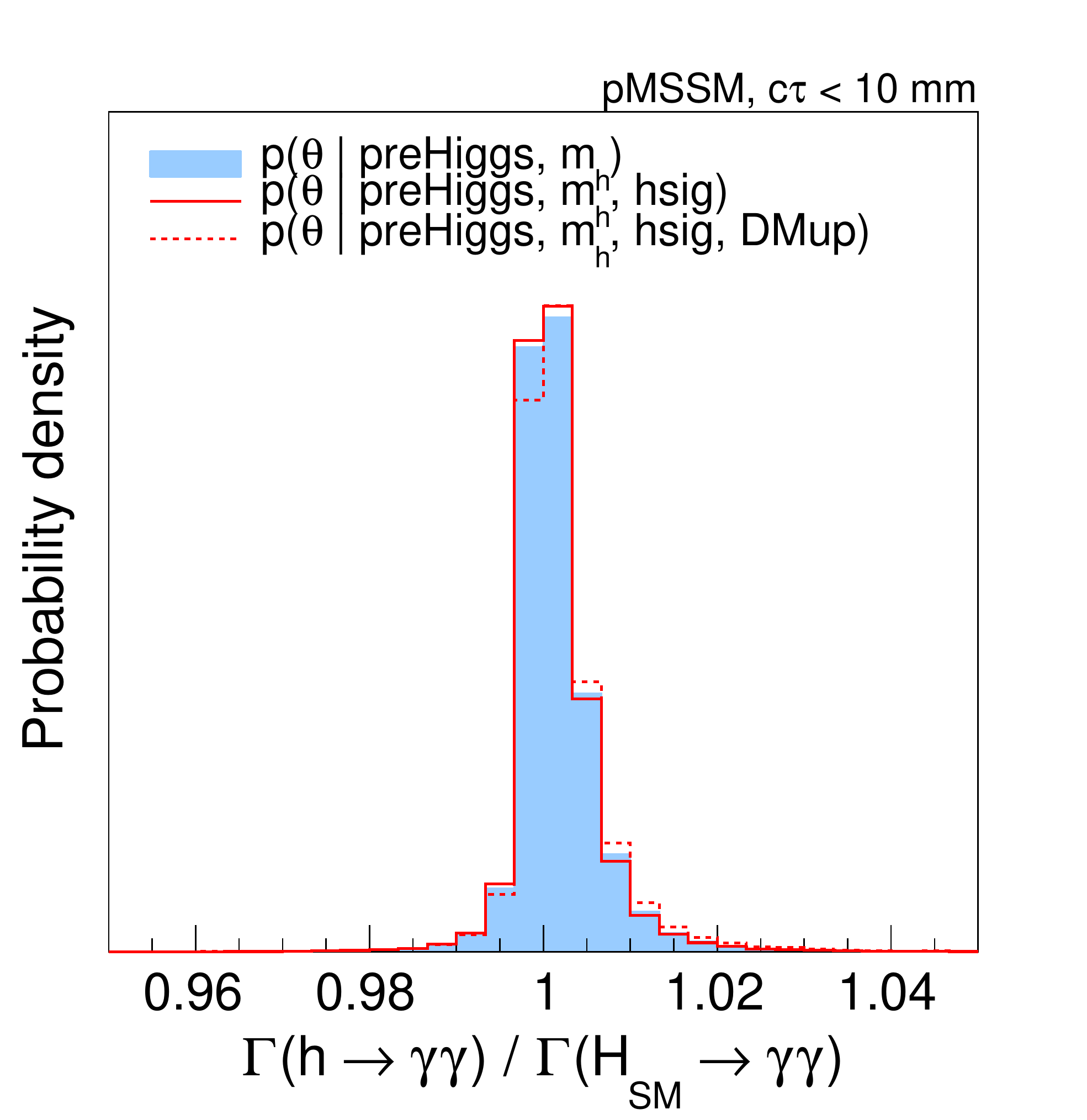}
\includegraphics[width=0.30\linewidth]{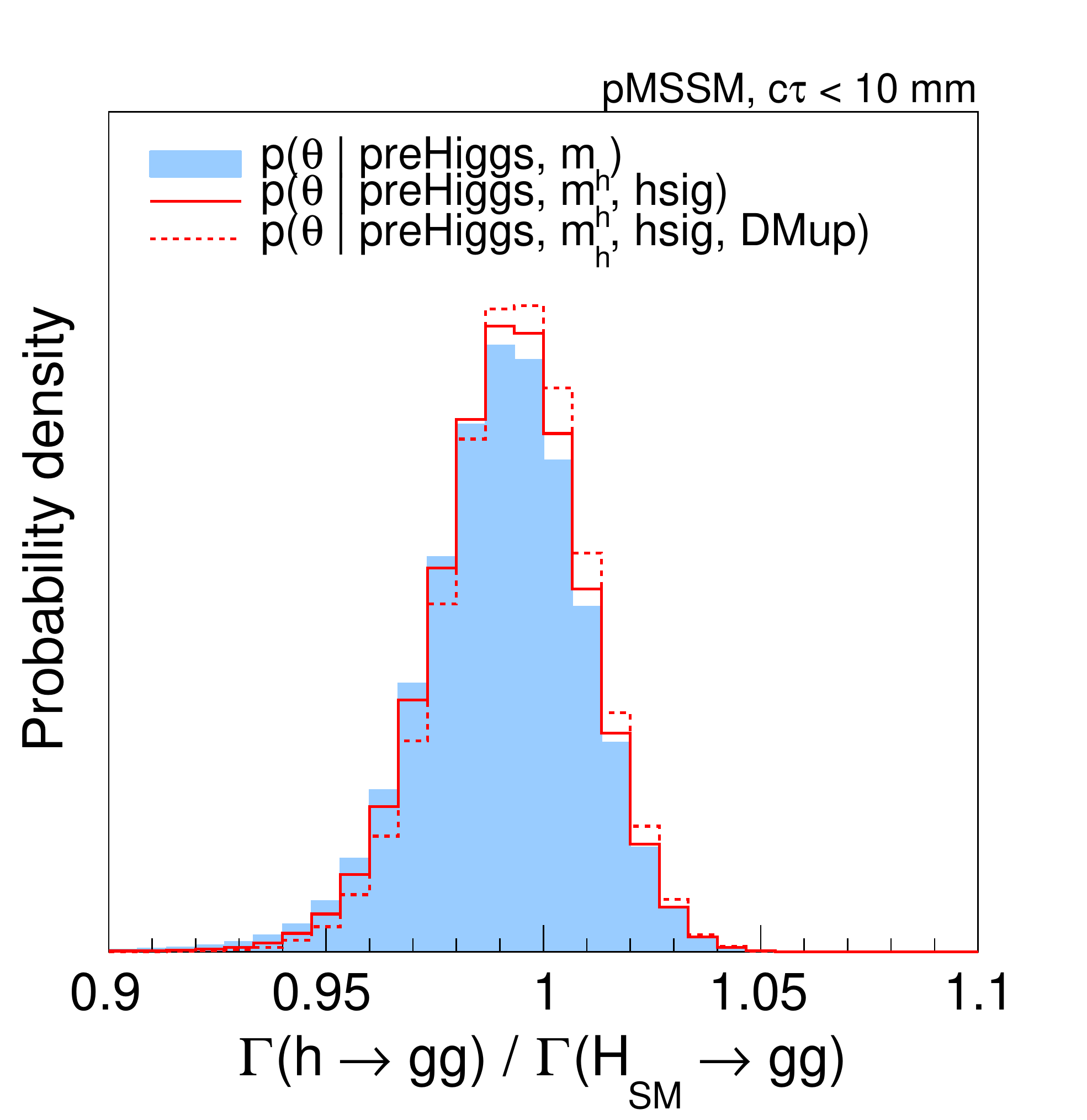}
\includegraphics[width=0.30\linewidth]{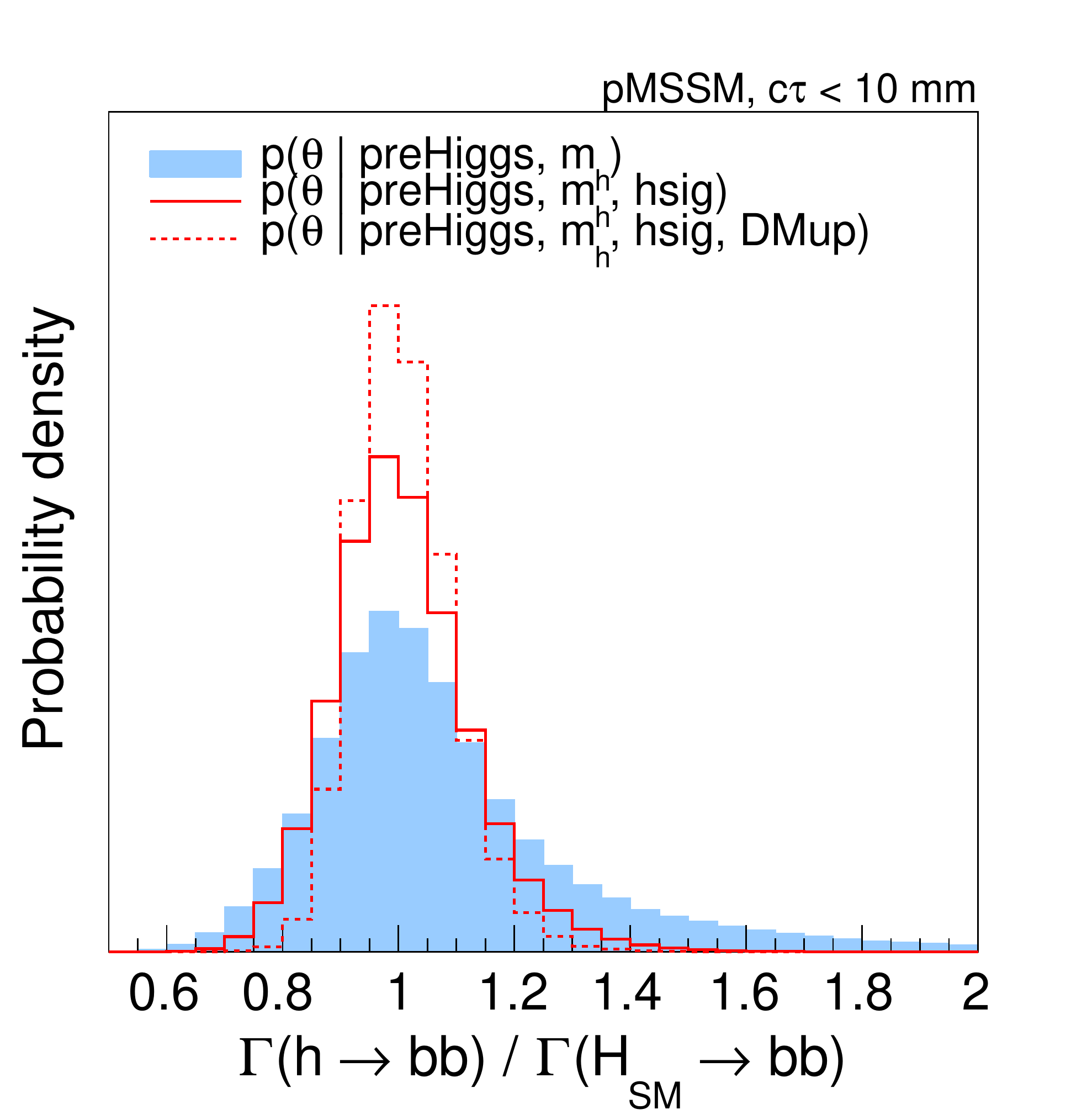}
\caption{Marginalized 1D posterior densities as in Fig.~\ref{fig:likehiggs1}, in the top row for BR$(h\to b\bar b)$ and $\Gamma_h$, in the bottom row for $\Gamma(h\to Y)/\Gamma(H_{\rm SM}\to Y)$ with, from left to right, $Y=\gamma\gamma$, $gg$ and $b\bar b$.}
\label{fig:hwidth}
\end{center}
\end{figure}

The posterior distributions of the $h$ signal strengths in the various channels are shown in Fig.~\ref{fig:likehiggs3}. 
The red line-histograms correspond of course to the constraints which we used as experimental input. 
For the $\gamma\gamma$, $ZZ$ and $\tau\tau$ final states, we find signal strengths of about 
$1 \pm 0.15$ after the Higgs signal requirements, 
and about $1 \pm 0.10$ after the DM requirements, at 95\% Bayesian Credibility (BC).
For the $b\bar b$ final state, the distribution is much narrower than required by observations---we find that $\mu(Vh \to b\bar{b})$ is restricted to the 95\% BC interval $\mu(Vh \to b\bar{b}) \in [0.91,1.09]$ after Higgs signal requirements, and $[0.94,1.06]$ after DM requirements. This is an indirect effect of the constraint on $\br(h\to b\bar b)$ and the total $h$ width, $\Gamma_h$, in order to have large enough signal in the other channels, see Fig.~\ref{fig:hwidth}. 
Interestingly, the constraints from the DM side narrow the signal strength distributions even more around the SM value of 1 because the higgsino mass $\mu$ tends to take on small values to fulfill the relic density requirement, leading to smaller $\Delta_b$.

Figure~\ref{fig:hwidth} also shows posterior distributions of  
$r_Y\equiv \Gamma(h\to Y)/\Gamma(H_{\rm SM}\to Y)$ for $Y=\gamma\gamma$, $gg$ and $b\bar b$. 
These ratios are equivalent to the ratios of the coupling strengths squared;    
$r_{\gamma\gamma}=C_\gamma^2$, $r_{gg}=C_g^2$, $r_{bb}=C_D^2$ in the notation of \cite{Belanger:2013xza}.
Our results for $r_Y$ can be compared to those for the neutralino LSP case in Ref.~\cite{Cahill-Rowley:2013vfa}. 
We observe that in our case
$r_{\gamma\gamma}$ peaks sharply at 1, the 95\% BC interval being [0.99,\,1.01], 
while $r_{gg}$ shows a wider distribution with a 95\% BC interval of [0.96,\,1.02].  
(The picture does not change if we remove the $c\tau$ cut.). 
%These features are in sharp contrast to 
%the much wider spread of $r_{\gamma\gamma}\approx 0.95-1.15$, peaking slightly above 1, and to the upper limit of $r_{gg}\lesssim 0.96$ found in \cite{Cahill-Rowley:2013vfa}. 
These features are different from those in \cite{Cahill-Rowley:2013vfa}, where 
the $r_{\gamma\gamma}$ distribution peaks within $r_{\gamma\gamma}\approx 1$--$1.05$, 
and $r_{gg}$ exhibits an upper limit of $r_{gg}\lesssim 0.97$. 
Also, the $r_{bb}$ distribution is quite different. 
Some differences are of course expected as the distributions in \cite{Cahill-Rowley:2013vfa} come from 
a flat  random sampling and thus do not have the statistical meaning that underlies our approach. 
More importantly, however, the SM calculation of {\tt HDECAY} employed in \cite{Cahill-Rowley:2013vfa}  
includes additional radiative corrections which are not present  in the MSSM calculation.\footnote{We thank 
Ahmed Ismail and Matthew Cahill-Rowley for communication on this matter.} 
In our case, we avoid this problem by taking the MSSM decoupling 
case as the SM limit for computing  $\Gamma(H_{\rm SM}\to Y)$, {\it cf.}\ Section~\ref{sec:higgslikeli}.   Of course, the $r_Y$ are not directly measurable at the LHC.  They become measurable only if it can be determined that the $h$ has no invisible (\eg\ $h\to\cnone\cnone$) or unseen (\eg\ $h\to 4\tau$) decay modes.

\FloatBarrier

\bigskip

%%% heavy higgses %%%

\begin{figure}[t!]
\begin{center}
\includegraphics[width=0.3\linewidth]{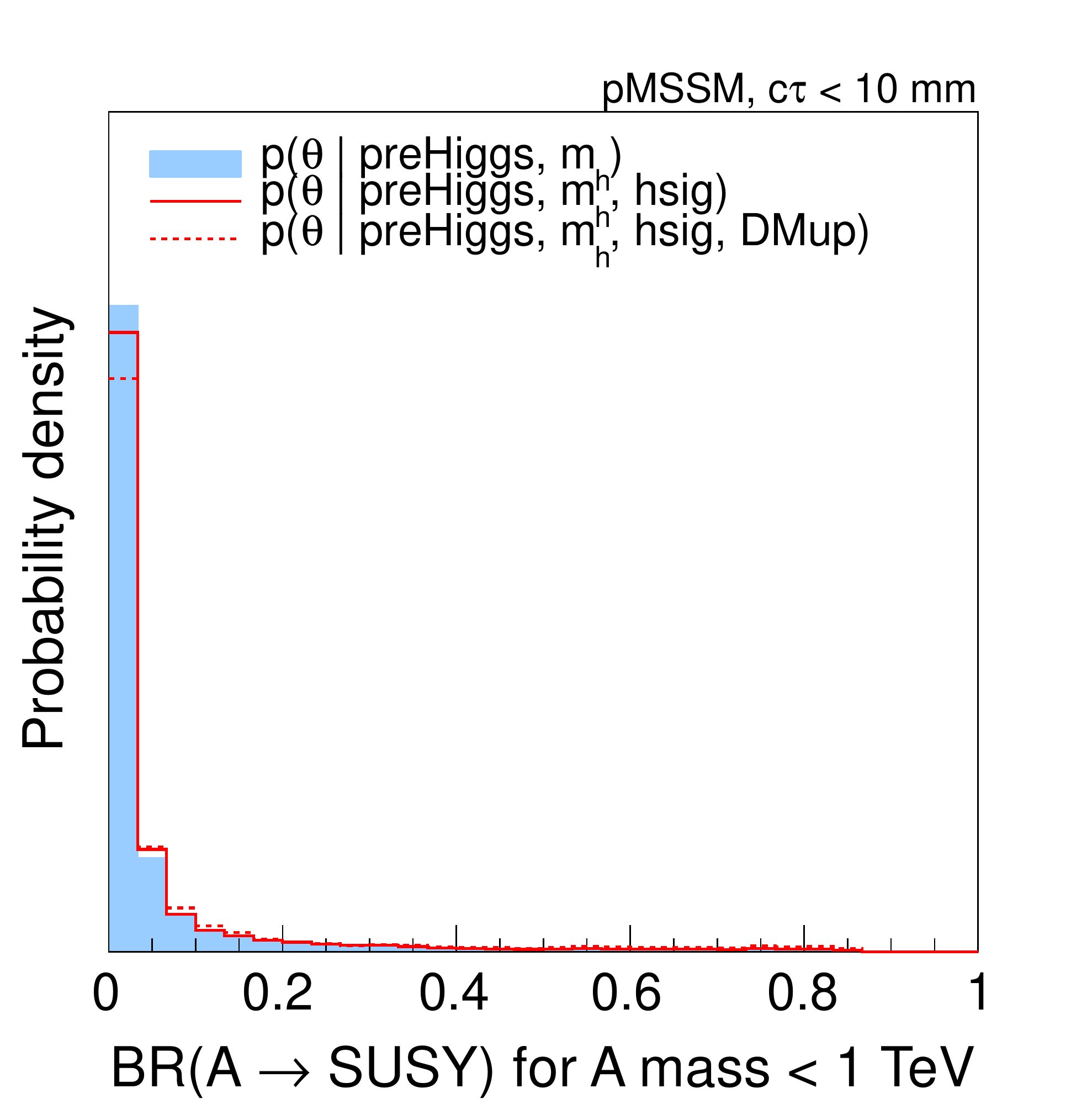}
\includegraphics[width=0.3\linewidth]{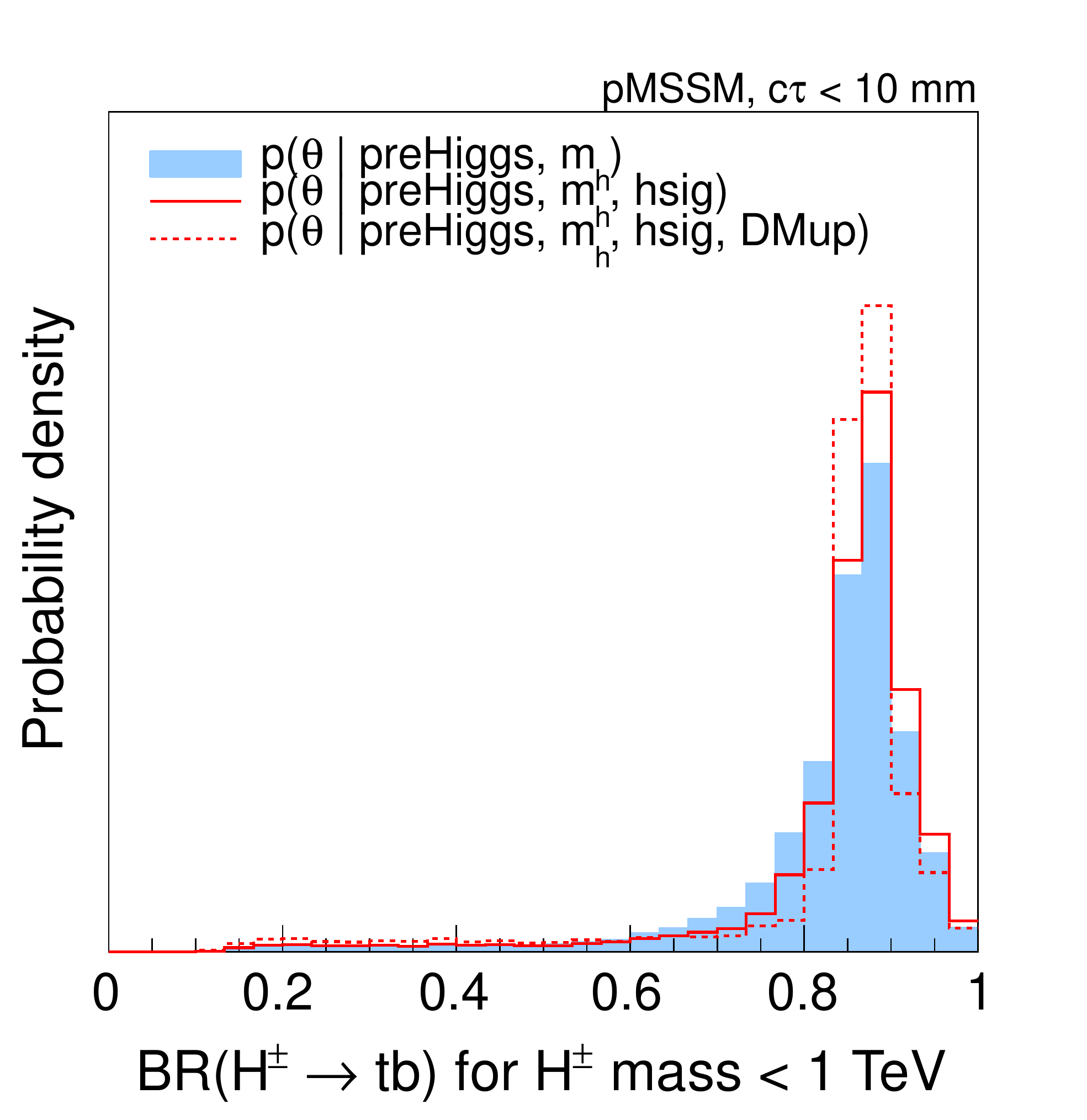}
\includegraphics[width=0.3\linewidth]{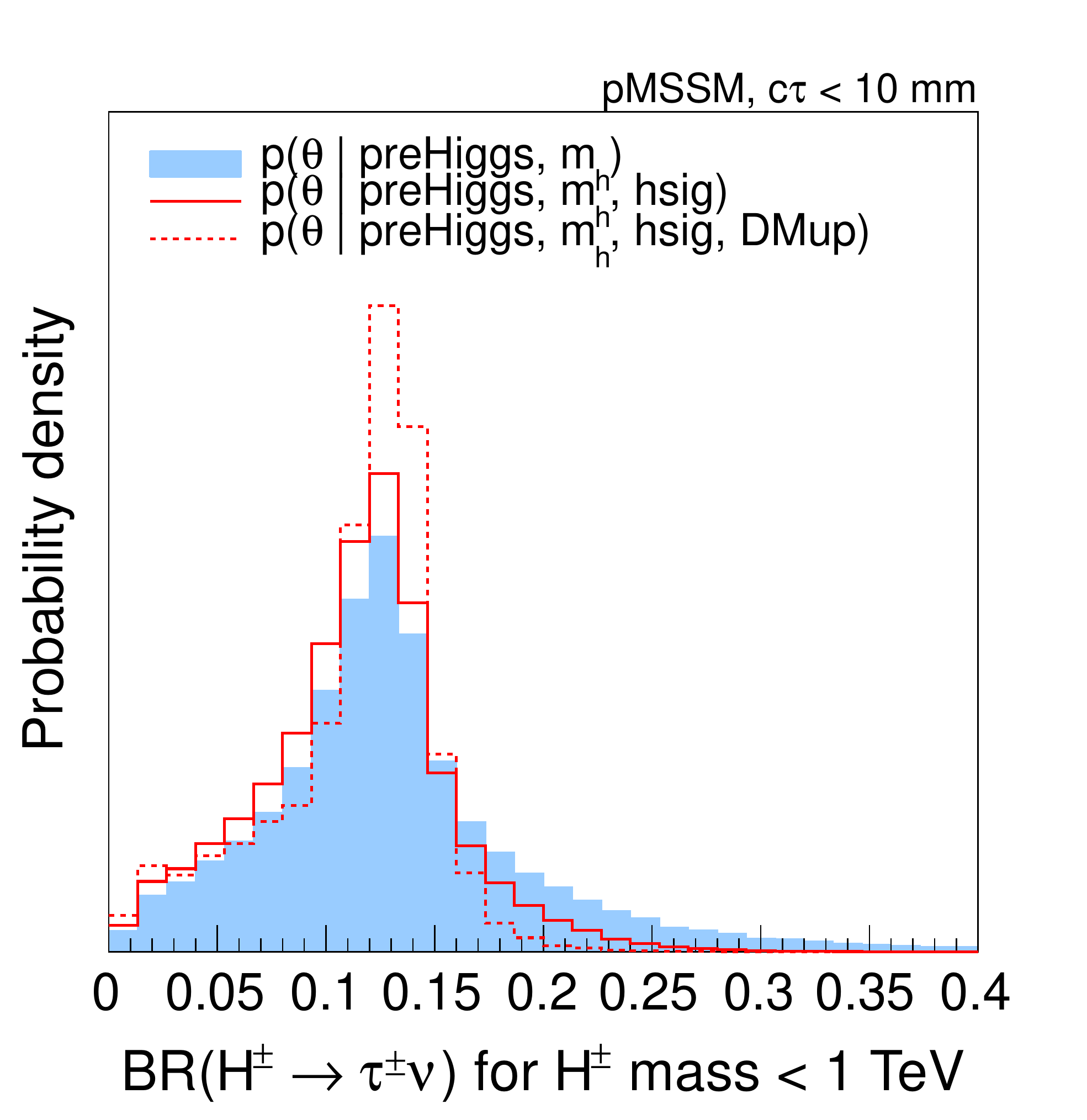}
\caption{Marginalized 1D posterior densities as in Fig.~\ref{fig:likehiggs1}, here for the branching ratios of the heavy MSSM Higgses $A$ and $H^\pm$ with masses below 1 TeV.}
\label{fig:heavierHiggses}
\end{center}
\end{figure}

Our procedure also allows us to derive predictions for the heavier MSSM Higgs states $H$, $A$ and $H^\pm$, as illustrated in Figs.~\ref{fig:heavierHiggses2D} and \ref{fig:heavierHiggses}. First, in the $\tan \beta$ versus $m_A$ plane, we show that the current CMS limit~\cite{CMS-PAS-HIG-13-021} interpreted in the $m_h^{\rm max}$ scenario has a negligible effect on our distributions, since after imposing constraints from low-energy observables and from Higgs measurements the likely region corresponds to $A$ masses above 500 GeV and moderate $\tan \beta$. (This observation remains valid when dark matter requirements are taken into account; in all cases we have checked that the current limits on $H \to ZZ$ are always satisfied.) We also show $\sigma(gg,b\bar b \to H,A) \times {\rm BR}(H,A \to \tau\tau)$ at $\sqrt s = 14~$TeV as a function of $m_{H,A}$, using {\tt SusHi\_1.1.1}~\cite{Harlander:2012pb} for the computation of the cross sections in the approximation of decoupled stops and sbottoms.\footnote{Neglecting contributions from stops and sbottoms in the computation of $gg, b\bar b \to H,A$ is a good approximation in most cases since the posterior densities of $m_{\tilde t_1}$ and $m_{\tilde b_1}$ peak around 2~TeV.}
These plots show that the signals from the CP-odd and CP-even Higgs bosons are very similar and that for high masses the dominant process is almost always $b\bar b \to H,A$ (see, in particular, the bottom right plot), where for a given mass $\sigma(b\bar b \to H,A)$ spans over about an order of magnitude due to its strong dependence on $\tan \beta$. Typical $\sigma \times \br$ values are of the order of 0.1 to 100~fb for $m_{H,A} < 1$~TeV and therefore most of this region should be probed during the next run of the LHC at 13--14~TeV.

Some more properties of the heavy Higgses (for masses $< 1$~TeV) are shown in Fig.~\ref{fig:heavierHiggses}. We see that the decay branching fraction of $A$ into SUSY particles is often very small because most of the supersymmetric partners generally lie at the (multi-)TeV scale. Concretely, the probability for $\br(A \to {\rm SUSY}) > 10\%$ is only 1.6\% after the Higgs signal likelihood (2.1\% after DM requirement). Compared to the preHiggs distributions, decays into SUSY particles are however slightly enhanced by the Higgs likelihood and dark matter requirements because $\mu$, and hence neutralino and chargino masses, are pushed to lower values. Also shown are the dominant decay modes of the charged Higgs: $H^{\pm} \to tb$ and $H^{\pm} \to \tau^{\pm} \nu$. The dominance of hadronic decays over leptonic ones is strengthened when Higgs measurements are taken into account since small values of $m_A$ and large values of $\tan \beta$ are then disfavored.

%\FloatBarrier

%----------------------------------------------------------------------------------------------------------------------------------
\subsection{Impact of the $\bm c\tau$ cut}\label{sec:GreenPlots}
%----------------------------------------------------------------------------------------------------------------------------------

\begin{figure}[t!]
\begin{center}
\includegraphics[width=0.25\linewidth]{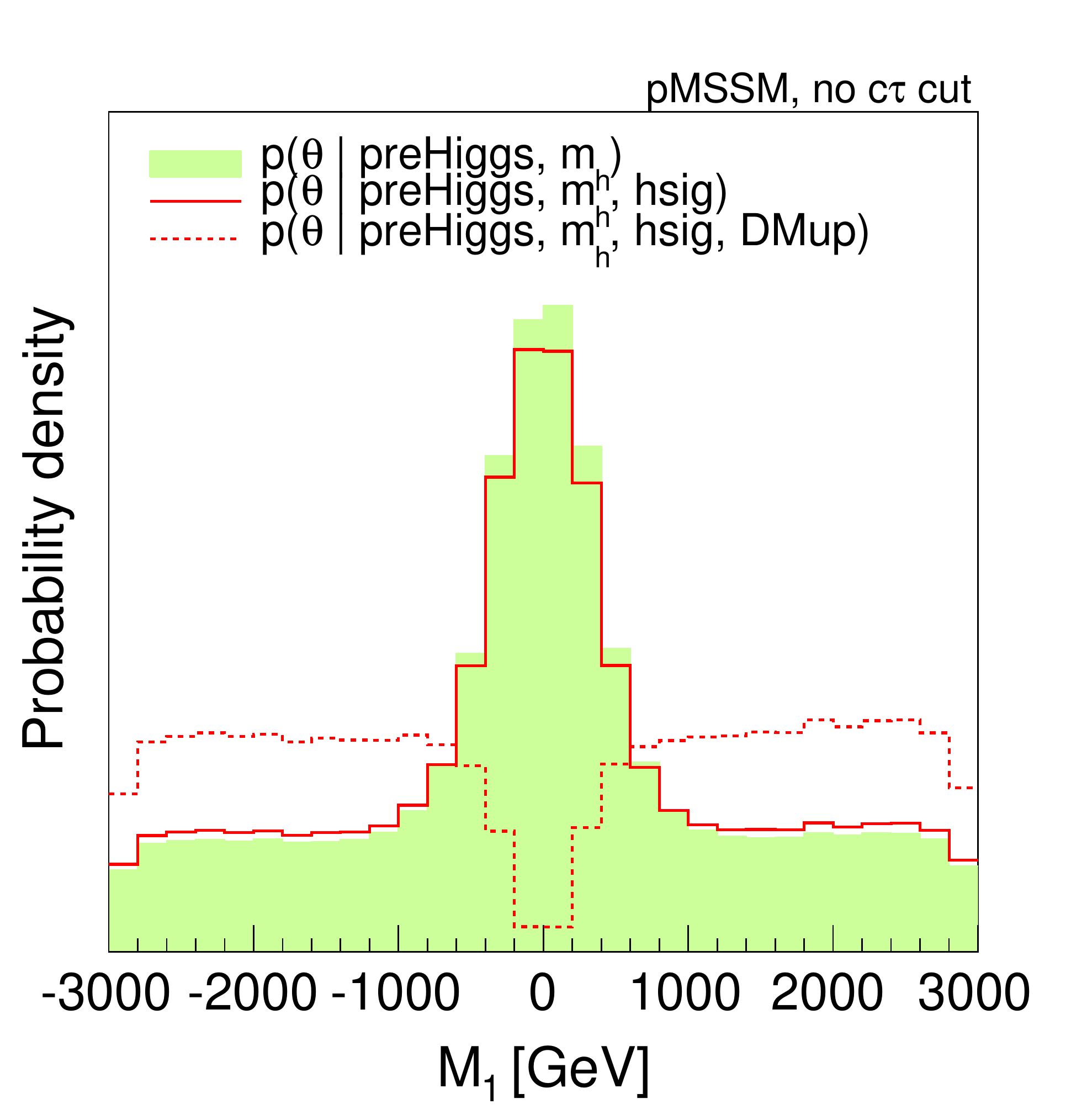}\includegraphics[width=0.25\linewidth]{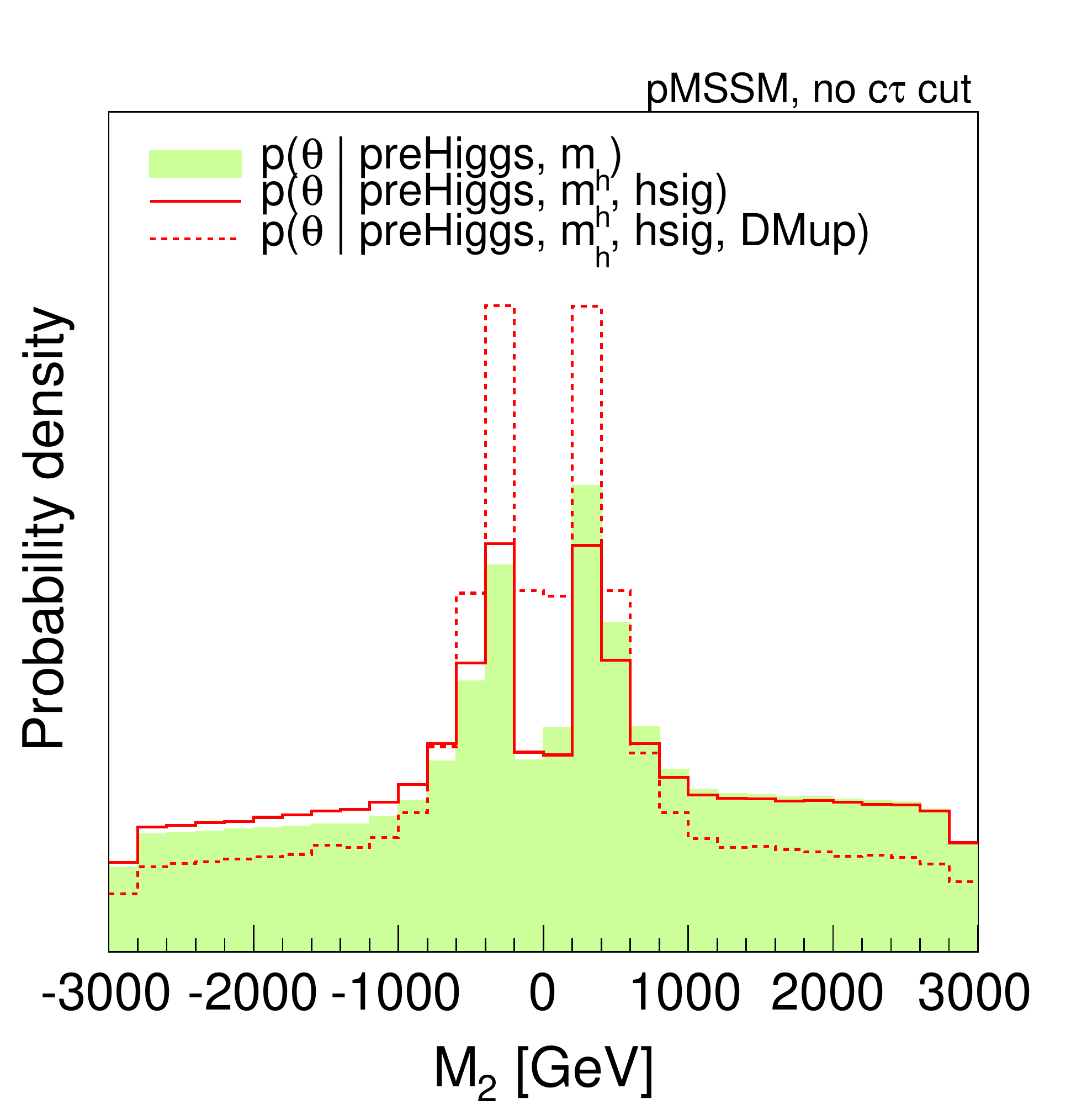}\includegraphics[width=0.25\linewidth]{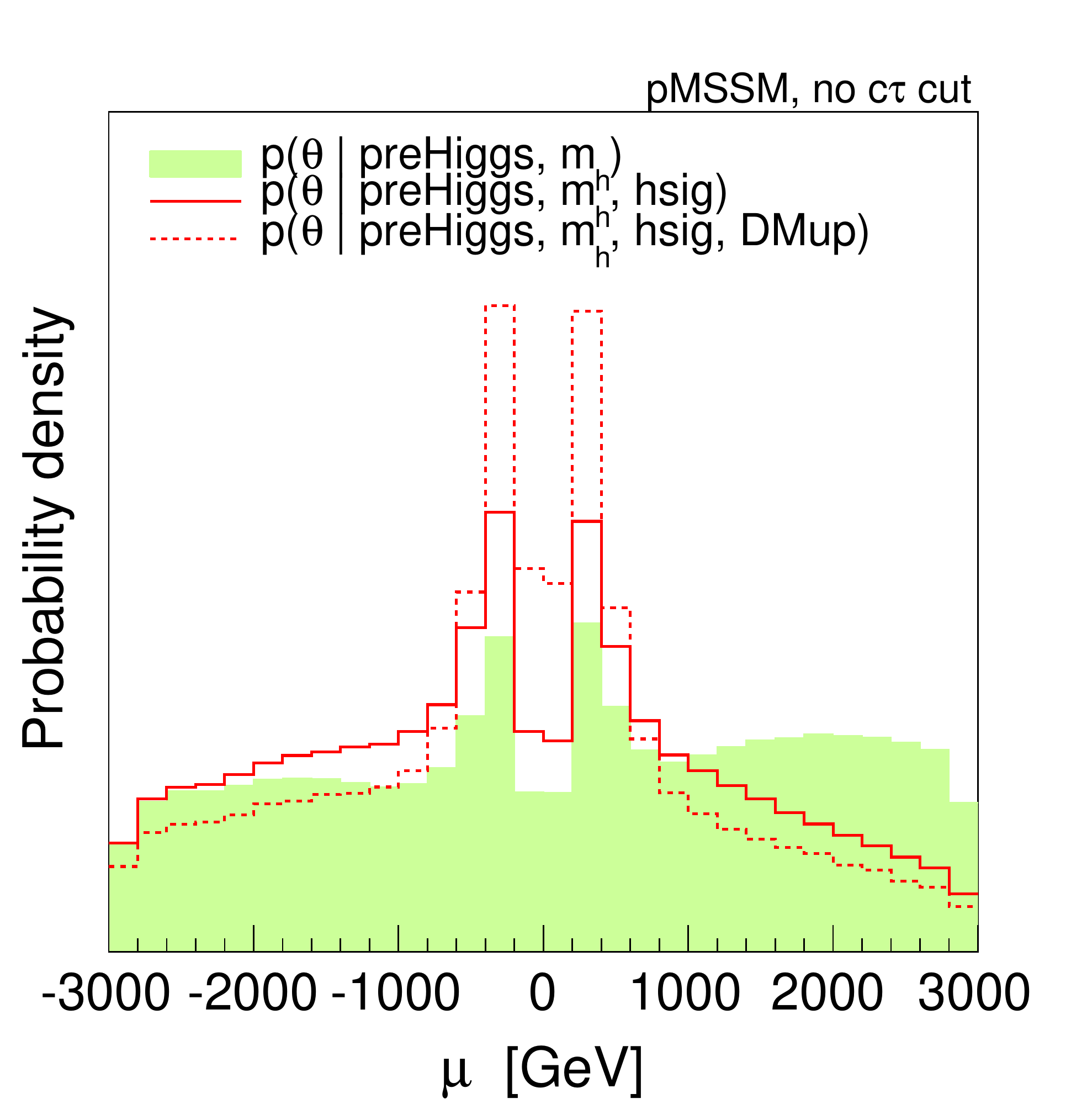}\includegraphics[width=0.25\linewidth]{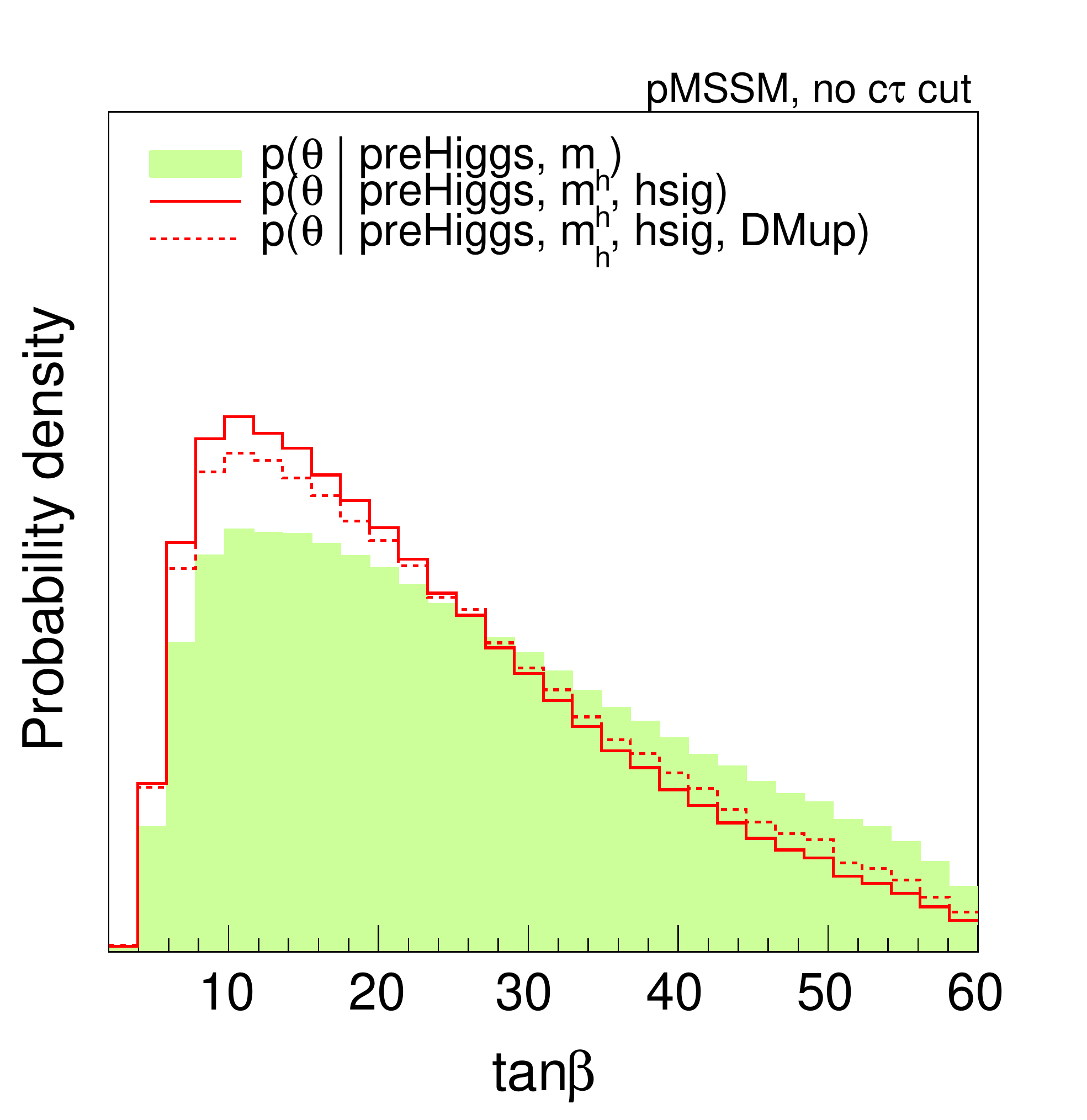}\\
\includegraphics[width=0.25\linewidth]{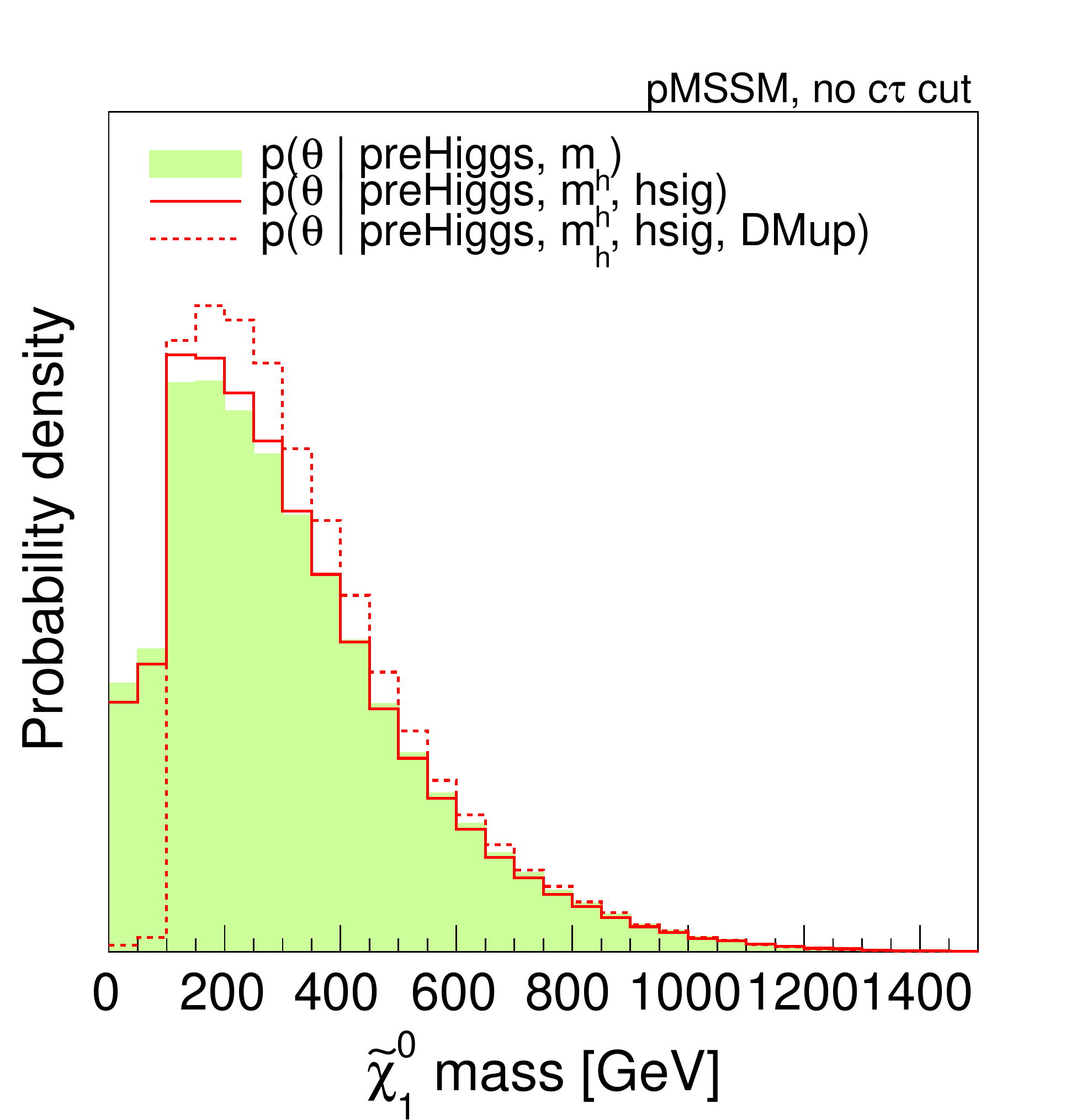}\includegraphics[width=0.25\linewidth]{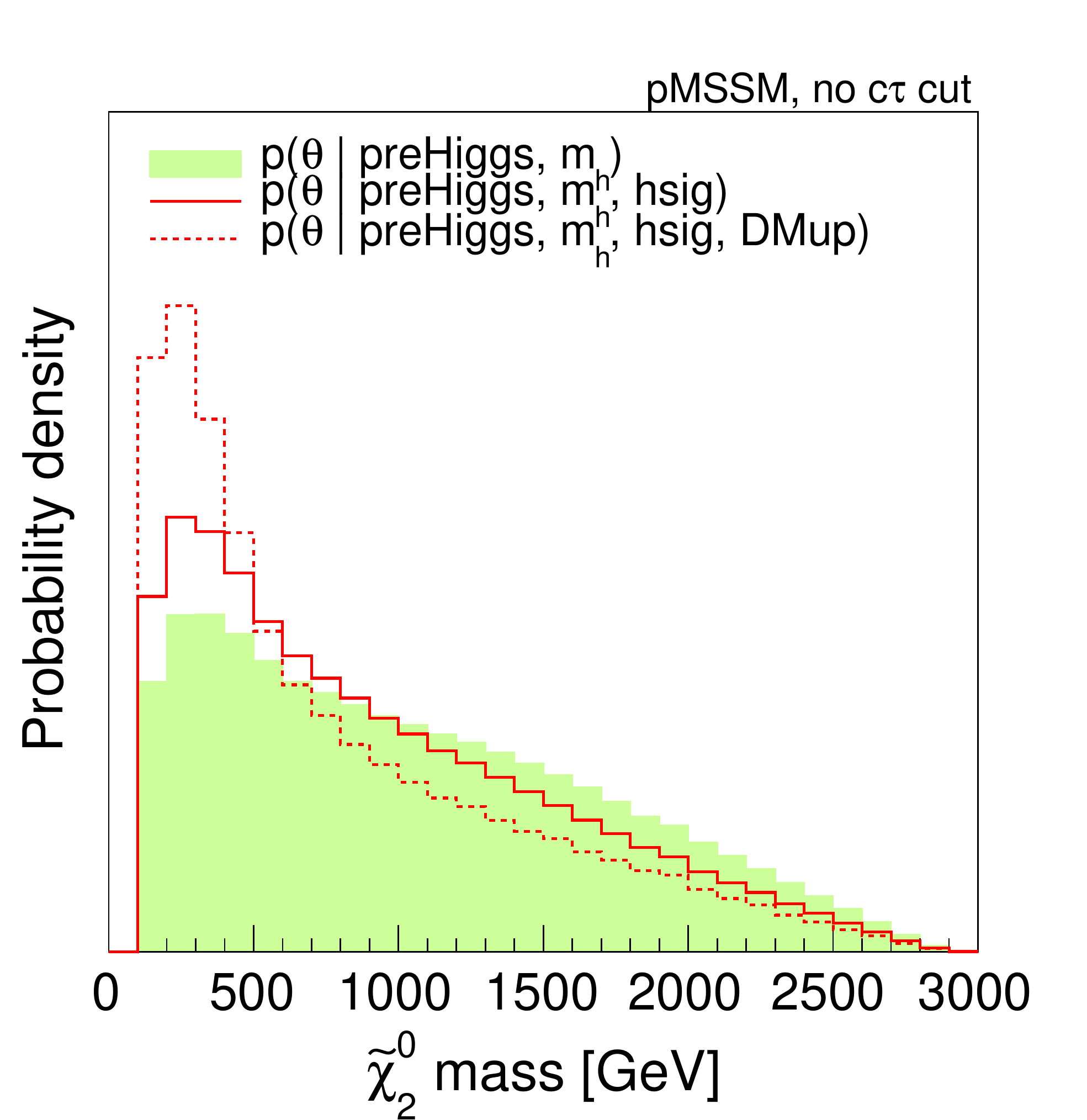}\includegraphics[width=0.25\linewidth]{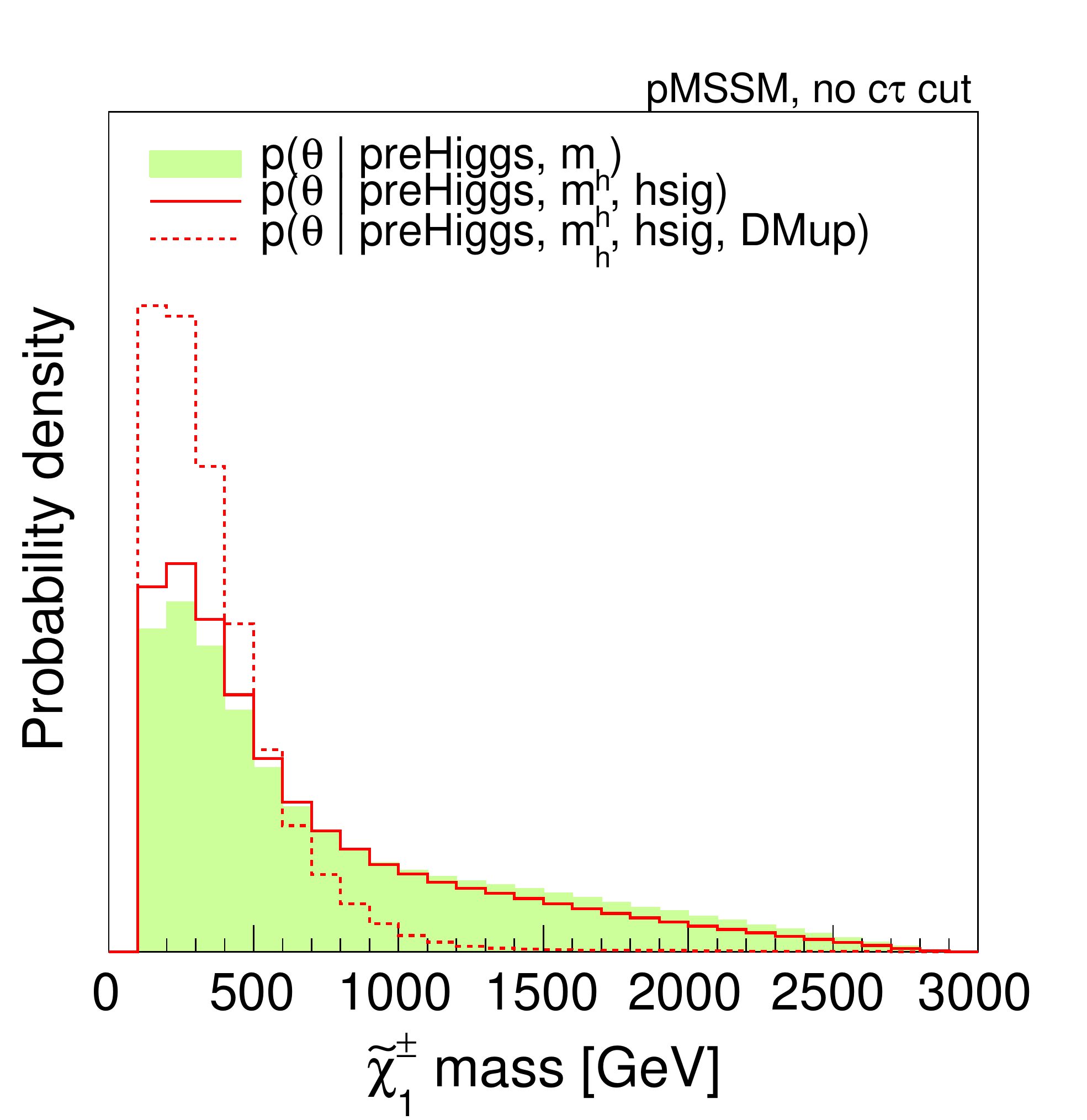}\includegraphics[width=0.25\linewidth]{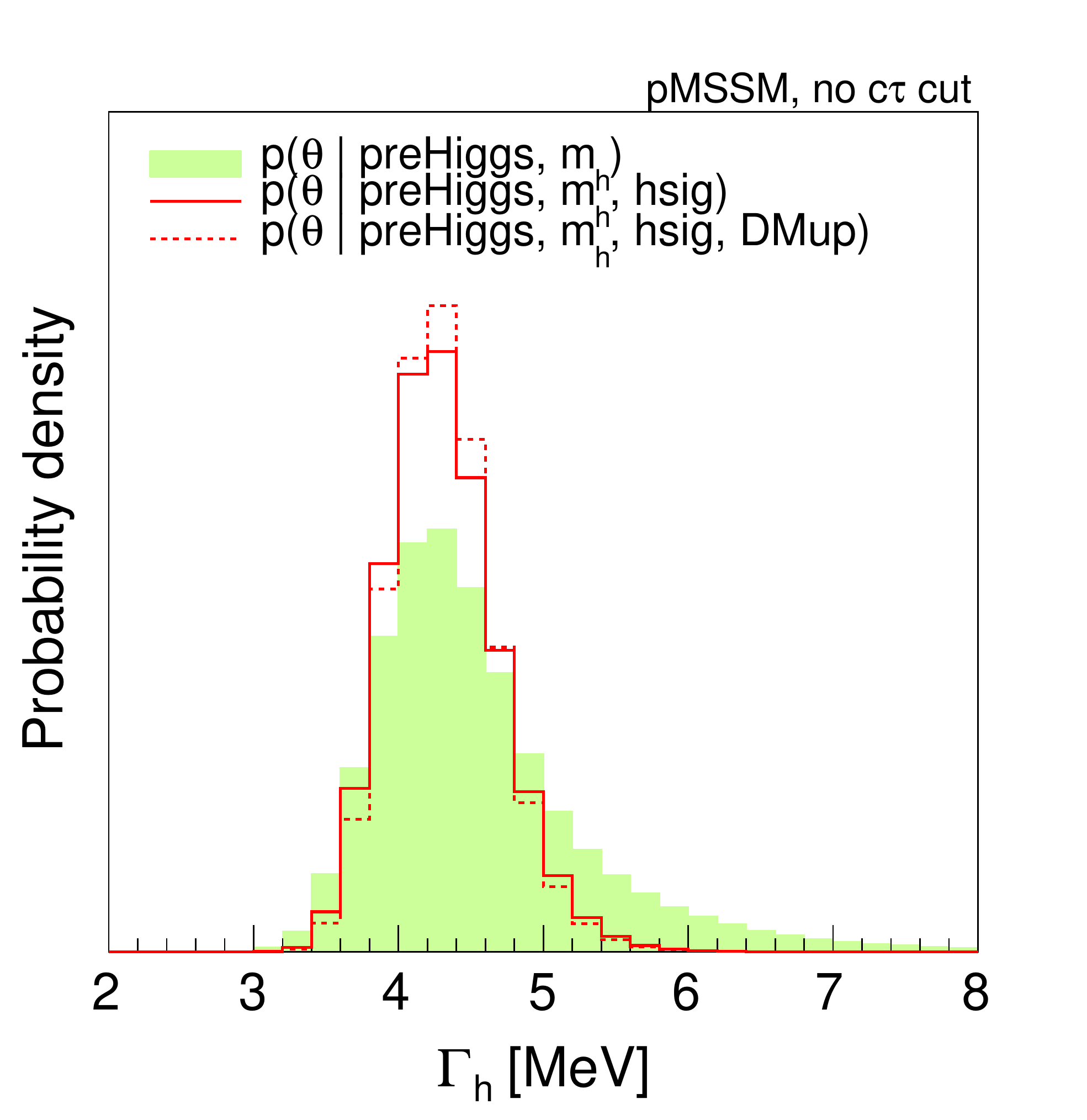}
\caption{Marginalized 1D posterior densities for selected parameters and derived quantities without the prompt chargino requirement.
The green histograms show the distributions  based on the ``preHiggs'' measurements 
of Table~\ref{tab:preHiggs} plus requiring in addition $m_h\in [123,\,128]$~GeV, but without the $c\tau$ cut. 
The solid red lines are the distributions when taking into account in addition the measured Higgs signal strengths 
in the various channels, as well as the limits from the heavy MSSM Higgs searches. The dashed red lines include 
in addition an upper limit on the neutralino relic density and the recent direct DM detection limit from LUX.}
\label{fig:likehiggsNoCtau}
\end{center}
\end{figure}

We saw from the plots in Section~\ref{sec:YellowPlots} that the ``prompt chargino'' requirement has 
a strong effect on some of the distributions, above all on that of the wino mass parameter $M_2$. 
The influence on $\mu$ and $M_1$ is less dramatic but still quite strong. As a consequence, it is mostly the 
chargino and neutralino masses (and their gaugino--higgsino composition) which are affected by the 
$c\tau<10$~mm requirement. To assess the impact of this cut,
the relevant posterior densities {\em without} the $c\tau$ cut are shown in Fig.~\ref{fig:likehiggsNoCtau}. 
Comparing these plots with their equivalents in Fig.~\ref{fig:sampling1} of Section~\ref{sec:BluePlots}, 
we see that, as expected, %without the $c\tau$ cut, 
in both  the ``preHiggs+$m_h$'' and the ``preHiggs+$m_h$+hsig'' distributions, light charginos and neutralinos are more preferred. The effect is more pronounced for the $\tilde\chi^\pm_1$ and  $\tilde\chi^0_2$ than for the $\tilde\chi^0_1$. Note also that the preference for smaller $\mu$ through the Higgs signal strength measurements remains. Finally, note that the DM upper limits largely overrule the effect of the $c\tau$ cut: the red dashed line histograms are almost the same with or without the $c\tau$ cut. The exception is the  $\tan\beta$ distribution. (The $\tilde\chi^0_1$, $\tilde\chi^0_2$, $\tilde\chi^\pm_1$ mass differences can however be smaller  without the $c\tau$ cut.) 
The posterior densities of other quantities, which do not directly depend on $M_1$, $M_2$ or $\mu$ show hardly any sensitivity to the $c\tau$ cut. In particular our conclusions about the Higgs signals remain unchanged. 

\begin{figure}[h!]
\begin{center}
\includegraphics[width=0.35\linewidth]{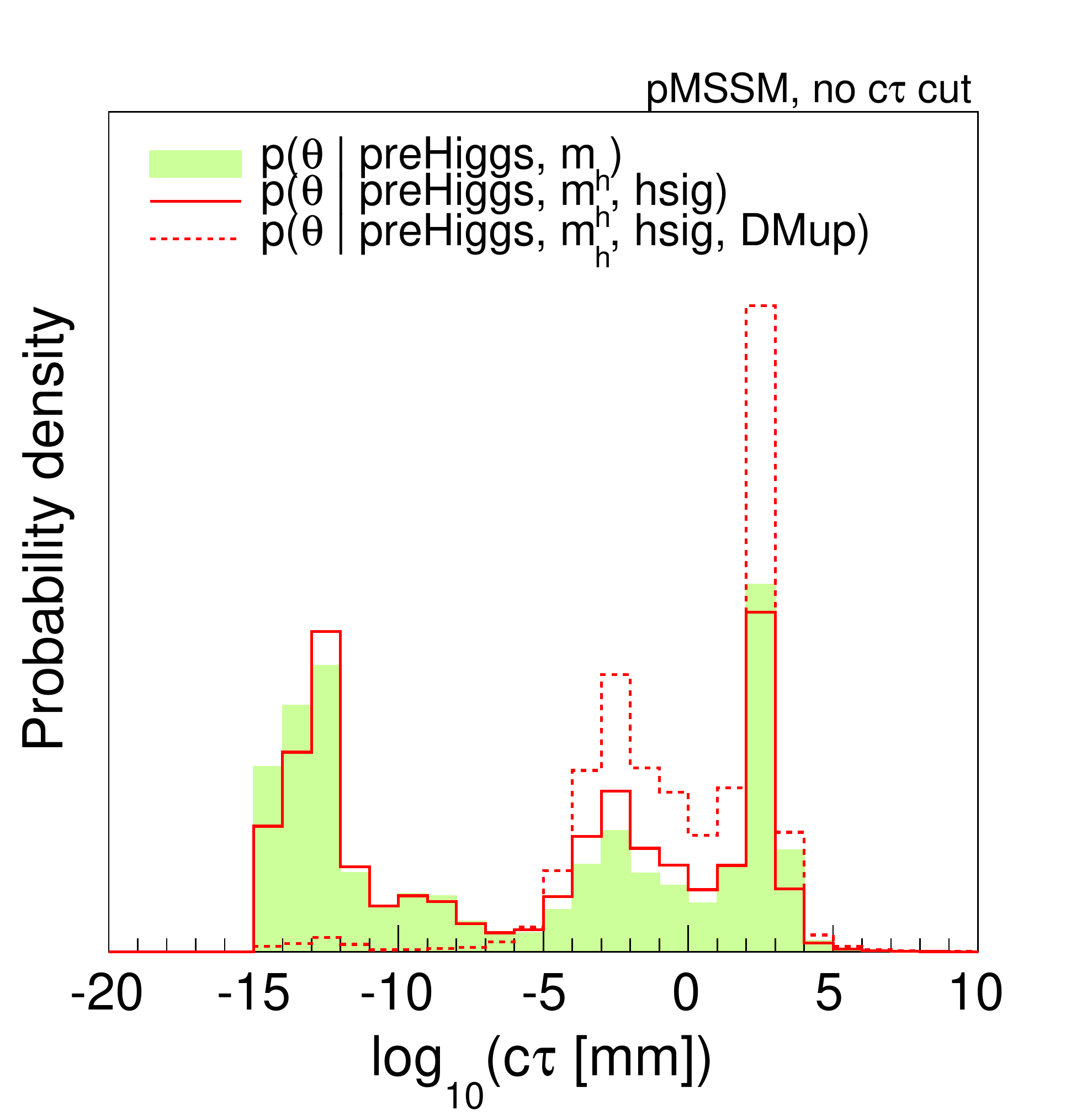}
\caption{Marginalized 1D posterior density of the average $\tilde\chi^\pm_1$ lifetime, $c\tau$ in mm. Color codes as in Fig.~\ref{fig:likehiggsNoCtau}.}
\label{fig:ctau}
\end{center}
\end{figure}

It is of course also interesting to ask how likely it is at all to have a long-lived chargino. 
To this end we show in Fig.~\ref{fig:ctau} the marginalized posterior density of the average $\tilde\chi^\pm_1$ lifetime. 
We find that the probability of $c\tau>10$~mm is 28\%, 25\% and 47\% at the 
``preHiggs+$m_h$'', ``preHiggs+$m_h$+hsig'', and  ``preHiggs+$m_h$+hsig+DMup'' levels, respectively. 

\FloatBarrier

%----------------------------------------------------------------------------------------------------------------------------------
\subsection{Interplay with dark matter searches}
%----------------------------------------------------------------------------------------------------------------------------------

\begin{figure}[t!]
\begin{center}
\includegraphics[width=0.3\linewidth]{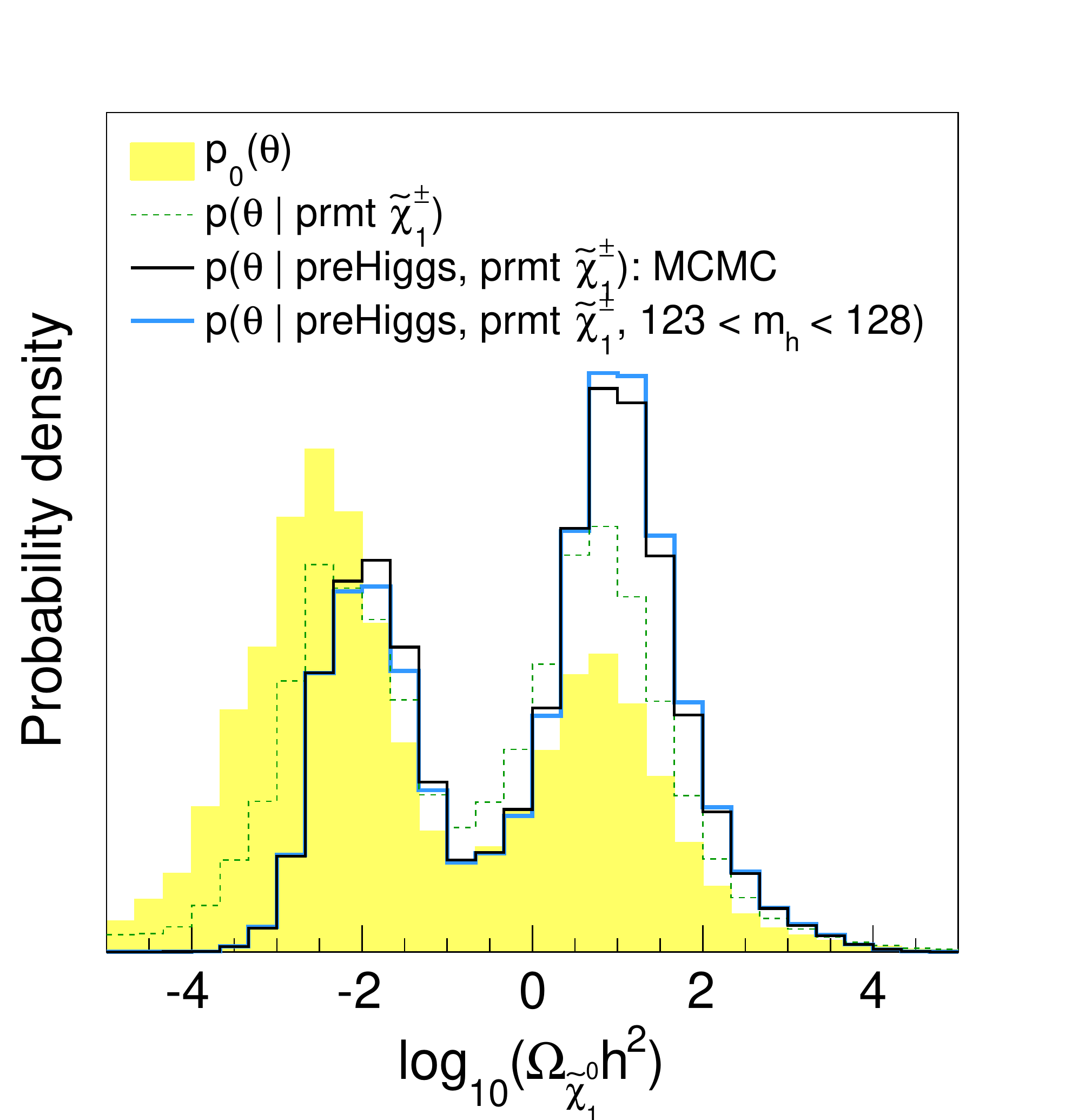}
\includegraphics[width=0.3\linewidth]{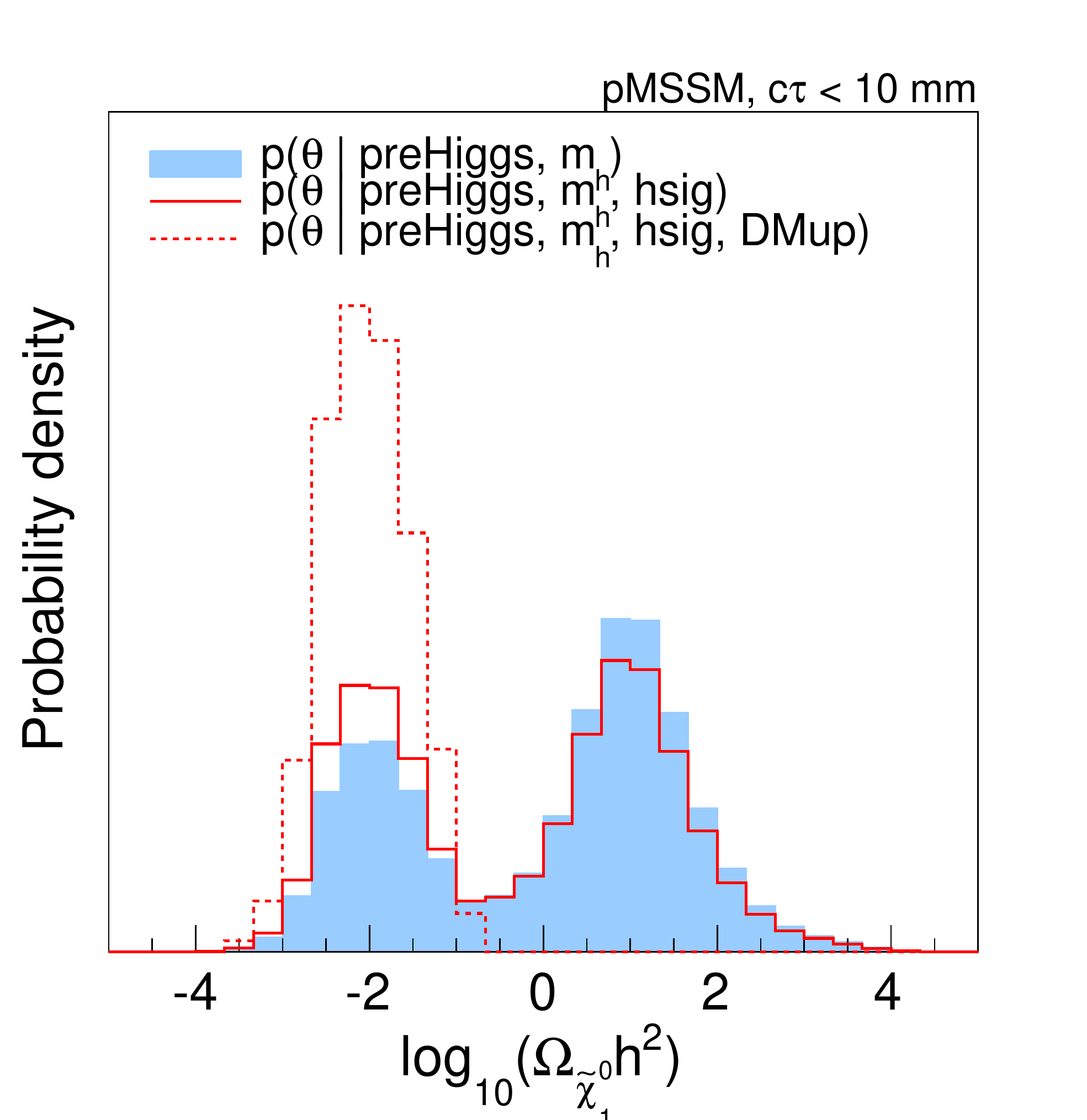}\includegraphics[width=0.3\linewidth]{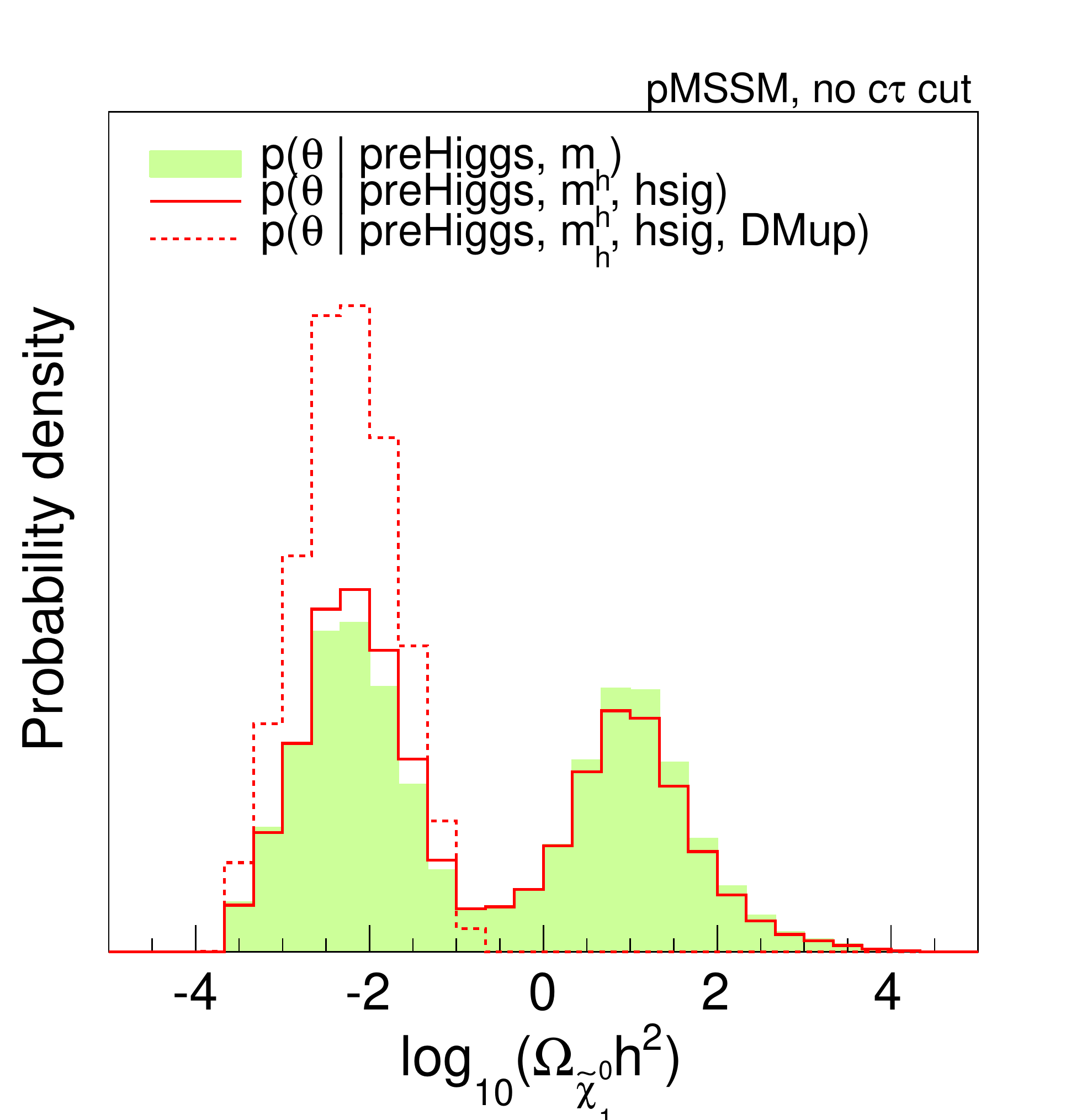}\\
\includegraphics[width=0.3\linewidth]{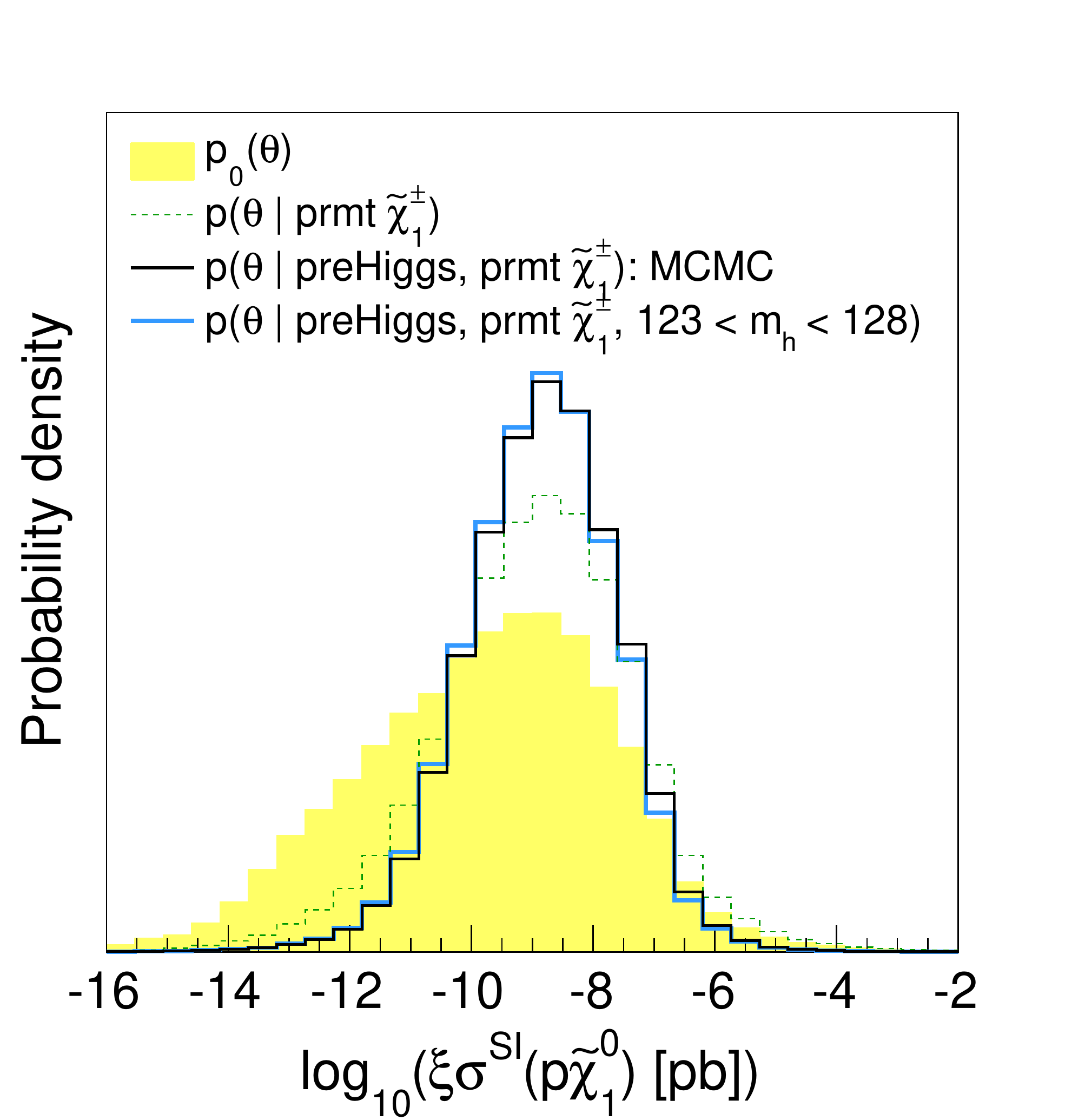}
\includegraphics[width=0.3\linewidth]{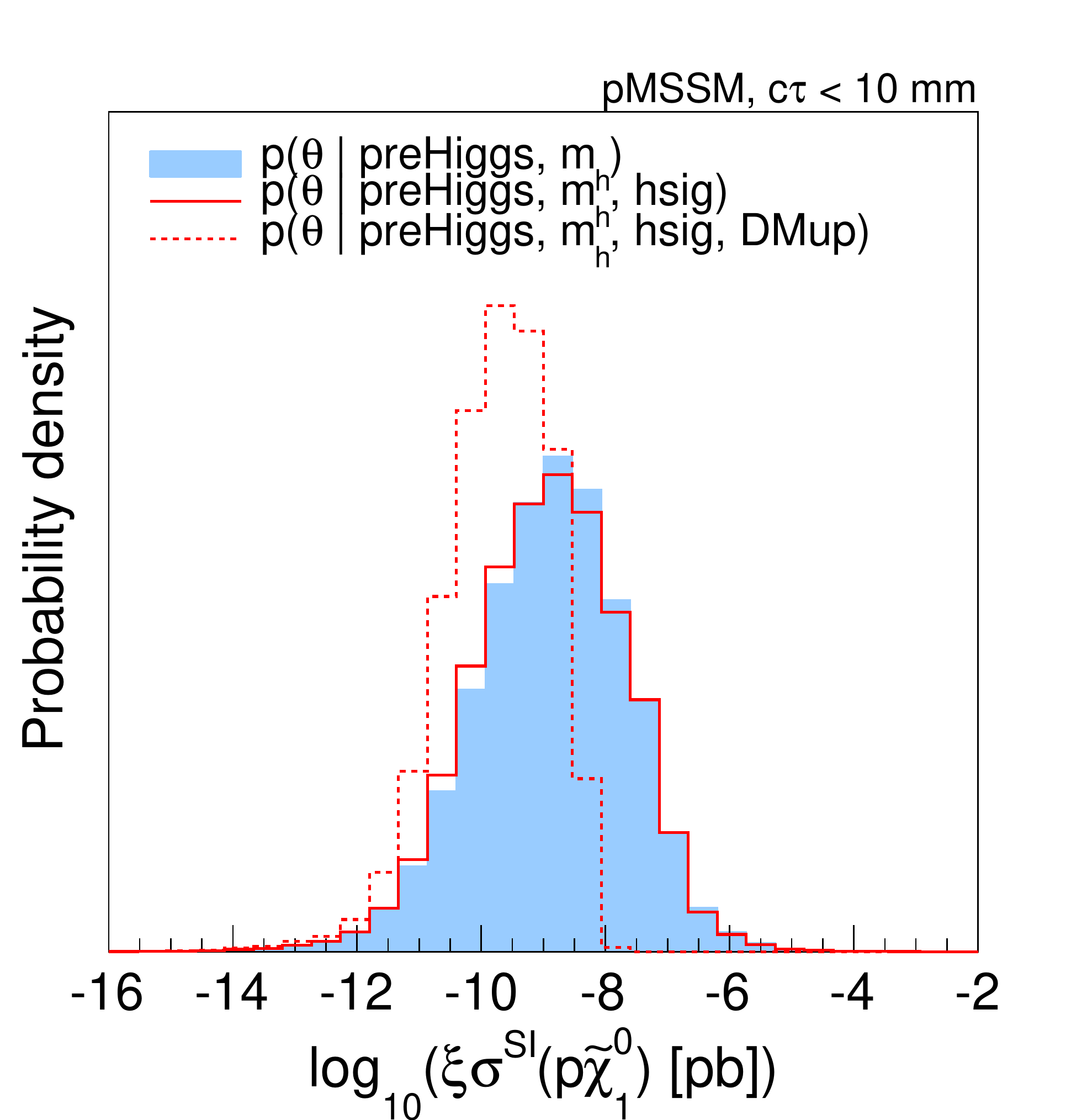}\includegraphics[width=0.3\linewidth]{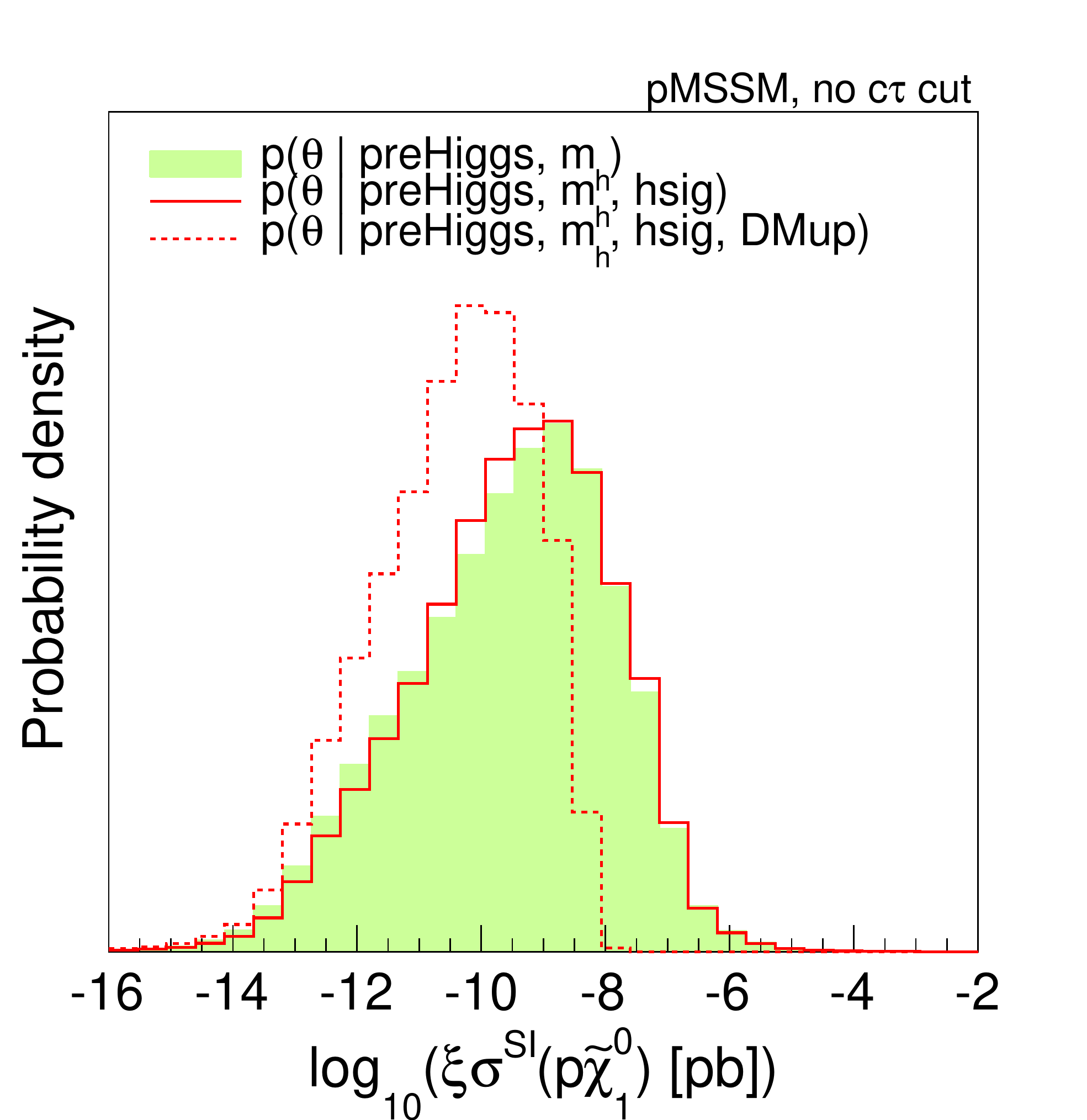}
\caption{Marginalized 1D posterior densities for dark matter quantities. 
Color codes as in Fig.~\ref{fig:sampling1} (left), Fig.~\ref{fig:likehiggs1} (middle) and Fig.~\ref{fig:likehiggsNoCtau}
(right).  }
\label{fig:darkmatter1D}
\end{center}
\end{figure}

\begin{figure}[h!]
\begin{center}
\includegraphics[width=0.38\linewidth]{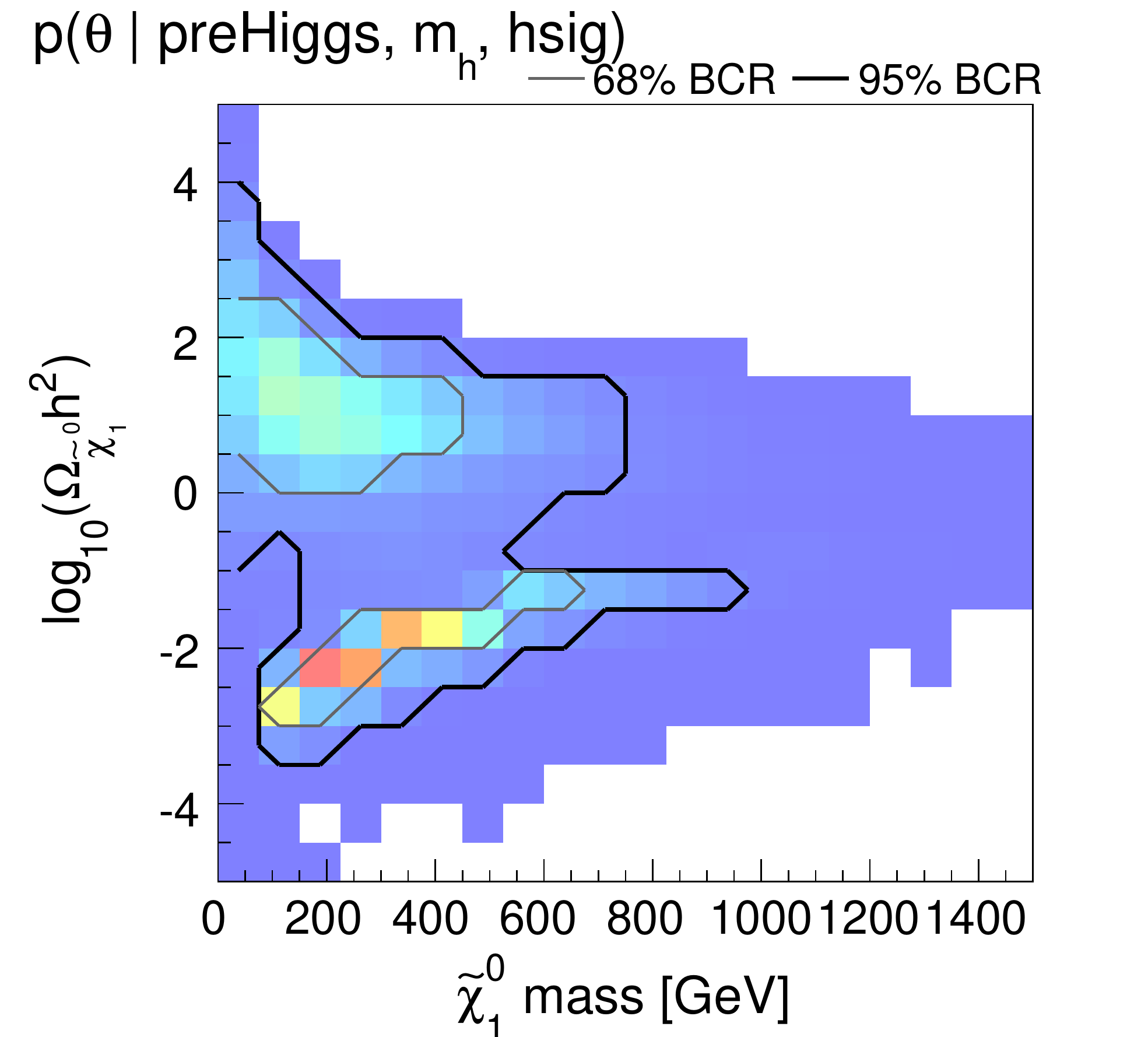}
\includegraphics[width=0.38\linewidth]{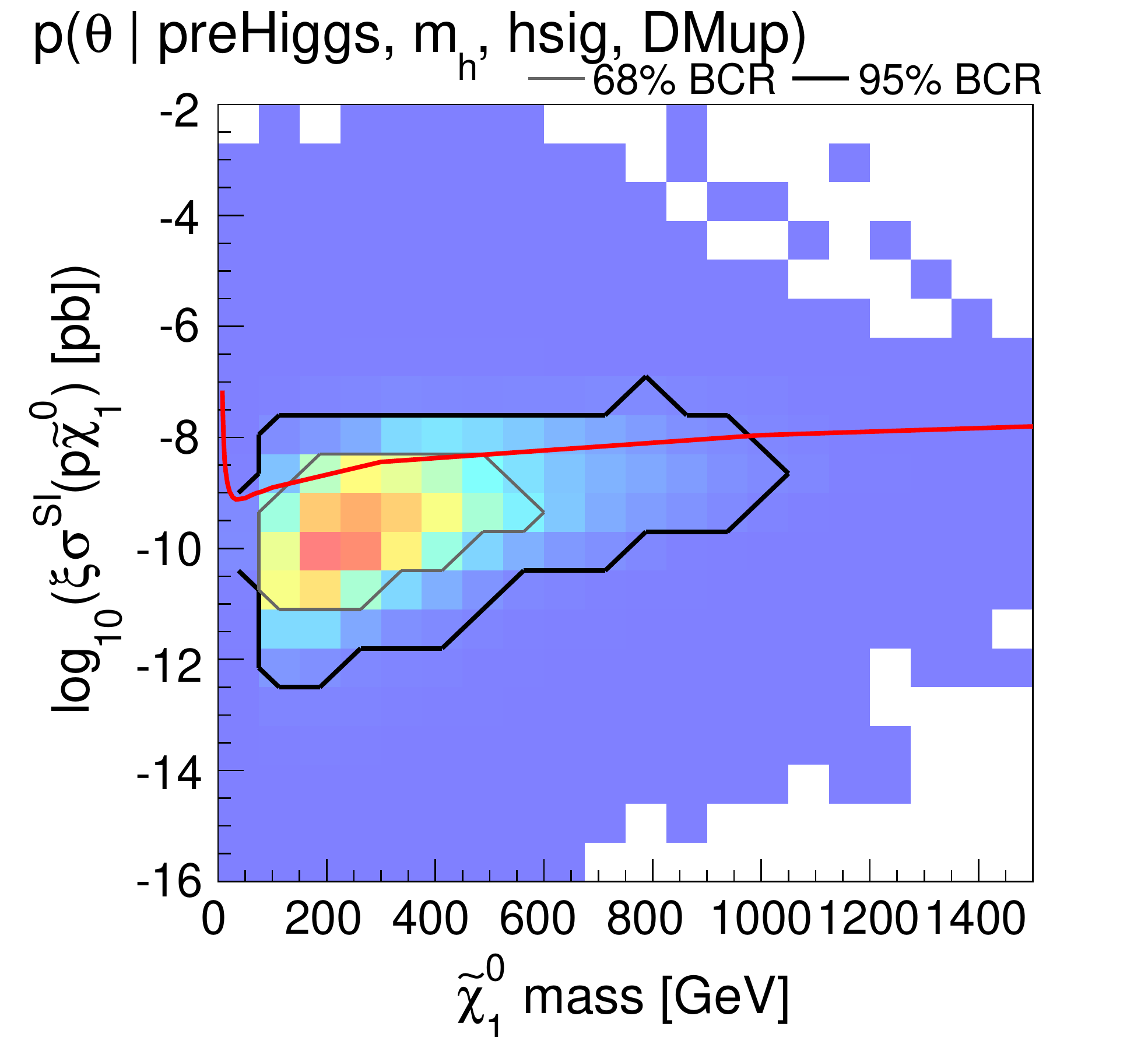}
\caption{Marginalized 2D posterior densities for dark matter quantities. The probability density is represented by color shading, ranging from low values in blue to high values in red. The grey and black lines are contours of 68\% and 95\% Bayesian Credibility, respectively. The red line in the right plot is the 90\%~CL limit from LUX.  }
\label{fig:darkmatter2D}
\end{center}
\end{figure}

As discussed above, the dark matter requirements ({\it i.e.}, imposing upper limits on the relic density and on 
the spin-independent scattering cross section) have a significant impact on the MSSM parameters and masses, 
and even on the $h$ signal strengths. 
In this subsection, we now focus on dark matter observables themselves. 
Results for the neutralino relic density $\omhsq$ and the re-scaled spin-independent scattering cross section 
$\xi\sigsi$ are shown in Figs.~\ref{fig:darkmatter1D} and\ \ref{fig:darkmatter2D}. 

Let us start the discussion with the 1D distributions of $\log_{10}(\omhsq)$, shown in the upper row of plots in Fig.~\ref{fig:darkmatter1D}. Already the $p_0(\theta)$ distribution shows a two-peak structure with the minimum actually lying near the cosmologically preferred value $\omhsq\approx 0.1$. This distribution is shifted to significantly higher values by the preHiggs constraints. Concretely, at preHiggs level, the probability for $\omhsq<0.14$ is 36\% (53\%) 
with (without) the prompt chargino requirement. This hardly changes when including also the requirement of $m_h=123-128$~GeV: $p(\omhsq<0.14)\simeq 34\%$ (53\%) in this case.  The Higgs signal likelihood has a larger effect, shifting the distribution towards lower $\omhsq$. This is mainly due to the preference for smaller $\mu$ induced by the 
Higgs signal likelihood. The effect is thus less pronounced without the $c\tau$ cut (RH-side plot) than with the $c\tau$ cut (middle plot). Concretely, we find $p(\omhsq<0.14)\simeq 43\%$ (57\%) with (without) the $c\tau$ cut. 
The peak at high $\omhsq$ values is of course completely removed by the DMup constraints. The probability of lying within the Planck window defined by $\omhsq=0.119\pm 0.024$ ($0.024$ being the $2\sigma$ error, dominated by theory uncertainties) is, for all three of the above cases, $\sim 1.1\%$ with the $c\tau$ cut and $\sim 0.9\%$ without the $c\tau$ cut.

Turning to the predictions for direct dark matter detection, we observe that the preHiggs constraints limit the probability of having very small values of $\xi\sigsi$. This is true with and without the $c\tau$ cut, though the effect is larger with the $c\tau$ cut. The latter is due to the fact that the prompt chargino requirement removes the pure wino-LSP scenarios which have extremely small $\omhsq$ and $\xi\sigsi$ (recall that $\xi=\omhsq/0.119$). Requiring consistency with the Higgs signal strengths has only a small effect, somewhat preferring smaller values of $\xi\sigsi$ 
because of the larger LSP higgsino component. 

The 2D distributions of $\omhsq$ and $\xi\sigsi$ versus the $\tilde\chi^0_1$ mass are shown in Fig.~\ref{fig:darkmatter2D}. We observe that on the one hand the neutralino LSP can have mass up to 1 TeV at 95\% BC without conflicting with the DM constraints. Very low neutralino masses, on the other hand, are severely constrained by DM requirements. Note, moreover, that %$\omhsq\approx 0.1$ prefers LSP masses in the 150--500~GeV range. 
the most likely values lie around $m_{\tilde\chi^0_1}\approx200$--$300$~GeV, $\omhsq\approx 10^{-2}$ and $\xi\sigsi\approx 10^{-10}$~pb.

\FloatBarrier

%----------------------------------------------------------------------------------------------------------------------------------
\subsection{Consequences of future $h$ signal strength measurements}
%----------------------------------------------------------------------------------------------------------------------------------

\begin{figure}[h!]
\begin{center}
\includegraphics[width=0.3\linewidth]{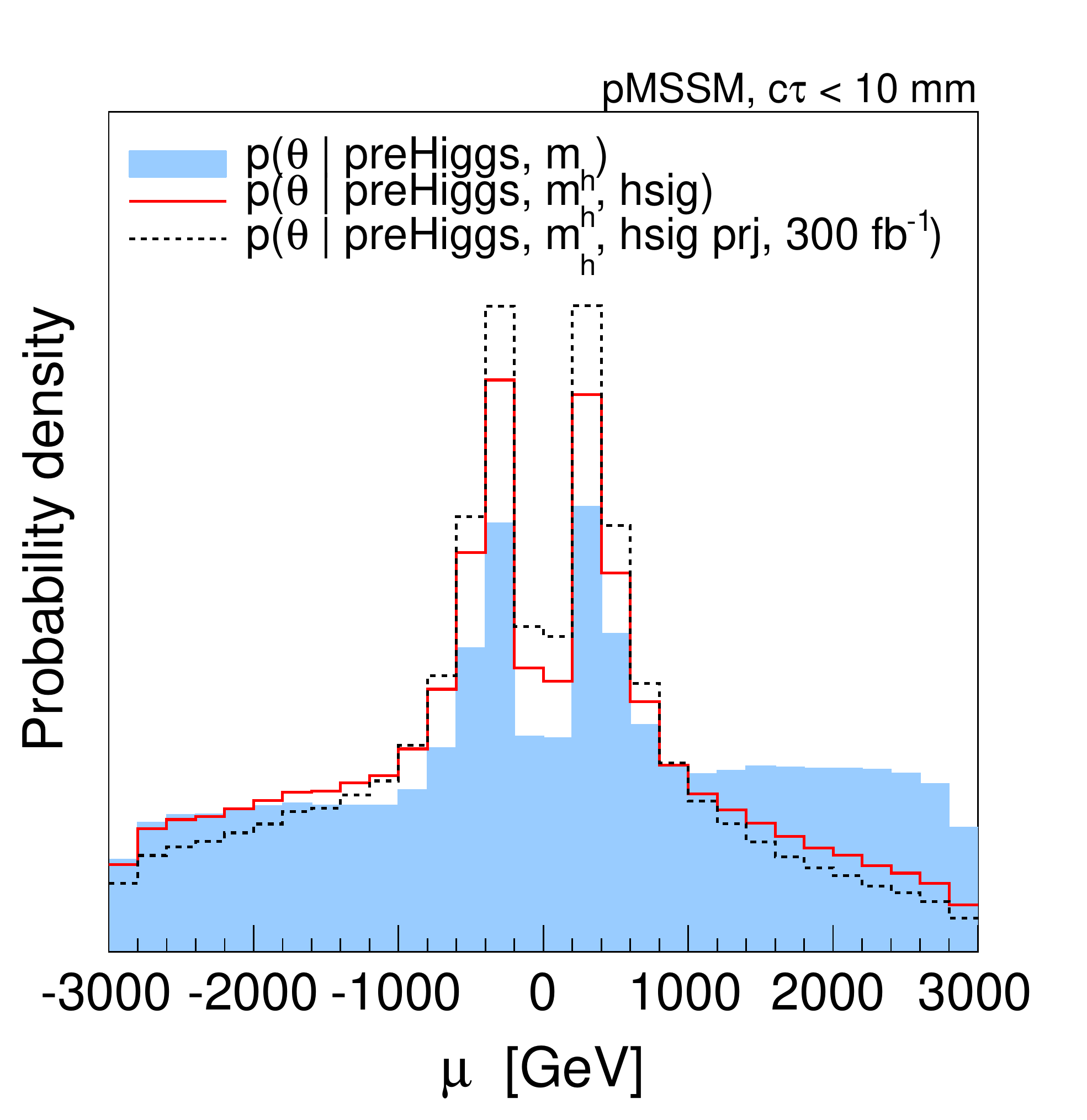}
\includegraphics[width=0.3\linewidth]{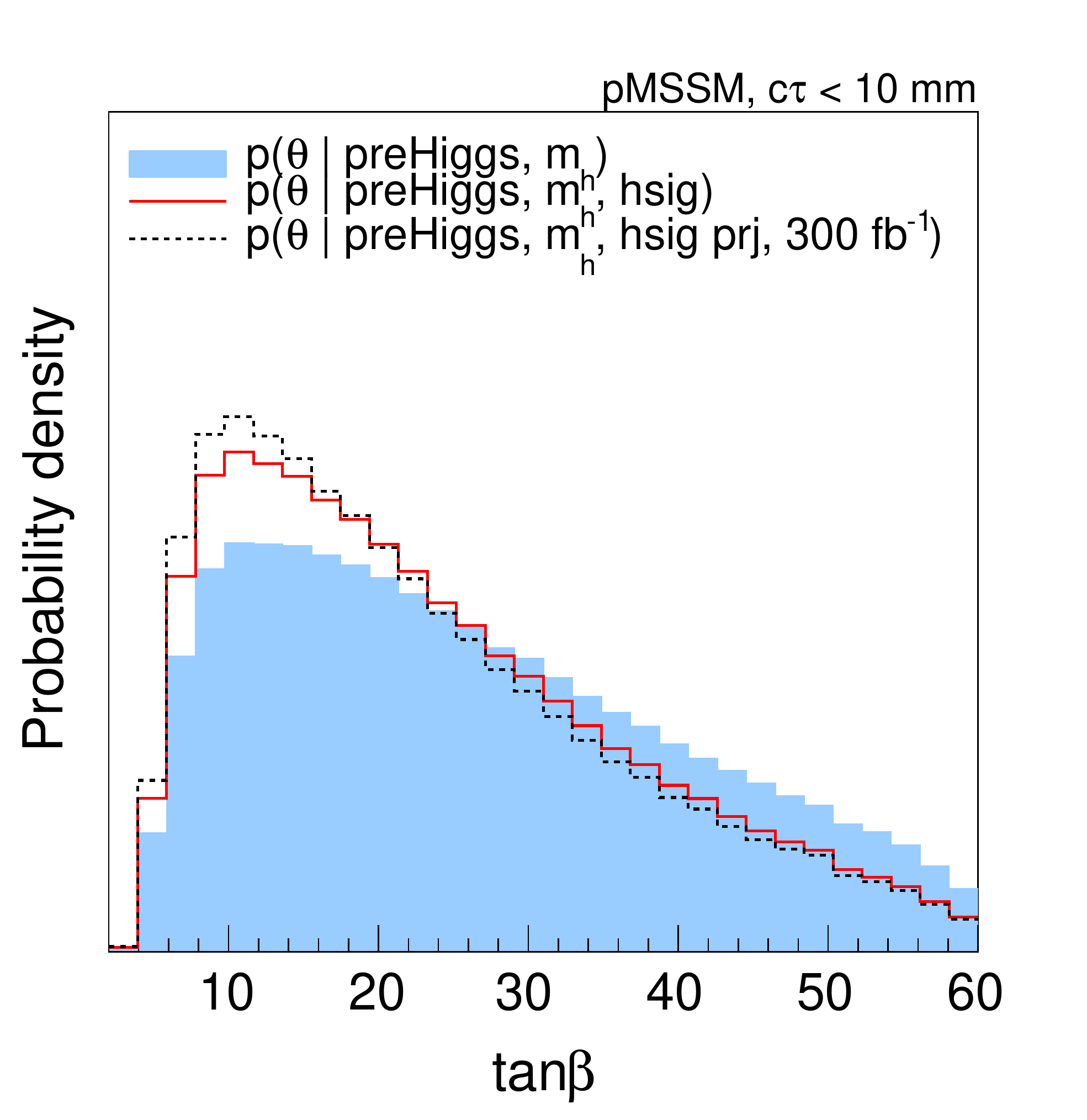}
\includegraphics[width=0.3\linewidth]{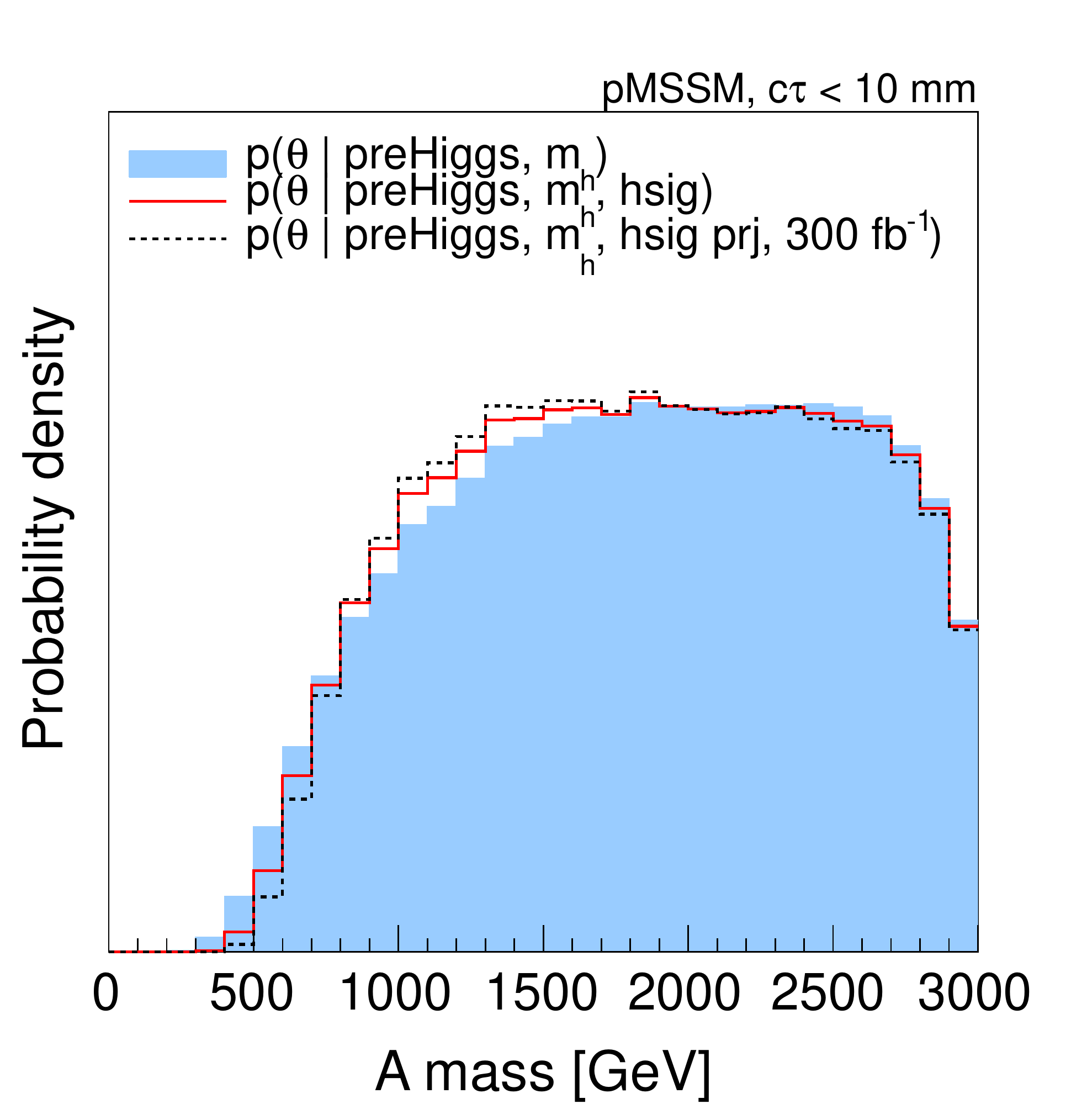}
\caption{Marginalized 1D posterior densities for some MSSM parameters, showing the effect of all $h$ signal strengths being $\approx 1$ with uncertainties as expected for 300$\fbi$ of data at 14~TeV, {\it cf.}\ Eq.~(\ref{eq:projection}).}
\label{fig:proj1d}
\end{center}
\end{figure}

It is also interesting to consider what happens if, with precision data at the next run of the LHC, 
the Higgs signal strengths have an even narrower probability distribution around unity. 
We estimate the precision attainable with 300$\fbi$ at 14~TeV based on \cite{ATLAS-collaboration:1484890,CMS-NOTE-2012-006}
\begin{eqnarray}
 &\mu({\rm ggF+ttH},\gamma\gamma) = 1 \pm 0.1\,,  \qquad 
    \mu({\rm VBF+VH},\gamma\gamma) = 1 \pm 0.3\,,  \nonumber \\
 &\mu({\rm ggF+ttH},VV) = 1 \pm 0.1\,, \qquad 
    \mu({\rm VBF+VH},VV) = 1 \pm 0.6\,,  \nonumber \\
 &\mu({\rm ggF+ttH},b\bar b) = 1 \pm 0.6\,, \qquad 
    \mu({\rm VBF+VH},b\bar b) = 1 \pm 0.2\,,  \nonumber  \\
 &\mu({\rm ggF+ttH},\tau\tau) = 1 \pm 0.2\,, \qquad 
    \mu({\rm VBF+VH},\tau\tau) = 1 \pm 0.2 \,.
\label{eq:projection}
\end{eqnarray}
%
%As illustrated in Fig.~\ref{fig:proj1d}, if the Higgs signal remains SM-like, the effect already observed on some SUSY parameters are only slightly strengthen by more precise measurements.
The effect of these hypothetical results is illustrated in Fig.~\ref{fig:proj1d}. We conclude that if the Higgs signal remains SM-like (but with smaller uncertainties), the effects already observed on some SUSY parameters are only slightly strengthened by more precise measurements.

\begin{figure}[t!]
\begin{center}
\includegraphics[width=0.3\linewidth]{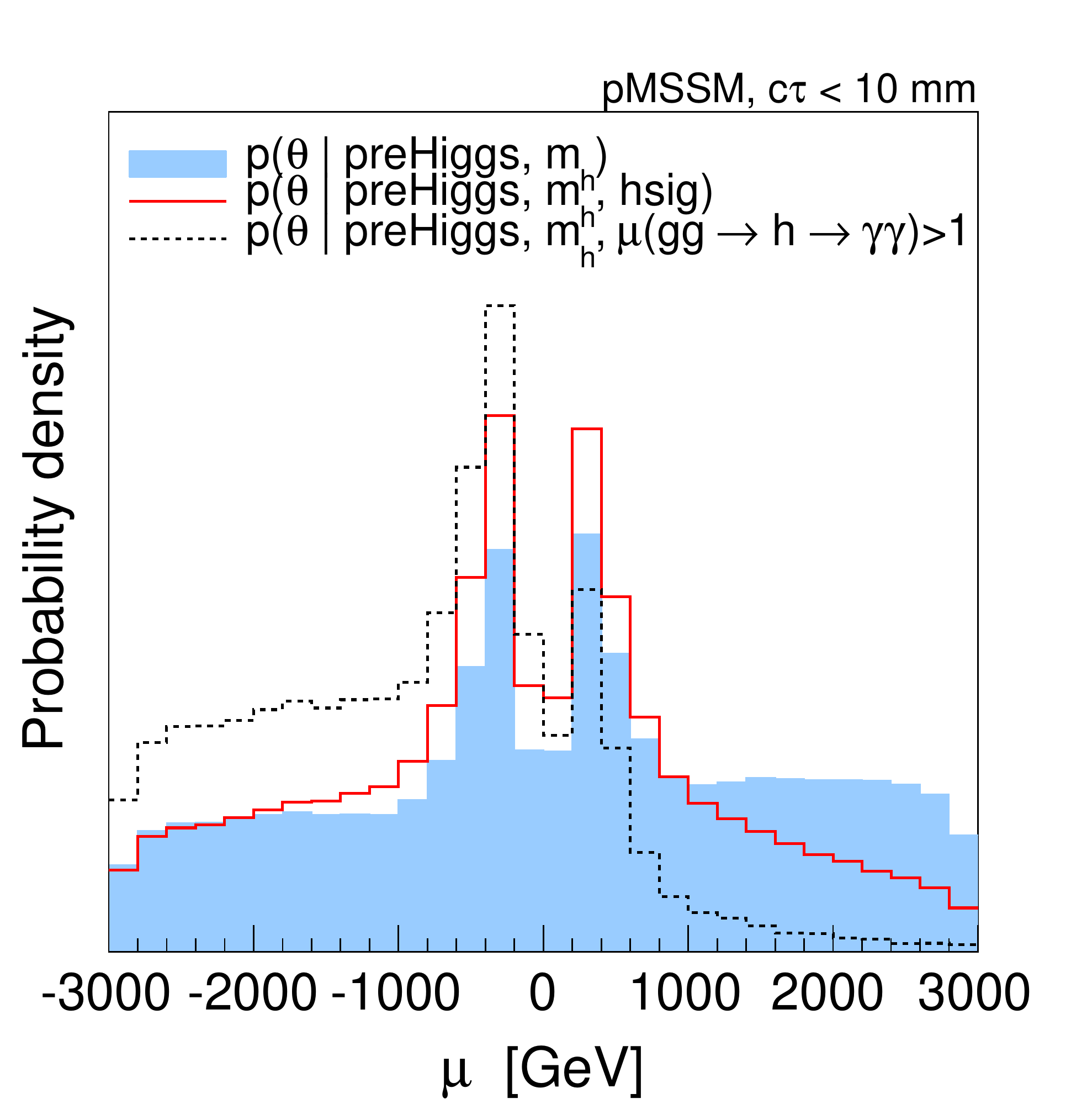}
\includegraphics[width=0.3\linewidth]{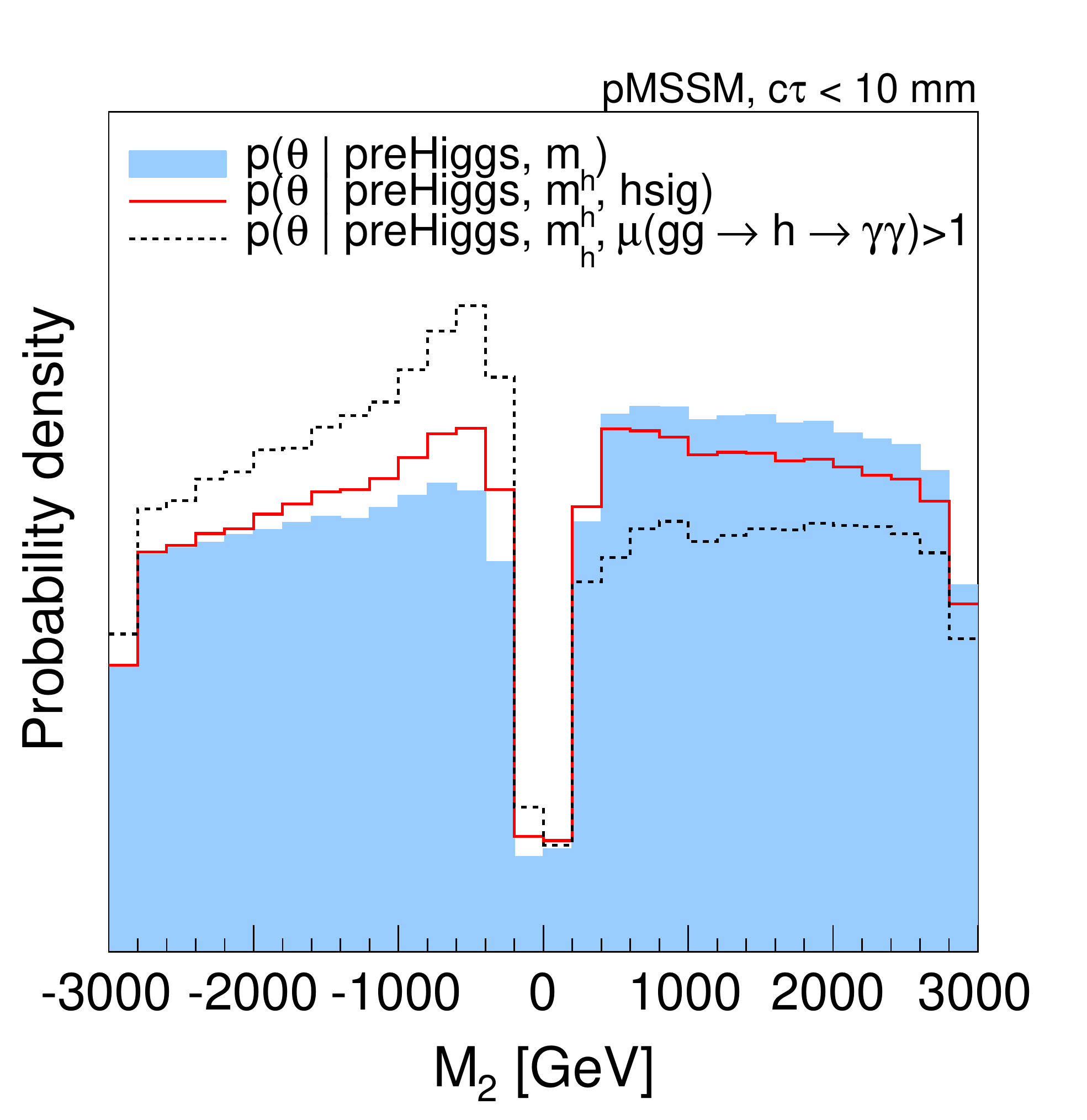}
\includegraphics[width=0.3\linewidth]{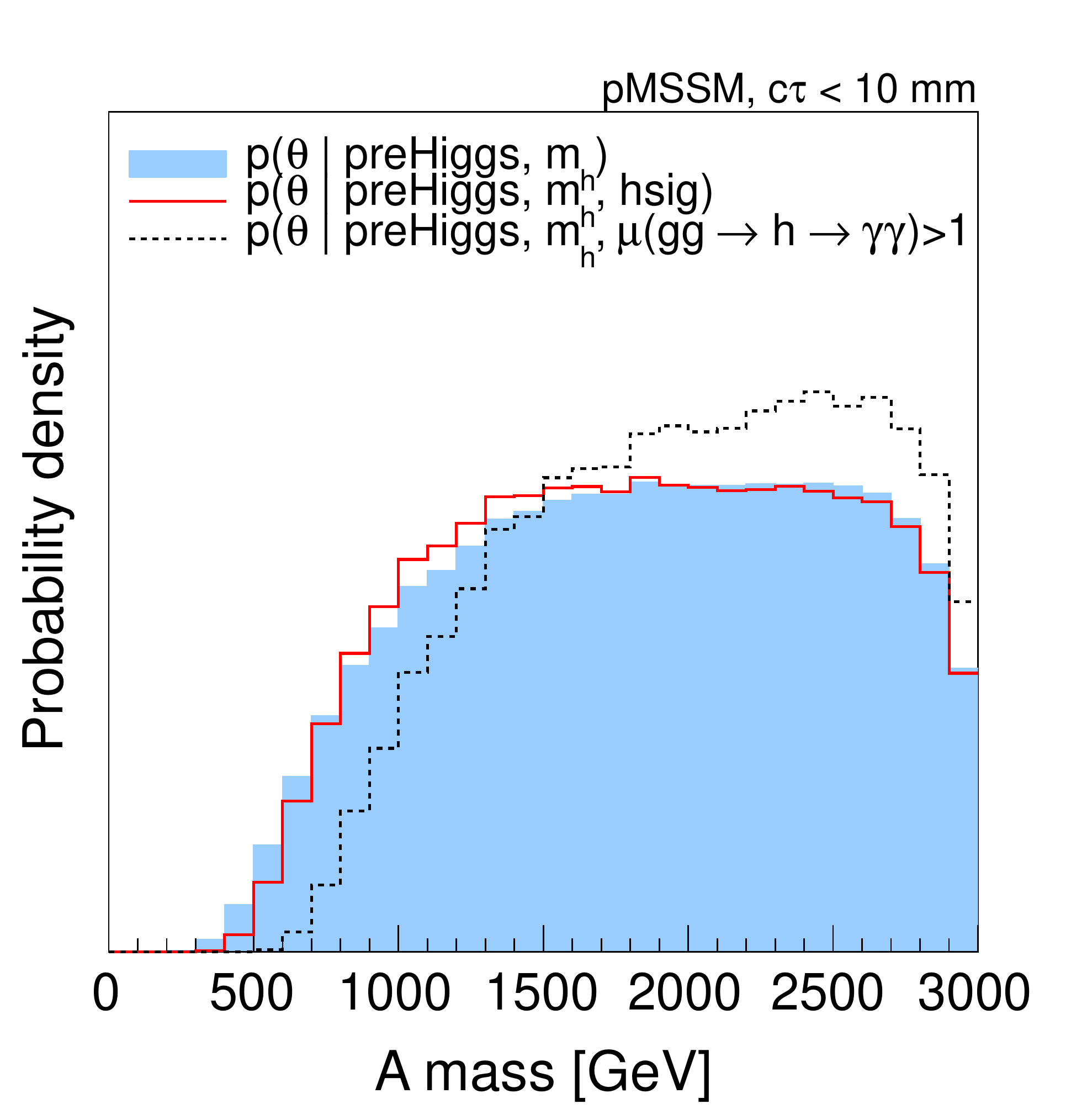}\\
\includegraphics[width=0.3\linewidth]{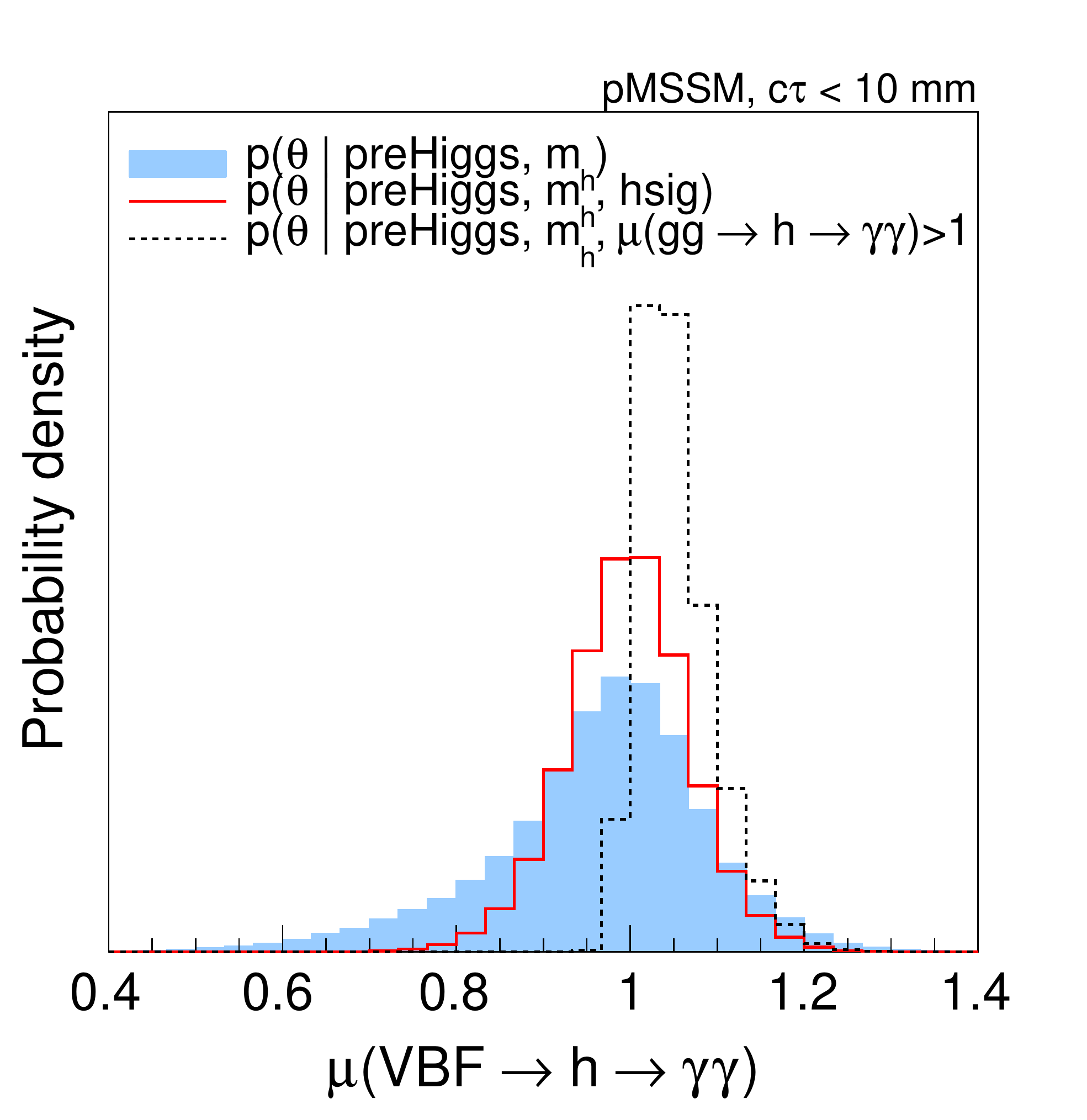}
\includegraphics[width=0.3\linewidth]{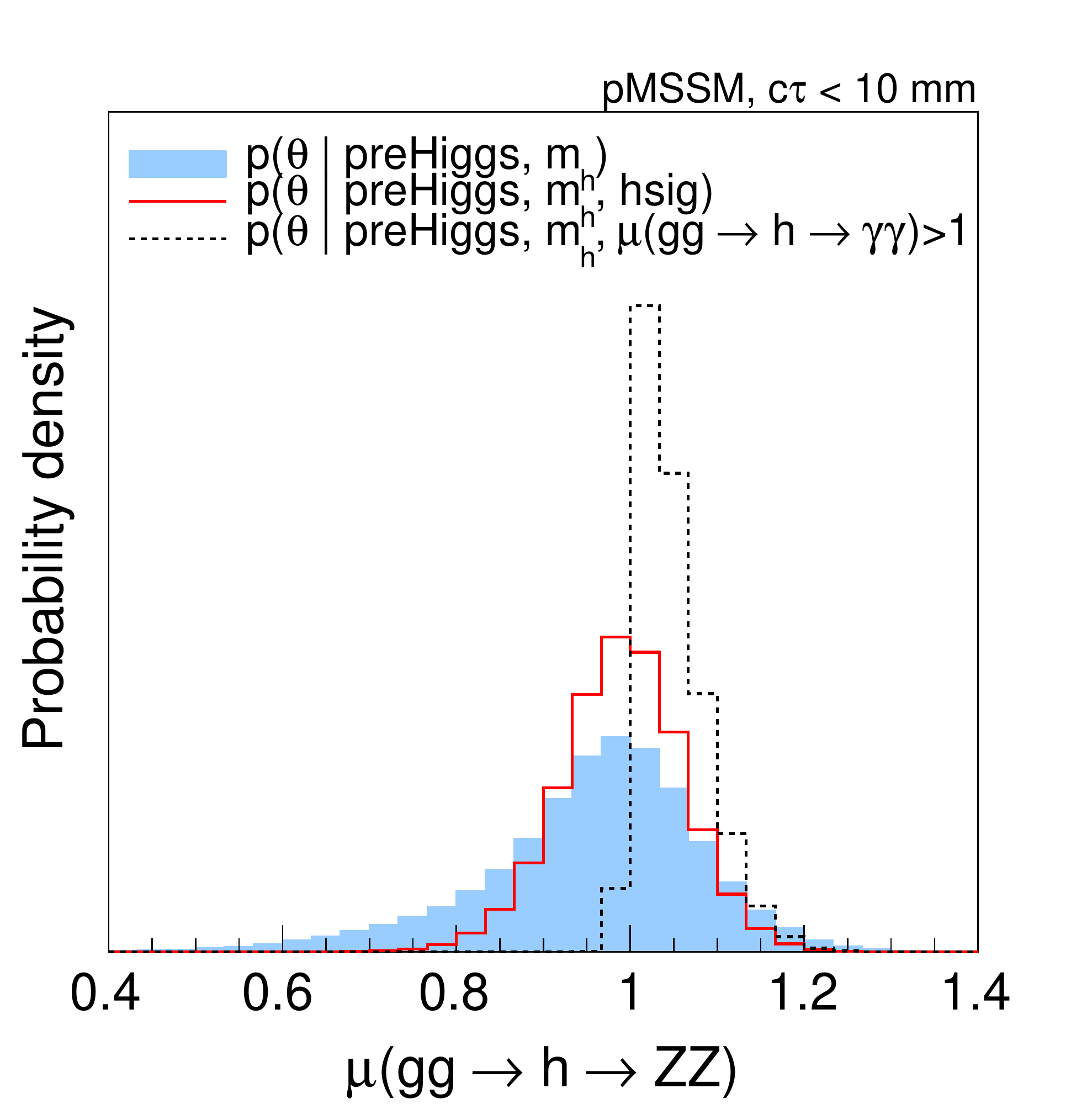}
\includegraphics[width=0.3\linewidth]{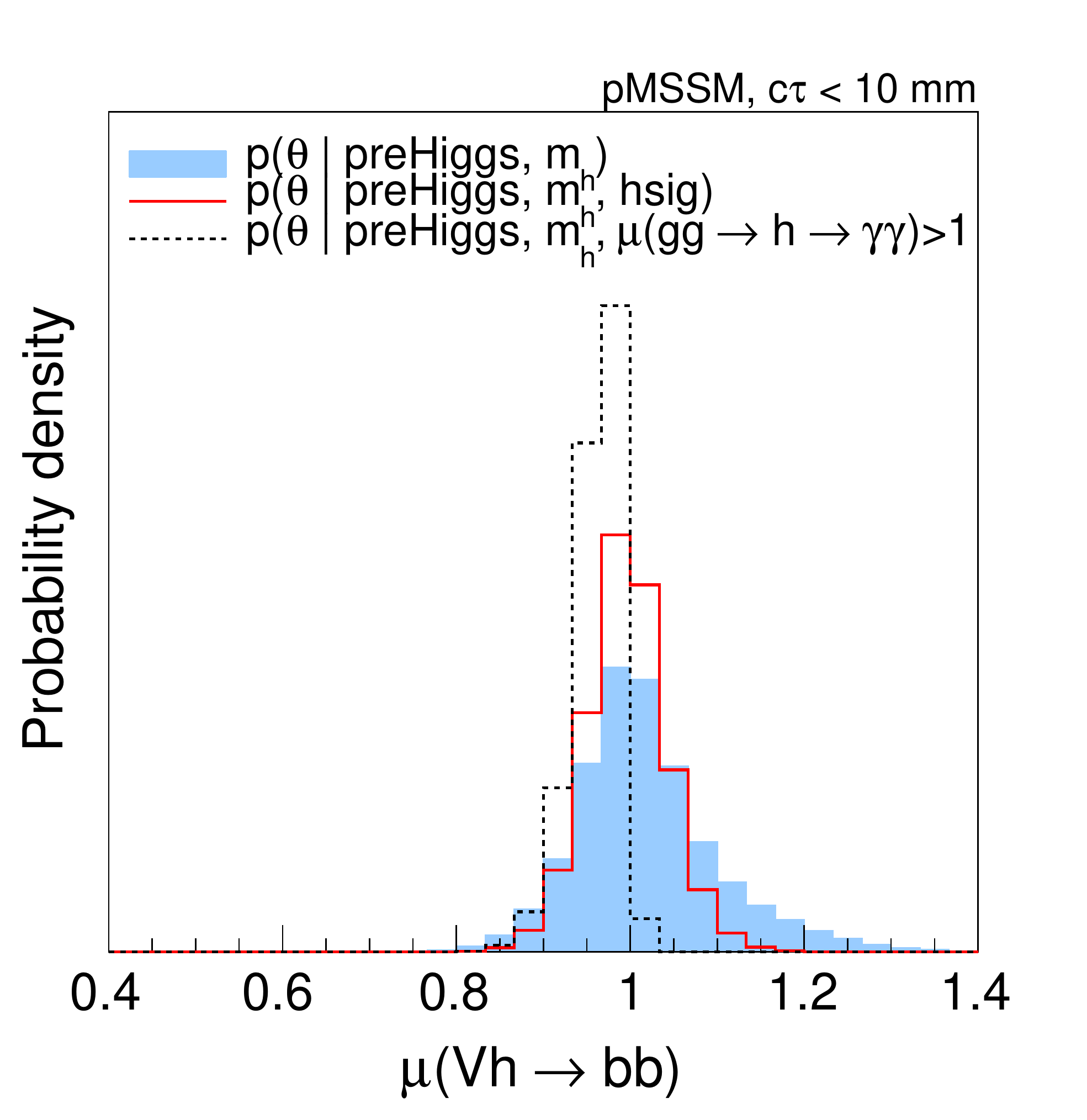}
\caption{Marginalized 1D posterior densities for  selected MSSM parameters and $h$ signal strengths, showing the effect of a hypothetical future determination of $\mu(gg\to h\to\gamma\gamma)>1$. }
\label{fig:mugt1}
\end{center}
\end{figure}

The picture is quite different should the signal strength finally turn out to be larger than one. For illustration, we assume  
$\mu({\rm ggF},\gamma\gamma)>1$ and show in Fig.~\ref{fig:mugt1} the impact on some other quantities.
As we have seen, $\Delta_b < 0$ corresponds to a suppression of $h \to b\bar b$ and, hence, to the enhancement of all other signal strengths. This is how one obtains $\mu({\rm ggF},\gamma\gamma)>1$ in our case. This leads to a strong preference for $\mu < 0$ and to an associated asymmetry for the  $M_2$ distribution. Moreover, strong evidence for $\mu({\rm ggF},\gamma\gamma)>1$ would strongly disfavor a CP-odd Higgs lying close to the current CMS bound because of the impact of $m_A$ on the tree-level coupling $hbb$.
Finally, $\mu({\rm ggF},\gamma\gamma)>1$ would also imply a preference for an enhancement of the diphoton signal in VBF production, as well as an enhancement of the $ZZ$ mode in both ggF and VBF. This is accompanied at the same time by the expected suppression of $Vh\to b\bar b$. Nonetheless, signal strength values close to 1 are still the most likely ones.

\FloatBarrier

%=======================================================================================
\section{Conclusions}
%=======================================================================================

We have performed a Bayesian analysis of the pMSSM taking into account the latest LHC results 
on the Higgs signal at $125.5$ GeV in addition to relevant low-energy observables and LEP constraints. 
We find that the requirement of obtaining the right $m_h$ strongly favors $|X_t/M_{\rm SUSY}|\approx 2$, 
\ie\ near-maximal (but not maximal) stop mixing.  Coincidently, such near-maximal mixing is also favored 
by naturalness arguments~\cite{Wymant:2012zp}. 

The constraints from the Higgs signal strengths in the various production$\times$decay modes, on the other hand, 
have an important influence on the posterior distributions of $\mu$ and $\tan\beta$, and hence on the electroweak-ino 
spectrum. 
Concretely, low values of $\mu$ and $\tan\beta\approx 10$ are favored. 
This is mainly due to radiative corrections to the bottom Yukawa coupling, which are proportional 
to $\mu\tan\beta$ and can significantly modify the total Higgs width. As a consequence, 
$\chitz$ and $\chipm$ masses below about 500~GeV are favored, as are LSPs with 
a significant higgsino fraction. 
While there is of course still a substantial tail at large masses, these results suggest that the Higgs data yield 
a certain preference for natural-SUSY-like scenarios. 

Regarding the heavy Higgs states, $H$ and $A$, we find that $m_{H,A}\gtrsim 500$~GeV mostly due to B-physics 
constraints. The 125 GeV Higgs data give only a small additional constraint; they mostly affect the heavy Higgses 
through their effect on $\tan\beta$. 
The limits from direct searches for $H,A\to \tau\tau$ at 7--8~TeV are less sensitive. 
If $m_A\lsim 1$~TeV, prospects for discovery of $H$ and $A$ at the next LHC run are substantial.  
Because $\tanb\gtrsim 10$ is preferred, we find that $b\anti b\to H,A$ typically dominates (by about a factor of 30) over  gluon fusion, with $\sigma(b\anti b \to H,A)\br(H,A\to \tau\tau)$ of the order of a few fb. 

 We have also explored the impact of DM limits associated with  $\omhsq$ and $\xi\sigsi$ on the Higgs bosons in the pMSSM context as well as the impact of the Higgs precision data on these same DM observables.  The most probable values for $\omhsq$ lie in the vicinity of $10^{-2}$, implying that DM would not consist entirely of the $\cnone$ 
(or that the missing abundance of $\cnone$ is substituted by non-thermal production). 
 The probability for obtaining $\omhsq$ within the Planck window is only of order 1\%: 
to get the correct annihilation rate, the $\cnone$ has to have a carefully balanced composition, 
or a  mass that is fine-tuned with respect to the $A$ or co-annihilating sparticles. Imposing the upper limit on $\omhsq$, we find $m_{\cnone}\in
[100,\,760]$~GeV and $\xi\sigsi\gtrsim 3.5\times 10^{-12}$~pb at 95\%
BC.

While we have not taken into account the recent LHC limits from direct SUSY searches, we have checked that 
our conclusions do not change when requiring gluino and squark masses above 1~TeV. The conclusions drawn 
from the Higgs sector are thus orthogonal to those from the SUSY searches. 
In particular, 
this makes our results directly comparable to the pMSSM interpretation of the CMS SUSY searches at 7--8~TeV~\cite{SUS-12-030,SUS-13-020}.

The 13--14~TeV run of the LHC will provide increased precision for Higgs measurements as well as 
a higher reach for SUSY particles.  
Particularly relevant in point of view of an interplay between Higgs and SUSY results is  
an improved sensitivity for higgsinos, gluinos and 3rd generation squarks.  
It will be interesting to see if a tension between Higgs results and SUSY limits arises or if there is a convergence as a result of the discovery of,  \eg,  light charginos and neutralinos.  Last but not least, if the Higgs boson is found in the end to have an enhanced $h\to \gam\gam$ rate compared to the SM, implications for $\mu$ and $M_2$ are substantial, $m_A$ is shifted to higher values and $\mu(Vh\to Vb\anti b)$ is suppressed --- allowing for some possibility of verifying consistency with or creating tension within the pMSSM.

%%%%%%%%%%%%%%%%%%%%%%%%%%%%%%%%%%%%%%%%%%%%%%%%%%%%%%%%%

\section{Acknowledgements}

We are deeply indebted to Sezen Sekmen for her extensive and crucial
contributions during the first part of the project. Without her, this
work would not have been possible. We strongly acknowledge her
collaboration.

This work was supported in part by the US DOE grant DE-FG03-91ER40674 and by IN2P3 under contract PICS FR--USA No.~5872. 
J.F.G. and S.K.\ also thank the Aspen Center for Physics, supported by the National Science Foundation under Grant No. PHYS-1066293, for hospitality and a great working atmosphere.

%%%%%%%%%%%%%%%%%%%%%%%%%%%%%%%%%%%%%%%%%%%%%%%%%%%%%%%%%

\providecommand{\href}[2]{#2}\begingroup\raggedright\endgroup

%%%%%%%%%%%%%%%%%%%%%%%%%%%%%%%%%%%%%%%%%%%%%%%%%%%%%%%%%

\end{document}